\documentclass[preprint,showpacs,showkeys,aps,prd,amsfonts,eqsecnum,nofootinbib]{revtex4} 
\usepackage{graphicx,epsfig,axodraw}

\begin{document}
\preprint{CUMQ/HEP 134}
%
%
\title{\Large $t \rightarrow cg,c\gamma,cZ$ IN THE
LEFT-RIGHT SUPERSYMMETRIC MODEL}
\author{Mariana Frank}\email[]{mfrank@vax2.concordia.ca}
\author{Ismail Turan}\email[]{ituran@physics.concordia.ca}
\affiliation{Department of Physics, Concordia University, 7141 Sherbrooke St.
West, Montreal, Quebec, CANADA H4B 1R6}
\date{\today}

\begin{abstract}
We analyse top quark flavor violating decays into a charm quark and a 
gluon, photon or $Z$ boson in a supersymmetric model incorporating 
left-right symmetry. We include loop calculations involving 
contributions from scalar quarks, gluinos, charginos and 
neutralinos. We perform the calculations first assuming the minimal 
(flavor diagonal) scalar quark scenario, and then allowing for 
arbitrary mixing between the second and third generation of scalar 
quarks, in both the up and down sectors.  In each case we present separately the contributions from gluino, chargino and neutralino loops and compare their respective strengths. In the flavor-diagonal case, the branching ratio cannot exceed $10^{-5}~(10^{-6})$ for the  gluon (photon/$Z$ boson); while for the unconstrained (flavor non-diagonal) case the same branching ratios can reach almost $10^{-4}$ for the gluon, $10^{-6}$ for the photon and $10^{-5}$ for the $Z$ boson, all of which are slightly below the expected reach of LHC.%
\end{abstract}
\vskip -2.5cm
\maketitle
\section{Introduction}\label{sec:intro}

Flavor changing interactions in general, and of top quark in 
particular, are a promising test ground for new physics.   In the 
Standard Model (SM) processes such as $t \rightarrow c g, \gamma, Z$ 
are absent at tree-level and highly suppressed by the Glashow-Iliopoulos-Maiani (GIM) mechanism 
at one loop. The branching ratios predicted in the SM \cite{eilam:91} are of order of 
$10^{-10}; 10^{-12}; 10^{-13}-10^{-12}$ for the decays $t \rightarrow c g; c\gamma; cZ$
and thus far from present and future reaches of 
$e^+e^-$ colliders or the LHC. Models beyond the SM predict branching 
ratios which are orders of magnitude larger than this, and thus an 
experimental signal can be interpreted as a signal for new physics. 
An added bonus of top quark phenomenology is that it provides a window into the electroweak 
breaking mechanism, since the top mass is close to the scale of symmetry breaking.  

 Previous analyses have explored several possibilities of enhancing 
flavor violating decays, mostly within supersymmetry. $t \rightarrow c V ~(V=g,~\gamma,~Z)$
have also been explored in two Higgs doublet models \cite{eilam:91,two higgs}, technicolor models \cite{technicolor}, top-color assisted technicolor models \cite{toptechnicolor}, models with extra vector singlets \cite{vectors} and supersymmetry with \cite {MSSM,couture:1995,Couture:1997ep,Lopez:1997xv,nonuniversal,Liu:2004qw,delepine}, and without \cite{not R}, R parity.
While supersymmetry 
appears to provide the most convincing scenario for new physics, most 
analyses concentrate on the minimal supersymmetric standard model 
(MSSM). Analyses have been performed initially in the constrained 
MSSM, where flavor changing interactions are driven by flavor 
universal supersymmetry breaking soft terms including 
left, right and intergenerational scalar quark mixing. Li {\it et al.} evaluated one-loop SUSY QCD and EW contributions \cite{MSSM} and obtained branching ratios of $10^{-6}$ for $t \rightarrow c g$ and $10^{-8}$ for $t \rightarrow c \gamma, Z$. Couture {\it et al.} included  the left-handed squark mixing in their calculation \cite{couture:1995} and obtained branching ratios of  $10^{-5}$ for the gluon,  $10^{-7}$ for the photon and  $10^{-6}$ for the $Z$ boson. Inclusion of the right-handed squark mixing  \cite{Couture:1997ep},  did not change branching ratios significantly. Lopez {\it et al.} further refined the calculations in the flavor diagonal case by including contributions from the neutralino quark-squark loops; in this case the branching ratios obtained were $10^{-5}$ for $t \rightarrow  c g$ and $10^{-7}$ for $t \rightarrow c \gamma, Z$. In the non-universal case several authors \cite{nonuniversal} found branching ratios of $10^{-5}$ for $t \rightarrow  c g$ and $10^{-6}$ for $t \rightarrow c \gamma, Z$.  

In  the unconstrained MSSM, assumptions about universality of soft 
supersymmetry breaking terms and new sources of flavor violation are 
included in the scalar quark matrices. Liu {\it et al.} looked at the fully unconstrained MSSM and found decay branching ratios which can reach $10^{-4}, 10^{-6}$ for $t \rightarrow c g$,  and $t \rightarrow c \gamma, Z$ respectively; while Del\'{e}pine and Khalil \cite {delepine} found, for the unconstrained MSSM with a light stop, branching ratios of $10^{-5}$ for the gluon decay and $10^{-6}$ for the photon. Note that not all authors use the same values for relevant parameters. Experimental limits on SUSY masses have decome more stringent, in particular, the gluino mass is taken to be higher in later papers, which effects the estimated branching ratios.

 However, not much work has been done on 
flavor changing interactions in the top quark decays in 
scenarios beyond the MSSM. We present here such an analysis, based on 
extending the SM group $SU(3)_C \times SU(2)_L \times U(1)_Y$ to a 
left-right symmetric group $SU(3)_C \times SU(2)_L \times SU(2)_R 
\times U(1)_{B-L}$ (LRSUSY). This gauge symmetry allows for the seesaw 
mechanism within a supersymmetric scenario and predicts neutrino masses and 
mixing naturally. Flavor violation in $b$ decays have shown possible 
enhancement of results over the the MSSM ones: in particular 
restrictions on intergenerational left-left (LL), right-right (RR), and 
left-right (LR) mixing of scalar quarks is more restrictive than in the 
case of MSSM \cite{Frank:2002nj}.
 
 In this work, we present a full investigation of the 
flavor changing two body decays of the top quark in the LRSUSY. We investigate 
both the constrained case, in which the only source of flavor 
violation comes from the Cabibbo-Kobayashi-Maskawa  (CKM)  matrix in the quark sector (with the 
proviso that we assume it to be the same for left- and right-handed 
quarks), as well as in the unconstrained model, in which soft 
symmetry breaking parameters are allowed to induce flavor-dependent 
mixing in the squark mass matrix. Such a flavor dependent scenario is 
motivated by neutrino oscillations, which indicate the presence of 
two large mixing angles (between the first and second, and second and 
third generations). If LRSUSY occurs in nature as an intermediate 
symmetry in breaking down from a SUSY GUT scenario, such as $SO(10)$, 
large neutrino mixing angles in the lepton mass matrices may appear 
not in the down quark mass matrices (because the right-handed charged 
current interaction is broken down at the unification scale), but in 
the squark mass matrices \cite{Chang:2002mq}. The new mixing might affect $b$ \cite{Harnik:2002vs} as well as 
$t$ flavor physics.

Previous studies of unconstrained supersymmetric models have been based on either the mass insertion \cite{Hall:1985dx} or on the mass eigenstate method \cite{Harnik:2002vs,Besmer:2001cj}. In the mass insertion framework \cite{Hall:1985dx} one chooses a basis for fermion and sfermion states in which all the couplings of these particles to neutral gauginos are flavor diagonal. Flavor changes in the squark sector arise from the non-diagonality of the squark propagators.
 Off-diagonal elements mix squark flavors for both left- and right-handed squarks.  In the mass eigenstate method \cite{Besmer:2001cj}, squark mass matrices are given in the super-CKM basis, and are diagonalized by rotating the superfields.  In this basis, the up-squark and down-squark mass matrices are correlated by this rotation and thus not independent. Potential new sources of flavor violation arise from couplings of quarks and squarks to gauginos. This method has the advantage that, when the off-diagonal elements in the squark mass matrices become large, the method is still valid, unlike the mass insertion which is a perturbation-based expansion. We will return to these considerations when performing specific evaluations, and chose the mass eigenstate method for its greater flexibility.
 
Our paper is organized as follows: In Section \ref{sec:model} we present 
the general framework of the left-right supersymmetric model. 
The squark sector is then 
discussed in Section \ref{sec:FCLRSUSY} in both the flavor diagonal
and non-diagonal scenarios, concentrating especially on the flavor changing 
mixing between the second and third generations. In Section \ref{sec:tcV}, we 
give the full analytical expressions contributing to $t\to c\,V~(V=g,~\gamma,~Z)$ 
decays by providing expressions for the gluino, chargino and neutralino contributions separately. 
The numerical analysis of the decays is given in Section \ref{sec:Numreslt} where we  
concentrate on individual and relative contributions of gluinos, charginos and 
neutralinos to the branching ratios of the decays. The discussion is carried out in 
the flavor diagonal and non-diagonal scenarios.
Section \ref{sec:conclusion} is devoted to our summary and conclusion. The chargino and neutralino mass 
matrices and their diagonalization procedures are summarized in Appendix \ref{sec:AppendixMasses}. 
Appendix \ref{sec:AppendixFynmVert} includes the relevant Feynman rules.

\section{\bf Description of the LRSUSY Model}\label{sec:model}

The minimal supersymmetric left-right model is based on the gauge group
$SU(3)_C \times SU(2)_L \times SU(2)_R \times U(1)_{B-L}$ \cite{history, frank1}. The matter
fields of this model consist of three families of quark and lepton chiral
superfields with the following transformations under the gauge group:
\begin{eqnarray}
Q_L^i &=&\left( \begin{array}{c} u_L^i \\ d_L^i \end{array} \right)  \sim \left ( 3, 2, 1, 1/3 \right ),~~
Q_R^i =\left( \begin{array}{c} d_R^i \\ u_R^i \end{array} \right) \sim \left ( 3,1, 2, -1/3 \right ),\nonumber \\
L_L^i& = &\left( \begin{array}{c} \nu_L^i \\ e_L^i \end{array} \right) \sim\left( 1,2, 1, -1 \right),~~
L_R^i= \left( \begin{array}{c} e_R^i \\ \nu_R^i \end{array} \right) \sim \left ( 1,1, 2, 1 \right ),
\end{eqnarray}
where the numbers in the brackets represent the quantum numbers under
$SU(3)_C \times SU(2)_L \times SU(2)_R \times U(1)_{B-L}$. The Higgs
sector consists of bidoublet and triplet Higgs superfields:
\begin{eqnarray}
\displaystyle
\Phi_i &&\hspace*{-0.3cm}= \left( \begin{array}{cc} \phi^0_{1i} &\phi^+_{2i} \\ \phi^-_{1i} & \phi^0_{2i} \end{array} \right) \;\;\;\;\;\; \sim \left (1,2,2,0 \right), ~~~  \mbox { }(i=u,d), \nonumber \\
\Delta_{L} &&\hspace*{-0.3cm} = \left(\begin{array}{cc}
\frac {\Delta_L^-}{\sqrt{2}}&\Delta_L^0\\
\Delta_{L}^{--}&-\frac{\Delta_L^-}{\sqrt{2}}
\end{array}\right) \sim (1,3,1,-2),~~~\delta_{L}  =
\left(\begin{array}{cc}
\frac {\delta_L^+}{\sqrt{2}}&\delta_L^{++}\\
\delta_{L}^{0}&-\frac{\delta_L^+}{\sqrt{2}}
\end{array}\right) \sim (1,3,1,2),\nonumber \\
\Delta_{R} &&\hspace*{-0.3cm} =
\left(\begin{array}{cc}
\frac {\Delta_R^-}{\sqrt{2}}&\Delta_R^0\\
\Delta_{R}^{--}&-\frac{\Delta_R^-}{\sqrt{2}}
\end{array}\right) \sim (1,1,3,-2),~~~\delta_{R}  =
\left(\begin{array}{cc}
\frac {\delta_R^+}{\sqrt{2}}&\delta_R^{++}\\
\delta_{R}^{0}&-\frac{\delta_R^+}{\sqrt{2}}
\end{array}\right) \sim (1,1,3,2).
\end{eqnarray}
The bidoublet  Higgs superfields appear in all LRSUSY  to
implement the $SU(2)_L \times U(1)_{Y}$ symmetry breaking and to generate
a CKM mixing matrix. Supplementary Higgs
representations are needed to break left-right symmetry spontaneously:
either doublets or triplets would achieve this, but triplet Higgs bosons
 $\Delta_L$ and $\Delta_R$ are chosen to facilitate the seesaw
mechanism \cite{mohapatra}. In supersymmetry, additional
triplet superfields $\delta_L, ~\delta_R$ are introduced to cancel triangle
gauge anomalies in the fermionic sector. The most general superpotential
involving these superfields in LRSUSY is:
\begin{eqnarray}
\label{superpotential}
W & = & {\bf Y}_{Q}^{(i)} Q_L^T\Phi_{i}i \tau_{2}Q_R + {\bf Y}_{L}^{(i)}
L_L^T \Phi_{i}i \tau_{2}L_R + {\bf Y}_{LR}(L_L^T i \tau_{2} \delta_L L_L +
L_R^{T} i \tau_{2}
\Delta_R L_R) \nonumber \\
& & + \mu_{LR}\left [Tr (\Delta_L  \delta_L +\Delta_R
\delta_R)\right] + \mu_{ij}Tr(i\tau_{2}\Phi^{T}_{i} i\tau_{2} \Phi_{j})
+W_{NR}
\end{eqnarray}
where ${\bf Y}_Q$ and  ${\bf Y}_L$ are
the Yukawa couplings
for the quarks  and leptons, respectively and ${\bf Y}_{LR}$ is the coupling for the triplet Higgs bosons.  The parameters $\mu_{ij}$ and $\mu_{LR}$ are the Higgs mass parameters. Left right symmetry requires all ${\bf Y}$-matrices to be Hermitean in the generation space  and  ${\bf Y}_{LR}$ matrix to be symmetric. 
Here $W_{NR}$ denotes (possible) non-renormalizable terms arising 
from higher scale
physics or Planck scale effects \cite{recmohapatra}. The presence of these terms
insures that, when the SUSY breaking scale is above $M_{W_{R}}$, the
ground state is R-parity conserving. In addition, the
Lagrangian also includes soft supersymmetry breaking terms as well as  $F$- and $D$-terms:
\begin{eqnarray}
\label{eq:soft}
{\cal L}_{\rm soft}&=&\left[ {\bf A}_{Q}^{i}{\bf Y}_{Q}^{(i)}{\tilde Q}_L^T\Phi_{i}
i\tau_{2}{\tilde Q}_R+ {\bf A}_{L}^{i}{\bf Y}_{L}^{(i)}{\tilde L}_L^T \Phi_{i}
i\tau_{2}{\tilde L}_R + {\bf A}_{LR} {\bf Y}_{LR}({\tilde L}_L^T i \tau_{2}
\delta_L{\tilde  L}_L + {\tilde L}_R^{T}i\tau_{2} \Delta_R{\tilde L}_R)
\right.
\nonumber
\\  & +&\left.{ m}_{\Phi}^{(ij) 2}
\Phi_i^{\dagger}  \Phi_j \right] + \left[( m_{L_L}^2)_{ij}{\tilde 
L}_{Li}^{\dagger}{\tilde
L}_{Lj}+ (m_{L_R}^2)_{ij}{\tilde L}_{Ri}^{ \dagger}{\tilde L}_{Rj} 
\right]- M_{LR}^2 \left[
Tr(  \Delta_R  \delta_R)+ Tr(  \Delta_L
  \delta_L) \right.\nonumber \\
&+ &\left. {\rm H.c.}\right] - [B \mu_{ij} \Phi_{i} \Phi_{j}+{\rm H.c.}] +\left[( m_{Q_L}^2)_{ij}{\tilde 
Q}_{Li}^{\dagger}{\tilde
Q}_{Lj}+ (m_{Q_R}^2)_{ij}{\tilde Q}_{Ri}^{\dagger}{\tilde Q}_{Rj} 
\right]
\end{eqnarray}
where ${\bf A}_{Q}, {\bf A}_{L}, {\bf A}_{LR}$ are trilinear scalar couplings.
The LR symmetry is broken spontaneously to
$U(1)_{em}$ through
non-zero vacuum  expectation values $(\rm VEV's)$ of the Higgs fields. These values are:
\begin{eqnarray}
\langle \Phi_u \rangle&&\hspace*{-0.3cm} = \left (\begin{array}{cc}
\kappa_u&0\\0&0
\end{array}\right),\hspace*{1cm}\langle \Phi_d \rangle = \left (\begin{array}{cc}
0&0\\0&\kappa_d
\end{array}\right), \hspace*{0.9cm}\langle \Delta_{L} \rangle = 0, \nonumber \\
\langle \,\delta_{L} \rangle&&\hspace*{-0.3cm} = 0
~,\hspace*{2.1cm}~\langle \,\Delta_{R} \rangle = \left
(\begin{array}{cc} 0&v_{\Delta_R}\\0&0
\end{array}\right),\hspace*{0.6cm} ~\langle \,\delta_{R} \rangle = \left
(\begin{array}{cc} 0&0\\v_{\delta_R}&0
\end{array}\right).
\nonumber
\end{eqnarray}
where we already set to zero the $CP$-violating
phase in the mixing of $W_{L}$ and $W_{R}$. The non-zero 
Higgs $\rm VEV's$ break both parity and $SU(2)_{R}$.
In the first stage of breaking, the right-handed gauge bosons, $W_{R}$ and
$Z_{R}$ acquire masses proportional to $v_{\Delta _R}, v_{\delta_R}$ and become much heavier
than the SM (left-handed) gauge bosons $W_{L}$ and $Z_{L}$, which in turn pick
up masses
proportional to $\kappa_{u}$ and $\kappa_{d}$ at the second stage of
breaking.

Fermionic partners of gauge and Higgs bosons mix. In LRSUSY there are six 
singly-charged
charginos, corresponding to $\tilde\lambda_{L}$,
$\tilde\lambda_{R}$, $\tilde\phi_{u}$,
$\tilde\phi_{d}$, $\tilde\Delta_{L}$, and
$\tilde\Delta_{R}$.
The model also has eleven neutralinos, corresponding to
$\tilde\lambda_{Z}$,
$\tilde\lambda_{Z^{\prime}}$,
$\tilde\lambda_{V}$,  $\tilde\phi_{1u}^0$, $\tilde\phi_{2u}^0$,
$\tilde\phi_{1d}^0$,  $\tilde\phi_{2d}^0$, $\tilde\Delta_{L}^0$,
$\tilde\Delta_{R}^0$,  $\tilde\delta_{L}^0$, and
$\tilde\delta_{R}^0$.  The doubly charged Higgs
and Higgsinos do not
affect quark phenomenology, but the neutral and singly charged
components do, through
mixings in the chargino and neutralino mass matrices.

The supersymmetric sources of flavor violation of interest in $t$ decays in the LRSUSY model
come from either the
Yukawa potential or the trilinear scalar coupling.

The interaction of fermions with scalar (Higgs) fields relevant for the quark sector has  the following
form
\begin{eqnarray}
\label{eq:yukawa}
{\cal L}_Y&=& {\bf Y}_Q^u\bar{Q}_L \Phi_u Q_R + {\bf Y}_Q^d \bar{Q}_L \Phi_d
Q_R\,+\,{\rm H.c.}
\end{eqnarray}
We present below a detailed analysis of flavor violation in the model in both the flavor-diagonal and non-diagonal case, before proceeding with calculation of the branching ratio
of $t\rightarrow c V$.

\section{\bf Flavor changing in the LRSUSY }\label{sec:FCLRSUSY}

When a basis, called the super-CKM, in which the quark states are diagonal, is used to 
express the squark mass matrix, flavor non-diagonal entries naturally arise. So, our 
analysis is focused on the effects of such scenarios in rare top quark decays in 
comparison with the flavor diagonal case. 

In the interaction (flavor) basis, $(\tilde{Q}_L^{i}, \tilde{Q}_R^{i})$, the
squared-mass matrix for squarks is
\begin{equation}
{\cal M}_f^2= \left( \begin{array}{cc}
                              m_{f\,LL}^2+F_{f\,LL}+D_{f\,LL} &
(m_{f\,LR}^2)+F_{f\,LR} \\
                              (m_{f\,LR}^2)^{\dagger}+F_{f\,RL} &
m_{f\,RR}^2+F_{f\,RR}+D_{f\,RR}
                         \end{array}
                  \right).
\end{equation}
The F-terms are diagonal in the flavor space, $F_{f\,LL,f\, RR}=m_f^2$,
$(F_{d\,LR})_{ij}=\mu
(m_{d_i} \tan \beta ) {\bf 1}_{ij}$,  $(F_{u \, LR})_{ij}=\mu
(m_{u_i} \cot \beta ) {\bf 1}_{ij}$, where $\tan \beta=\kappa_u/\kappa_d$ is defined.
The D-terms are also flavor-diagonal
\begin{eqnarray}
D_{f \, LL}&=&m_Z^2 \cos 2 \beta (T_{3f}-Q_f \sin^2 \theta_{\rm W}) {\bf 1}_{3\times3},
\nonumber \\
D_{f \, RR}&=&m_Z^2 \cos 2 \beta Q_f \sin^2 \theta_{\rm W} {\bf 1}_{3\times3}.
\end{eqnarray}

In the universal case, one has $(m_{f\,LL,f\,RR}^2)_{ij}=m^2_{\tilde{Q}_{L,R}} \delta_{ij}$, $m_{f
\, LR}^2=A_f^* m_f$. 
To reduce the number of free parameters, we also consider the following parameters
to be universal, 
$(m^2_{\tilde{Q}_{L,R}})_{ij}={\rm M_{susy}^2} \delta_{ij}$, $A_{d,ij}=A
\delta_{ij}$ and $A_{u,ij}=A \delta_{ij}$. Note that due to the invariance under both $SU(2)_L$ and $SU(2)_R$, 
$m_{u\,LL,u\,RR}^2$ in the up sector can not be specified independently from the matrix $m_{d\,LL,d\,RR}^2$, in the down sector. 
They are related as $m_{u\,LL,u\,RR}^2 = K_{\rm CKM} \left(m_{d\,LL,d\,RR}^2\right) K_{\rm CKM}^{\dagger}$, where $K_{\rm CKM}$ is the CKM matrix.\footnote{For simplicity, we assume $K_{\rm CKM}^L=K_{\rm CKM}^R$ which is a conservative choice and doesn't require new mixing angles in the right-handed quark matrices \cite{Frank:2002nj}.}

The mass(-squared) matrix in the universal case for the U-type squarks then reduces to, in block form
\begin{eqnarray}
{\cal M}_{U_k}^2= \left( \begin{array}{cc}
{\rm M_{susy}^2}+ m_{u_k}^2+m_Z^2(T_{3u}-Q_u \sin^2 \theta_{\rm W}) \cos 2\beta & m_{u_k} (A+\mu \cot \beta) \\
m_{u_k} (A+\mu \cot \beta)          &\hspace*{-0.6cm}         {\rm M_{\rm susy}^2}+m_{u_k}^2++m_Z^2 Q_u \sin^2 \theta_{\rm W} \cos 2 \beta
                         \end{array}
                  \right),\nonumber
\end{eqnarray}
and, for the D-type squarks, to
\begin{eqnarray}
{\cal M}_{D_k}^2= \left( \begin{array}{cc}
{\rm M_{\rm susy}^2}+ m_{d_k}^2+m_Z^2(T_{3d}-Q_d \sin^2 \theta_{\rm W}) \cos 2\beta & m_{d_k} (A+\mu \tan \beta)\\
m_{d_k} (A+\mu \tan \beta) &\hspace*{-0.6cm} {\rm M_{\rm susy}^2}+ m_{d_k}^2+m_Z^2 Q_d \sin^2 \theta_{\rm W} \cos2 \beta
                         \end{array}
                  \right).\nonumber
\end{eqnarray}
The corresponding mass eigenstates are defined as
\begin{equation}
\left( \begin{array}{c}{\tilde Q}_{L}\\
{\tilde Q}_{R}\end{array}\right)
=\left( \begin{array}{c}\Gamma^{\dagger}_{Q\,L}\\
\Gamma^{\dagger}_{Q\,R}\end{array}\right)\tilde{q},\label{flavortomass}
\end{equation}
where $\Gamma_{Q\,L,Q\,R}$ are $6\times3$ mixing matrices and $ \tilde{q}$ is a $6\times1$ column vector.

These up- and down squark mass  matrices are  $6\times 6$,  but are written above in $2\times 2$ block form. 
In the flavor diagonal  scenario (constrained LRSUSY), each of these blocks has no non-zero off-diagonal elements.
That is, there is no intergenerational mixings for squarks and the only source of flavor mixing comes from the CKM
matrix. 

The way to induce off-diagonal entries radiatively is to consider the evolution of the squark and quark 
masses from the SUSY-breaking scale (where both quark and squark mass matrices are flavor diagonal in 
the same basis), down to the electroweak scale via renormalization group equations. Experimental bounds from 
$D^0-\bar{D}^0$ and $K^0-\bar{K}^0$ data \cite{Gabbiani:1996hi} involving the first generation are tightly constrained. The mixings between the second and third generations are, on the other hand, free \cite{Gabbiani:1996hi}. So, in our analysis we assume significant mixing between the second and third
generations in the up- and down-squark mass matrices and neglect those involving the first generation. 

The up-squark mass squared matrix in the $(u_L,c_L,t_L,u_R,c_R,t_R)$ basis can be written as
\begin{eqnarray}
{\cal M}_{U_k}^{2,\,FC}=  {\rm M_{susy}^2} \left( \begin{array}{cccccc}
    1     &    0                        &  0                     &                     0   &     0              &        0           \\
    0     &    1                        &  (\delta_U^{LL})_{23}  &                     0   &(\delta^{LR}_U)_{22}& \delta_U^{LR})_{23}\\
    0     &(\delta_U^{LL})_{32}         &  1                     &                     0   &(\delta_U^{LR})_{32}&(\delta^{LR}_U)_{33}\\
    0     & 0                           & 0                      &                     1   &     0              &   0                 \\  
    0     &(\delta^{LR}_U)_{22}         &(\delta_U^{RL})_{23}    &                     0   &   1                &(\delta_U^{RR})_{23} \\ 
    0     & (\delta_U^{RL})_{32}        &(\delta^{LR}_U)_{33}    &                     0   &(\delta_U^{RR})_{32}& 1                   \\ 
                         \end{array}
                  \right),\label{MUFC}
\end{eqnarray}
where $(\delta^{LR}_U)_{22}=m_c (A+\mu \cot \beta)/{\rm M_{\rm susy}^2}$, $(\delta^{LR}_U)_{33}=m_t (A+\mu \cot \beta)/{\rm M_{\rm susy}^2}$. Here we assume 
that all diagonal elements on the main diagonal are set to a common value ${\rm M_{\rm susy}^2}$ \footnote{We comment on the effect of relaxing this condition. Basically, there are non-negligible corrections coming to the 3rd and 6th diagonal 
entries of both the up- and down-squark matrices. The term to be added to unity in the up-sector is $m_t^2/{\rm M_{\rm susy}^2}$, and in the down-sector $m_b^2/{\rm M_{\rm susy}^2}$ with a contribution from the D-term. So, only in the up sector, the $M_{\rm susy}\sim m_t$ case can give sizable contributions. However, in the flavor non-diagonal scenario, we don't consider such small $M_{\rm susy}$ values.} and the flavor changing off-diagonal entries of each block are furthermore 
scaled with ${\rm M_{\rm susy}^2}$ to make the parameters dimensionless. Thus, these off-diagonal parameters are defined as
\begin{eqnarray}
(\delta_U^{LL})_{ij}&&\hspace{-0.3cm}=\frac{(m_{u\,LL}^2)_{ij}}{{\rm M_{\rm susy}^2}}\,,\;\;(\delta_U^{LR})_{ij}=\frac{(m_{u\,LR}^2)_{ij}}{{\rm M_{\rm susy}^2}}\,,\nonumber\\
(\delta_U^{RL})_{ij}&&\hspace*{-0.3cm}=\frac{(m_{u\,RL}^2)_{ij}}{{\rm M_{susy}^2}}\,,\;\;(\delta_U^{RR})_{ij}=\frac{(m_{u\,RR}^2)_{ij}}{{\rm M_{susy}^2}}\,,\;\;i\neq j=2,3.
\end{eqnarray}
Here we still take the diagonal elements of $m_{u\,LR,u\,RL}$ same as in the flavor diagonal case. The down sector case is completely analogous to the up sector. 

As mentioned in the Introduction, there are two approaches to compute the effects of the flavor changing parameters $\delta$'s to physical quantities. The first historical one is the mass insertion formalism \cite{Hall:1985dx} and the second and more recent one is the general mass eigenstate formalism \cite{Besmer:2001cj,Liu:2004qw}. 
In the mass insertion formalism, the $\delta$ terms represent mixing between chirality states of different squarks, and it is possible to compute the contributions of the first order flavor changing mass insertions perturbatively if one assumes smallness of the intergenerational mixing elements ($\delta$'s) when compared with the diagonal elements. Otherwise, second or even higher order mass insertions have to be taken into account, as is the case for Kaon decays \cite{Colangelo:1998pm}.  

The more recent approach, the general mass eigenstate formalism, is an alternative to the mass insertion formalism. Its advantage is that it allows for large off-diagonal elements where the mass insertion method is no longer applicable. In the general mass eigenstate formalism,  the mass matrix in Eq.~(\ref{MUFC}) (and the similar one in the down-sector) is diagonalized and the flavor changing parameters enter into our expressions through the matrix $\Gamma_{Q\,L,Q\,R}$ defined in Eq.~(\ref{flavortomass}). So, in the rare top decays $t\to cV$, the new flavor changing neutral currents show themselves in both gluino-squark-quark and neutralino-squark-quark couplings in the up-type squark loops and in the chargino-squark-quark coupling in the down-type squark loop. 

After giving expressions for the effective one-loop $tcV$ vertex in the next section, we analyse the flavor-diagonal case first. Then we consider
the effect of intergenerational mixings on the rate of the process $t
\rightarrow c V$.  We assume significant mixing between
the second and third
generations only, in both the up and down squark mass matrices.

\section{\bf $t \rightarrow c V$ in LRSUSY}\label{sec:tcV}
\begin{figure}[htb]
\begin{center}
\begin{picture}(200,90)(-30,0)
\SetWidth{0.8}
\ArrowLine(0,0)(20,0)
\Text(2,-7)[c]{$t$}
\ArrowLine(20,0)(40,0)
\DashArrowArcn(75,0)(35,180,0){3}
\Photon(20,0)(40,-20){3}{5}
\Text(25,-25)[l]{\small{$g,\gamma,Z$}}
\ArrowLine(40,0)(110,0)
\ArrowLine(110,0)(150,0)
\Text(-10,50)[l]{(a)}
\Text(143,-7)[c]{$c$}
\Text(75,8)[c]{\tiny{$\bf{\tilde g,\chi^0_n(\chi^+_a)}$}}
\Text(75,42)[c]{\tiny{$\bf{\tilde u_k(\tilde d_k)}$}}
\end{picture}
\begin{picture}(200,90)(-20,0)
\SetWidth{0.8}
\ArrowLine(0,0)(40,0)
\Text(10,-7)[c]{$t$}
\DashArrowArcn(75,0)(35,180,0){3}
\Text(136,28)[l]{\small{$g,\gamma,Z$}}
\ArrowLine(40,0)(110,0)
\ArrowLine(110,0)(130,0)
\ArrowLine(130,0)(150,0)
\Photon(130,0)(150,20){3}{5}
\Text(-10,50)[l]{(b)}
\Text(148,-7)[c]{$c$}
\Text(75,8)[c]{\tiny{$\bf{\tilde g,\chi^0_n(\chi^+_a)}$}}
\Text(75,42)[c]{\tiny{$\bf{\tilde u_k(\tilde d_k)}$}}
\end{picture}
\begin{picture}(200,90)(-30,15)
\SetWidth{0.8}
\ArrowLine(0,0)(40,0)
\Text(10,-7)[c]{$t$}
\DashArrowArcn(75,0)(35,180,90){3}
\DashArrowArcn(75,0)(35,90,0){3}
\Photon(75,35)(95,55){3}{5}
\Text(79,63)[l]{\small{$g,\gamma,Z$}}
\ArrowLine(40,0)(110,0)
\ArrowLine(110,0)(150,0)
\Text(-10,50)[l]{(c)}
\Text(143,-7)[c]{$c$}
\Text(75,8)[c]{\tiny{$\bf{\tilde g,\chi^0_n(\chi^+_a)}$}}
\Text(53,23)[l]{\tiny{$\bf{\tilde u_h(\tilde d_h)}$}}
\Text(105,23)[l]{\tiny{$\bf{\tilde u_k(\tilde d_k)}$}}
\end{picture}
\begin{picture}(200,90)(-20,15)
\SetWidth{0.8}
\ArrowLine(0,0)(40,0)
\Text(10,-7)[c]{$t$}
\DashArrowArcn(75,0)(35,180,0){3}
\Photon(75,0)(95,-25){3}{5}
\Text(80,-33)[l]{${g,\gamma,Z}$}
\ArrowLine(40,0)(75,0)
\ArrowLine(75,0)(110,0)
\ArrowLine(110,0)(150,0)
\Text(-10,50)[l]{(d)}
\Text(143,-7)[c]{$c$}
\Text(75,42)[c]{\tiny{$\bf{\tilde u_k(\tilde d_k)}$}}
\Text(57,-7)[c]{\tiny{$\bf{\tilde g,\chi^0_m(\chi^+_b)}$}}
\Text(92,7)[c]{\tiny{$\bf{\tilde g,\chi^0_n(\chi^+_a)}$}}
\end{picture}
\end{center}
\vspace*{0.9cm}
\caption{The one-loop Feynman Diagrams contributing to $t\to cg,\gamma,Z$ including gluino, chargino and neutralino loops. For each decay mode, not all four types of diagrams contribute. In $t\to cg$ case, diagram (d) for the chargino and neutralino loops doesn't contribute. In $t\to c\gamma$ case, diagram (d) for the gluino and neutralino loops doesn't contribute. In $t\to cZ$ case, diagram (d) for the gluino loop doesn't contribute.}\label{FeynFig}
\end{figure}
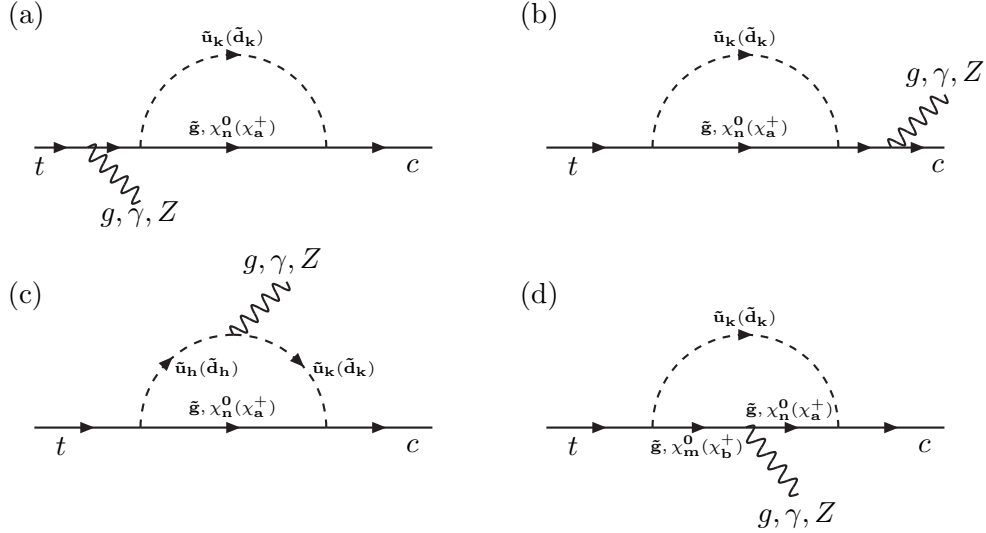
In this section we present the one-loop LRSUSY effective $tcV$ vertex by considering gluino, chargino and neutralino loops with up- and down- squarks, including the effects from left-right and intergenerational squark mixing. The relevant one-loop Feynman diagrams for $t\to cV \,(V=g,\gamma, Z)$ are shown in Fig.~\ref{FeynFig}. Since there are no flavor violating $tcV$ couplings in the Lagrangian, there is no tree level contribution. Under the assumption that the charm quark mass is negligible, the transition amplitude of the decay is given by
\begin{eqnarray}
{\cal M}(t\to c V)=\bar{u}(p-k)A_{V}^\mu u(p)\epsilon_{\mu}(k,\lambda),
\label{eq:amplitude}
\end{eqnarray}
where $p$ and $k$ are the momenta of the top-quark and gauge boson $V$, respectively; $\epsilon(k,\lambda)$ is the polarization vector of the gauge boson. The effective vertex $A_{V}^{\mu}$ can be expressed as 
\begin{eqnarray}
A_{V}^{\mu}=i \sum_{w=\tilde{g},\chi^+,\chi^0}\Big(\gamma^{\mu}\left(P_L F^w_{V\,L}+P_R F^w_{V\,R}\right)+\frac{2p^{\mu}}{m_t}\left(P_L T^w_{V\,L}+P_R T^w_{V\,R}\right)\Big),
\label{eq:vertex}
\end{eqnarray}
where $P_{L,R}=\left(1\mp \gamma_5\right)/2$ is the left (right) chirality projection operator, $F_{V\,L(R)}$ and $T_{V\,L(R)}$ are the form factors. In order to compute these form factors, we need to diagonalize the mass matrices for both chargino and neutralino sectors, which are given in Appendix \ref{sec:AppendixMasses}. Certain Feynman rules of the relevant vertices are presented in Appendix \ref{sec:AppendixFynmVert}. The calculations are carried out in the 't Hooft-Feynman gauge in $d$-dimensions. Below we give the explicit expressions of the form factors for each decay. We prefer to use the notation in Ref.~\cite{Couture:1997ep} which makes the cancellation of infinities more obvious when we set $d=4$. To put the form factors into compact form we have further used the identities given in Ref.~\cite{Lopez:1997xv}.
\subsection{\bf $t \rightarrow c g$ decay}\label{sec:tcg}
\subsubsection{Gluino contribution}\label{subsec:gg}
\begin{eqnarray}
F^{\tilde{g}}_{g\,L}&&\hspace*{-0.3cm}=\frac{g_s^3 T^{a\prime}}{8\pi^2}\sum_{k=1}^{6}\left[C_2[F]\left(\Gamma_{UL}^{k3}\Gamma_{UR}^{*k2}\left(C^{c\tilde{g}[kk]}_{\epsilon}-C_{\rm SE}^{\tilde{g}[k]}\right)+
\Gamma_{UR}^{k3}\Gamma_{UR}^{*k2}\left(C_{\rm SEG}^{\tilde{g}[k]}-C_{\rm SEG}^{\tilde{g}[k]}(m_t^2=0)\right)\frac{m_{\tilde{g}}}{m_t}\right)\right.\nonumber\\
&&\left.+\frac{C_2[G]}{2}\left(\Gamma_{UL}^{k3}\Gamma_{UR}^{*k2}\left(-C^{c\tilde{g}[kk]}_{\epsilon}+C^{d\tilde{g}[k]}_{\epsilon}+C_{\tilde{g}}^{[k]}+C_{k^2}^{\tilde{g}[k]}+C_t^{\tilde{g}[k]}\right)-
\Gamma_{UR}^{k3}\Gamma_{UR}^{*k2}C_{\tilde{g}}^{[k]}\frac{m_t}{m_{\tilde{g}}}\right)\right]\,,\nonumber\\
F^{\tilde{g}}_{g\,R}&&\hspace*{-0.3cm}=\frac{g_s^3 T^{a\prime}}{8\pi^2}\sum_{k=1}^{6}\left[C_2[F]\left(\Gamma_{UR}^{k3}\Gamma_{UL}^{*k2}\left(C^{c\tilde{g}[kk]}_{\epsilon}-C_{\rm SE}^{\tilde{g}[k]}\right)+
\Gamma_{UL}^{k3}\Gamma_{UL}^{*k2}\left(C_{\rm SEG}^{\tilde{g}[k]}-C_{\rm SEG}^{\tilde{g}[k]}(m_t^2=0)\right)\frac{m_{\tilde{g}}}{m_t}\right)\right.\nonumber\\
&&\left.+\frac{C_2[G]}{2}\left(\Gamma_{UR}^{k3}\Gamma_{UL}^{*k2}\left(-C^{c\tilde{g}[kk]}_{\epsilon}+C^{d\tilde{g}[k]}_{\epsilon}+C_{\tilde g}^{[k]}+C_{k^2}^{\tilde{g}[k]}+C_t^{\tilde{g}[k]}\right)-
\Gamma_{UL}^{k3}\Gamma_{UL}^{*k2}C_{\tilde g}^{[k]}\frac{m_t}{m_{\tilde{g}}}\right)\right]\,,\nonumber\\
T^{\tilde{g}}_{g\,L}&&\hspace*{-0.3cm}=-\frac{g_s^3 T^{a\prime}}{8\pi^2}\sum_{k=1}^{6}\Bigg[C_2[F]\Big(\Gamma_{UL}^{k3}\Gamma_{UL}^{*k2}C_{\tilde{g}\rm top}^{[kk]}-\Gamma_{UR}^{k3}\Gamma_{UL}^{*k2}C_{top}^{\tilde{g}[kk]}\Big)\Bigg.\nonumber\\
&&\Bigg.+\frac{C_2[G]}{2}\left(\Gamma_{UL}^{k3}\Gamma_{UL}^{*k2}\left(C_{\tilde{g}t}^{[k]}-C_{\tilde{g}\rm top}^{[kk]}-C_{\tilde g}^{[k]} \frac{m_t}{m_{\tilde{g}}}\right)
+\Gamma_{UR}^{k3}\Gamma_{UL}^{*k2}\left(C_t^{\tilde{g}[k]}+C_{\rm top}^{\tilde{g}[kk]}\right)\right)\Bigg]\,,\nonumber\\
T^{\tilde{g}}_{g\,R}&&\hspace*{-0.3cm}=-\frac{g_s^3 T^{a\prime}}{8\pi^2}\sum_{k=1}^{6}\Bigg[C_2[F]\Big(\Gamma_{UR}^{k3}\Gamma_{UR}^{*k2}C_{\tilde{g}\rm top}^{[kk]}-\Gamma_{UL}^{k3}\Gamma_{UR}^{*k2}C_{\rm top}^{\tilde{g}[kk]}\Big)\Bigg.\nonumber\\
&&\Bigg.+\frac{C_2[G]}{2}\left(\Gamma_{UR}^{k3}\Gamma_{UR}^{*k2}\left(C_{\tilde{g}t}^{[k]}-C_{\tilde{g}\rm top}^{[kk]}-C_{\tilde g}^{[k]} \frac{m_t}{m_{\tilde{g}}}\right)+\Gamma_{UL}^{k3}\Gamma_{UR}^{*k2}\left(C_t^{\tilde{g}[k]}+C_{\rm top}^{\tilde{g}[kk]}\right)\right)\Bigg]\,.
\label{eq:gg}
\end{eqnarray}
\subsubsection{Chargino contribution}\label{subsec:gc}
\begin{eqnarray}
F^{\chi^+}_{g\,L}&&\hspace*{-0.3cm}=-\frac{g_s g^2 T^{a\prime}}{16\pi^2}\sum_{k=1}^{6}\sum_{a=1}^{5}\left(G_{DL}^{*[a,k,2]}-H_{DR}^{*[a,k,2]}\right)\Bigg[\left(G_{DL}^{[a,k,3]}-H_{DR}^{[a,k,3]}\right)\left(C^{c\,\chi^+[akk]}_{\epsilon}-C^{\chi^+[ak]}_{\rm SE}\right)\Bigg.\nonumber\\
&&-\Bigg.\left(G_{DR}^{[a,k,3]}-H_{DL}^{[a,k,3]}\right)\left(C^{\chi^+[ak]}_{\rm SEG}-C^{\chi^+[ak]}_{\rm SEG}(m_t^2=0)\right)\frac{m^a_{\chi^+}}{m_t}\Bigg]\,,\nonumber\\
F^{\chi^+}_{g\,R}&&\hspace*{-0.3cm}=-\frac{g_s g^2 T^{a\prime}}{16\pi^2}\sum_{k=1}^{6}\sum_{a=1}^{5}\left(G_{DR}^{*[a,k,2]}-H_{DL}^{*[a,k,2]}\right)\Bigg[\left(G_{DR}^{[a,k,3]}-H_{DL}^{[a,k,3]}\right)\left(C^{c\,\chi^+[akk]}_{\epsilon}-C^{\chi^+[ak]}_{\rm SE}\right)\Bigg.\nonumber\\
&&-\Bigg.\left(G_{DL}^{[a,k,3]}-H_{DR}^{[a,k,3]}\right)\left(C^{\chi^+[ak]}_{\rm SEG}-C^{\chi^+[ak]}_{\rm SEG}(m_t^2=0)\right)\frac{m^a_{\chi^+}}{m_t}\Bigg]\,,\nonumber\\
T^{\chi^+}_{g\,L}&&\hspace*{-0.3cm}=-\frac{g_s g^2 T^{a\prime}}{16\pi^2}\sum_{k=1}^{6}\sum_{a=1}^{5}\left(G_{DR}^{*[a,k,2]}-H_{DL}^{*[a,k,2]}\right)\Bigg[\left(G_{DL}^{[a,k,3]}-H_{DR}^{[a,k,3]}\right)C_{\chi^+\rm top}^{[akk]}\Bigg.\nonumber\\
&&\Bigg.+\left(G_{DR}^{[a,k,3]}-H_{DL}^{[a,k,3]}\right)C^{\chi^+[akk]}_{\rm top}\Bigg]\,,\nonumber\\
T^{\chi^+}_{g\,R}&&\hspace*{-0.3cm}=-\frac{g_s g^2 T^{a\prime}}{16\pi^2}\sum_{k=1}^{6}\sum_{a=1}^{5}\left(G_{DL}^{*[a,k,2]}-H_{DR}^{*[a,k,2]}\right)\Bigg[\left(G_{DR}^{[a,k,3]}-H_{DL}^{[a,k,3]}\right)C_{\chi^+\rm top}^{[akk]}\Bigg.\nonumber\\
&&\Bigg.+\left(G_{DL}^{[a,k,3]}-H_{DR}^{[a,k,3]}\right)C^{\chi^+[akk]}_{\rm top}\Bigg]\,.
\end{eqnarray}
\subsubsection{Neutralino contribution}\label{subsec:gn}
\begin{eqnarray}
F^{\chi^0}_{g\,L}&&\hspace*{-0.3cm}=-\frac{g_s g^2 T^{a\prime}}{16\pi^2}\sum_{k=1}^{6}\sum_{n=1}^{9}\left(\sqrt{2}G_{UL}^{*[n,k,2]}+H_{UR}^{*[n,k,2]}\right)\Bigg[\left(\sqrt{2}G_{UL}^{[n,k,3]}+H_{UR}^{[n,k,3]}\right)\Bigg.\nonumber\\
&&\Bigg.\times\left(C^{c\chi^0[nkk]}_{\epsilon}-C^{\chi^0[nk]}_{\rm SE}\right)+\left(\sqrt{2}G_{UR}^{[n,k,3]}-H_{UL}^{[n,k,3]}\right)\left(C^{\chi^0[nk]}_{\rm SEG}-C^{\chi^0[nk]}_{\rm SEG}(m_t^2=0)\right)\frac{m^n_{\chi^0}}{m_t}\Bigg]\,,\nonumber\\
F^{\chi^0}_{g\,R}&&\hspace*{-0.3cm}=-\frac{g_s g^2 T^{a\prime}}{16\pi^2}\sum_{k=1}^{6}\sum_{n=1}^{9}\left(\sqrt{2}G_{UR}^{*[n,k,2]}-H_{UL}^{*[n,k,2]}\right)\Bigg[\left(\sqrt{2}G_{UR}^{[n,k,3]}-H_{UL}^{[n,k,3]}\right)\Bigg.\nonumber\\
&&\Bigg.\times\left(C^{c\chi^0[nkk]}_{\epsilon}-C^{\chi^0[nk]}_{\rm SE}\right)+\left(\sqrt{2}G_{UL}^{[n,k,3]}+H_{UR}^{[n,k,3]}\right)\left(C^{\chi^0[nk]}_{\rm SEG}-C^{\chi^0[nk]}_{\rm SEG}(m_t^2=0)\right)\frac{m^n_{\chi^0}}{m_t}\Bigg]\,,\nonumber\\
T^{\chi^0}_{g\,L}&&\hspace*{-0.3cm}=\frac{g_s g^2 T^{a\prime}}{16\pi^2}\sum_{k=1}^{6}\sum_{n=1}^{9}\left(\sqrt{2}G_{UR}^{*[n,k,2]}-H_{UL}^{*[n,k,2]}\right)\Bigg[\left(\sqrt{2}G_{UL}^{[n,k,3]}+H_{UR}^{[n,k,3]}\right)C_{\chi^0\rm top}^{[nkk]}\Bigg.\nonumber\\
&&\Bigg.-\left(\sqrt{2}G_{UR}^{[n,k,3]}-H_{UL}^{[n,k,3]}\right)C^{\chi^0[nkk]}_{\rm top}\Bigg]\,,\nonumber\\
T^{\chi^0}_{g\,R}&&\hspace*{-0.3cm}=\frac{g_s g^2 T^{a\prime}}{16\pi^2}\sum_{k=1}^{6}\sum_{n=1}^{9}\left(\sqrt{2}G_{UL}^{*[n,k,2]}+H_{UR}^{*[n,k,2]}\right)\Bigg[\left(\sqrt{2}G_{UR}^{[n,k,3]}-H_{UL}^{[n,k,3]}\right)C_{\chi^0\rm top}^{[nkk]}\Bigg.\nonumber\\
&&\Bigg.-\left(\sqrt{2}G_{UL}^{[n,k,3]}+H_{UR}^{[n,k,3]}\right)C^{\chi^0[nkk]}_{\rm top}\Bigg]\,.
\end{eqnarray}

\subsection{\bf $t \rightarrow c \gamma$ decay}\label{subsec:tggamma}
\subsubsection{Gluino contribution}\label{subsec:gammag}
\begin{eqnarray}
F^{\tilde{g}}_{\gamma\,L}&&\hspace*{-0.3cm}=\frac{g_s^2 g Q_u C_2[F] \sin{\theta_{\rm W}}}{8\pi^2}\sum_{k=1}^{6}\Bigg[\Gamma_{UL}^{k3}\Gamma_{UR}^{*k2}\left(C^{c\tilde{g}[kk]}_{\epsilon}-C_{\rm SE}^{\tilde{g}[k]}\right)\Bigg.\nonumber\\
&&\Bigg.+\Gamma_{UR}^{k3}\Gamma_{UR}^{*k2}\left(C_{\rm SEG}^{\tilde{g}[k]}-C_{\rm SEG}^{\tilde{g}[k]}(m_t^2=0)\right)\frac{m_{\tilde{g}}}{m_t}\Bigg]\,,\nonumber\\
F^{\tilde{g}}_{\gamma\,R}&&\hspace*{-0.3cm}=\frac{g_s^2 g Q_u C_2[F] \sin{\theta_{\rm W}}}{8\pi^2}\sum_{k=1}^{6}\Bigg[\Gamma_{UR}^{k3}\Gamma_{UL}^{*k2}\left(C^{c\tilde{g}[kk]}_{\epsilon}-C_{\rm SE}^{\tilde{g}[k]}\right)\Bigg.\nonumber\\
&&\Bigg.+\Gamma_{UL}^{k3}\Gamma_{UL}^{*k2}\left(C_{\rm SEG}^{\tilde{g}[k]}-C_{\rm SEG}^{\tilde{g}[k]}(m_t^2=0)\right)\frac{m_{\tilde{g}}}{m_t}\Bigg]\,,\nonumber\\
T^{\tilde{g}}_{\gamma\,L}&&\hspace*{-0.3cm}=\frac{g_s^2 g Q_u C_2[F] \sin{\theta_{\rm W}}}{8\pi^2}\sum_{k=1}^{6}\left(\Gamma_{UR}^{k3}\Gamma_{UL}^{*k2}C_{\rm top}^{\tilde{g}[kk]}-\Gamma_{UL}^{k3}\Gamma_{UL}^{*k2} C_{\tilde{g}\rm top}^{[kk]}\right)\,,\nonumber\\
T^{\tilde{g}}_{\gamma\,R}&&\hspace*{-0.3cm}=\frac{g_s^2 g Q_u C_2[F] \sin{\theta_{\rm W}}}{8\pi^2}\sum_{k=1}^{6}\left(\Gamma_{UL}^{k3}\Gamma_{UR}^{*k2}C_{\rm top}^{\tilde{g}[kk]}-\Gamma_{UR}^{k3}\Gamma_{UR}^{*k2} C_{\tilde{g}\rm top}^{[kk]}\right)\,.
\end{eqnarray}

\subsubsection{Chargino contribution}\label{subsec:gammac}
\begin{eqnarray}
F^{\chi^+}_{\gamma\,L}&&\hspace*{-0.3cm}=-\frac{g^3 \sin{\theta_{\rm W}}}{16\pi^2}\sum_{k=1}^{6}\sum_{a=1}^{5}\left(G_{DL}^{*[a,k,2]}-H_{DR}^{*[a,k,2]}\right)
\Bigg[\left(G_{DL}^{[a,k,3]}-H_{DR}^{[a,k,3]}\right)\left(Q_d C^{c\chi^+[akk]}_{\epsilon}\right.\Bigg.\nonumber\\
&&\Bigg.\left.- Q_u C^{\chi^+[ak]}_{\rm SE}+C^{d\chi^+[aak]}_{\epsilon}+C_{\chi^+}^{[aak]} +C_{k^2}^{\chi^+[aak]}+C_t^{\chi^+[aak]}\right)+\left(G_{DR}^{[a,k,3]}-H_{DL}^{[a,k,3]}\right)\Bigg.\nonumber\\
&&\Bigg.\times\left(C_{\chi^+}^{[aak]}\frac{m_t}{m_{\chi^+}^a}-Q_u\left( C_{\rm SEG}^{\chi^+[ak]}-C_{\rm SEG}^{\chi^+[ak]}(m_t^2=0)\right)\frac{m_{\chi^+}^a}{m_t}\right)\Bigg]\,,\nonumber\\
F^{\chi^+}_{\gamma\,R}&&\hspace*{-0.3cm}=-\frac{g^3 \sin{\theta_{\rm W}}}{16\pi^2}\sum_{k=1}^{6}\sum_{a=1}^{5}\left(G_{DR}^{*[a,k,2]}-H_{DL}^{*[a,k,2]}\right)
\Bigg[\left(G_{DR}^{[a,k,3]}-H_{DL}^{[a,k,3]}\right)\left(Q_d C^{c\chi^+[akk]}_{\epsilon}\right.\Bigg.\nonumber\\
&&\Bigg.\left.-Q_u C^{\chi^+[ak]}_{\rm SE}+C^{d\chi^+[aak]}_{\epsilon}+C_{\chi^+}^{[aak]} +C_{k^2}^{\chi^+[aak]}+C_t^{\chi^+[aak]}\right)+\left(G_{DL}^{[a,k,3]}-H_{DR}^{[a,k,3]}\right)\Bigg.\nonumber\\
&&\Bigg.\times\left(C_{\chi^+}^{[aak]}\frac{m_t}{m_{\chi^+}^a}-Q_u\left( C_{\rm SEG}^{\chi^+[ak]}-C_{\rm SEG}^{\chi^+[ak]}(m_t^2=0)\right)\frac{m_{\chi^+}^a}{m_t}\right)\Bigg]\,,\nonumber\\
T^{\chi^+}_{\gamma\,L}&&\hspace*{-0.3cm}=-\frac{g^3 \sin{\theta_{\rm W}}}{16\pi^2}\sum_{k=1}^{6}\sum_{a=1}^{5}\left(G_{DR}^{*[a,k,2]}-H_{DL}^{*[a,k,2]}\right)
\Bigg[\left(G_{DR}^{[a,k,3]}-H_{DL}^{[a,k,3]}\right)\left(Q_d C^{\chi^+[akk]}_{\rm top}\right.\Bigg.\nonumber\\
&&\Bigg.\left.+ C^{\chi^+[aak]}_{t}\right)+\left(G_{DL}^{[a,k,3]}-H_{DR}^{[a,k,3]}\right)\left(C_{\chi^+ t}^{[aak]}+Q_d C_{\chi^+\rm top}^{[akk]}-C_{\chi^+}^{[aak]}\frac{m_t}{m_{\chi^+}^a}\right)\Bigg]\,,\nonumber\\
T^{\chi^+}_{\gamma\,R}&&\hspace*{-0.3cm}=-\frac{g^3 \sin{\theta_{\rm W}}}{16\pi^2}\sum_{k=1}^{6}\sum_{a=1}^{5}\left(G_{DL}^{*[a,k,2]}-H_{DR}^{*[a,k,2]}\right)
\Bigg[\left(G_{DL}^{[a,k,3]}-H_{DR}^{[a,k,3]}\right)\left(Q_d C^{\chi^+[akk]}_{\rm top}\right.\Bigg.\nonumber\\
&&\Bigg.\left.+ C^{\chi^+[aak]}_{t}\right)+\left(G_{DR}^{[a,k,3]}-H_{DL}^{[a,k,3]}\right)\left(C_{\chi^+ t}^{[aak]}+Q_d C_{\chi^+\rm top}^{[akk]}-C_{\chi^+}^{[aak]}\frac{m_t}{m_{\chi^+}^a}\right)\Bigg]\,.
\end{eqnarray}

\subsubsection{Neutralino contribution}\label{subsec:gamman}
\begin{eqnarray}
F^{\chi^0}_{\gamma\,L}&&\hspace*{-0.3cm}=-\frac{g^3 Q_u \sin{\theta_{\rm W}}}{16\pi^2}\sum_{k=1}^{6}\sum_{n=1}^{9}\left(\sqrt{2}G_{UL}^{*[n,k,2]}+H_{UR}^{*[n,k,2]}\right)\Bigg[\left(\sqrt{2}G_{UL}^{[n,k,3]}+H_{UR}^{[n,k,3]}\right)\Bigg.\nonumber\\
&&\Bigg.\times\left(C^{c\chi^0[nkk]}_{\epsilon}-C_{\rm SE}^{\chi^0[nk]}\right)+\left(\sqrt{2}G_{UR}^{[n,k,3]}-H_{UL}^{[n,k,3]}\right)\left(C_{\rm SEG}^{\chi^0[nk]}-C_{\rm SEG}^{\chi^0[nk]}(m_t^2=0)\right)\frac{m_{\chi^0}^{n}}{m_t}\Bigg],\nonumber\\
F^{\chi^0}_{\gamma\,R}&&\hspace*{-0.3cm}=-\frac{g^3 Q_u \sin{\theta_{\rm W}}}{16\pi^2}\sum_{k=1}^{6}\sum_{n=1}^{9}\left(\sqrt{2}G_{UR}^{*[n,k,2]}-H_{UL}^{*[n,k,2]}\right)\Bigg[\left(\sqrt{2}G_{UR}^{[n,k,3]}-H_{UL}^{[n,k,3]}\right)\Bigg.\nonumber\\
&&\Bigg.\times\left(C^{c\chi^0[nkk]}_{\epsilon}-C_{\rm SE}^{\chi^0[nk]}\right)+\left(\sqrt{2}G_{UL}^{[n,k,3]}+H_{UR}^{[n,k,3]}\right)\left(C_{\rm SEG}^{\chi^0[nk]}-C_{\rm SEG}^{\chi^0[nk]}(m_t^2=0)\right)\frac{m_{\chi^0}^{n}}{m_t}\Bigg],\nonumber\\
T^{\chi^0}_{\gamma\,L}&&\hspace*{-0.3cm}=\frac{g^3 Q_u \sin{\theta_{\rm W}}}{16\pi^2}\sum_{k=1}^{6}\sum_{n=1}^{9}\left(\sqrt{2}G_{UR}^{*[n,k,2]}-H_{UL}^{*[n,k,2]}\right)\Bigg[\left(\sqrt{2}G_{UL}^{[n,k,3]}+H_{UR}^{[n,k,3]}\right)C_{\chi^0\rm top}^{[nkk]}\Bigg.\nonumber\\
&&\Bigg.-\left(\sqrt{2}G_{UR}^{[n,k,3]}-H_{UL}^{[n,k,3]}\right)C_{\rm top}^{\chi^0[nkk]}\Bigg]\,,\nonumber\\
T^{\chi^0}_{\gamma\,R}&&\hspace*{-0.3cm}=\frac{g^3 Q_u \sin{\theta_{\rm W}}}{16\pi^2}\sum_{k=1}^{6}\sum_{n=1}^{9}\left(\sqrt{2}G_{UL}^{*[n,k,2]}+H_{UR}^{*[n,k,2]}\right)\Bigg[\left(\sqrt{2}G_{UR}^{[n,k,3]}-H_{UL}^{[n,k,3]}\right)C_{\chi^0\rm top}^{[nkk]}\Bigg.\nonumber\\
&&\Bigg.-\left(\sqrt{2}G_{UL}^{[n,k,3]}+H_{UR}^{[n,k,3]}\right)C_{\rm top}^{\chi^0[nkk]}\Bigg]\,.
\end{eqnarray}

\subsection{\bf $t \rightarrow c Z$ decay}\label{subsec:tcZ}
\subsubsection{Gluino contribution}\label{subsec:Zg}
\begin{eqnarray}
F^{\tilde{g}}_{Z\,L}&&\hspace*{-0.3cm}=-\frac{g_s^2 g C_2[F]}{8\pi^2\cos{\theta_{\rm W}}}\sum_{k=1}^{6}\Bigg[\left(\Gamma_{UL}^{k3}\Gamma_{UR}^{*k2}C_{\rm SE}^{\tilde{g}[k]}-\Gamma_{UR}^{k3}\Gamma_{UR}^{*k2}\left(C_{\rm SEG}^{\tilde{g}[k]}-C_{\rm SEG}^{\tilde{g}[k]}(m_t^2=0)\right)\frac{m_{\tilde{g}}}{m_t}\right)\Bigg.\nonumber\\
&&\Bigg.\times\left(T_{3u}-Q_u\sin^2\theta_{\rm W}\right)-\sum_{h=1}^{6}\Gamma_{UL}^{h 3}\Gamma_{UR}^{*k 2}C^{c\tilde{g}[hk]}_{\epsilon}\left(T_{3u}\sum_{j=1}^{3}\Gamma_{UL}^{*h j}\Gamma_{UL}^{k j}-Q_u\delta^{hk}\sin^2\theta_{\rm W}\right)\Bigg]\,,\nonumber\\
F^{\tilde{g}}_{Z\,R}&&\hspace*{-0.3cm}=\frac{g_s^2 g C_2[F]}{8\pi^2\cos{\theta_{\rm W}}}\sum_{k=1}^{6}\Bigg[\left(\Gamma_{UR}^{k3}\Gamma_{UL}^{*k2}C_{\rm SE}^{\tilde{g}[k]}-\Gamma_{UL}^{k3}\Gamma_{UL}^{*k2}\left(C_{\rm SEG}^{\tilde{g}[k]}-C_{\rm SEG}^{\tilde{g}[k]}(m_t^2=0)\right)\frac{m_{\tilde{g}}}{m_t}\right)Q_u\sin^2\theta_{\rm W}\Bigg.\nonumber\\
&&\Bigg.+\sum_{h=1}^{6}\Gamma_{UR}^{h 3}\Gamma_{UL}^{*k 2}C^{c\tilde{g}[hk]}_{\epsilon}\left(T_{3u}\sum_{j=1}^{3}\Gamma_{UL}^{*h j}\Gamma_{UL}^{k j}-Q_u\delta^{hk}\sin^2\theta_{\rm W}\right)\Bigg]\,,\nonumber\\
T^{\tilde{g}}_{Z\,L}&&\hspace*{-0.3cm}=-\frac{g_s^2 g C_2[F]}{8\pi^2\cos\theta_{\rm W}}\sum_{k=1}^{6}\sum_{h=1}^{6}\left(\Gamma_{UL}^{h 3}\Gamma_{UL}^{*k2}C_{\tilde{g}\rm top}^{[hk]}-\Gamma_{UR}^{h 3}\Gamma_{UL}^{*k2}C_{\rm top}^{\tilde{g}[hk]}\right)\nonumber\\
&&\times\left(T_{3u}\sum_{j=1}^{3}\Gamma_{UL}^{*h j}\Gamma_{UL}^{k j}-Q_u\delta^{hk}\sin^2\theta_{\rm W}\right),\nonumber\\
T^{\tilde{g}}_{Z\,R}&&\hspace*{-0.3cm}=-\frac{g_s^2 g C_2[F]}{8\pi^2\cos\theta_{\rm W}}\sum_{k=1}^{6}\sum_{h=1}^{6}\left(\Gamma_{UR}^{h 3}\Gamma_{UR}^{*k2}C_{\tilde{g}\rm top}^{[hk]}-\Gamma_{UL}^{h 3}\Gamma_{UR}^{*k2}C_{\rm top}^{\tilde{g}[hk]}\right)\nonumber\\
&&\times\left(T_{3u}\sum_{j=1}^{3}\Gamma_{UL}^{*h j}\Gamma_{UL}^{k j}-Q_u\delta^{hk}\sin^2\theta_{\rm W}\right).
\end{eqnarray}

\subsubsection{Chargino contribution}\label{subsec:Zc}
\begin{eqnarray}
F^{\chi^+}_{Z\,L}&&\hspace*{-0.3cm}=\frac{g^3}{16\pi^2\cos{\theta_{\rm W}}}\sum_{k=1}^{6}\sum_{a=1}^{5}\left(G_{DL}^{*[a,k,2]}-H_{DR}^{*[a,k,2]}\right)\Bigg[\Bigg(\left(G_{DL}^{[a,k,3]}-H_{DR}^{[a,k,3]}\right)C_{\rm SE}^{\chi^+[ak]}\Big.\Bigg.\nonumber\\
&&\Bigg.\Bigg.+\left(G_{DR}^{[a,k,3]}-H_{DL}^{[a,k,3]}\right)
\left(C_{\rm SEG}^{\chi^+[ak]}-C_{\rm SEG}^{\chi^+[ak]}(m_t^2=0)\right)\frac{m_{\chi^+}^a}{m_t}\Bigg)\left(T_{3u}-Q_u \sin^2{\theta_{\rm W}}\right)\Bigg.\nonumber\\
&&\Bigg.-\sum_{h=1}^{6}\left(G_{DL}^{[a,h,3]}-H_{DR}^{[a,h,3]}\right)C_{\epsilon}^{c\chi^+[ahk]}\left(T_{3d}\sum_{j=1}^{3}\Gamma_{DL}^{*hj}\Gamma_{DL}^{kj}-Q_d\delta^{hk}\sin^2\theta_{\rm W}\right)\Bigg.\nonumber\\
&&\Bigg.+\sum_{b=1}^{5}\Bigg(\left(G_{DL}^{[b,k,3]}-H_{DR}^{[b,k,3]}\right)\Big[\left(C_{\epsilon}^{d\chi^+[abk]}+C_{k^2}^{\chi^+[abk]}-C_t^{\chi^+[abk]}\right)O_R^{ab}+C_{\chi^+}^{[abk]}O_L^{ab}\Big]\Bigg.\Bigg.\nonumber\\
&&\Bigg.\Bigg.+\left(G_{DR}^{[b,k,3]}-H_{DL}^{[b,k,3]}\right)\left(-C^{\prime[aak]}_{\chi^+t}O_L^{ab}+C^{\prime[bbk]}_{\chi^+t}O_R^{ab}+C_{\chi^+}^{[abk]}O_L^{ab}\frac{m_t}{m_{\chi^+}^{b}}\right)\Bigg)\Bigg]\,,\nonumber\\
F^{\chi^+}_{Z\,R}&&\hspace*{-0.3cm}=-\frac{g^3}{16\pi^2\cos{\theta_{\rm W}}}\sum_{k=1}^{6}\sum_{a=1}^{5}\left(G_{DR}^{*[a,k,2]}-H_{DL}^{*[a,k,2]}\right)\Bigg[\Bigg(\left(G_{DR}^{[a,k,3]}-H_{DL}^{[a,k,3]}\right)C_{\rm SE}^{\chi^+[ak]}\Big.\Bigg.\nonumber\\
&&\Bigg.\Bigg.+\left(G_{DL}^{[a,k,3]}-H_{DR}^{[a,k,3]}\right)
\left(C_{\rm SEG}^{\chi^+[ak]}-C_{\rm SEG}^{\chi^+[ak]}(m_t^2=0)\right)\frac{m_{\chi^+}^a}{m_t}\Bigg)Q_u\sin^2{\theta_{\rm W}}\Bigg.\nonumber\\
&&\Bigg.+\sum_{h=1}^{6}\left(G_{DR}^{[a,h,3]}-H_{DL}^{[a,h,3]}\right)C_{\epsilon}^{c\chi^+[ahk]}\left(T_{3d}\sum_{j=1}^{3}\Gamma_{DL}^{*hj}\Gamma_{DL}^{kj}-Q_d\delta^{hk}\sin^2\theta_{\rm W}\right)\Bigg.\nonumber\\
&&\Bigg.-\sum_{b=1}^{5}\Bigg(\left(G_{DR}^{[b,k,3]}-H_{DL}^{[b,k,3]}\right)\Big[\left(C_{\epsilon}^{d\chi^+[abk]}+C_{k^2}^{\chi^+[abk]}-C_t^{\chi^+[abk]}\right)O_L^{ab}+C_{\chi^+}^{[abk]}O_R^{ab}\Big]\Bigg.\Bigg.\nonumber\\
&&\Bigg.\Bigg.+\left(G_{DL}^{[b,k,3]}-H_{DR}^{[b,k,3]}\right)\left(-C^{\prime[aak]}_{\chi^+t}O_R^{ab}+C^{\prime[bbk]}_{\chi^+t}O_L^{ab}+C_{\chi^+}^{[abk]}O_R^{ab}\frac{m_t}{m_{\chi^+}^{b}}\right)\Bigg)\Bigg]\,,\nonumber\\
T^{\chi^+}_{Z\,L}&&\hspace*{-0.3cm}=-\frac{g^3}{16\pi^2\cos{\theta_{\rm W}}}\sum_{k=1}^{6}\sum_{a=1}^{5}\left(G_{DR}^{*[a,k,2]}-H_{DL}^{*[a,k,2]}\right)\Bigg[\sum_{h=1}^{6}\left(\left(G_{DL}^{[a,h,3]}-H_{DR}^{[a,h,3]}\right)C_{\chi^+\rm top}^{[ahk]}\right.\Bigg.\nonumber\\
&&\Bigg.\left.+\left(G_{DR}^{[a,h,3]}-H_{DL}^{[a,h,3]}\right)C_{\rm top}^{\chi^+[ahk]}\right)
\left(T_{3d}\sum_{j=1}^{3}\Gamma_{DL}^{*hj}\Gamma_{DL}^{kj}-Q_d\delta^{hk}\sin^2\theta_{\rm W}\right)\Bigg.\nonumber\\
&&\Bigg.-\sum_{b=1}^{5}\Bigg(\left(G_{DR}^{[b,k,3]}-H_{DL}^{[b,k,3]}\right)O_L^{ab}+\left(G_{DL}^{[b,k,3]}-H_{DR}^{[b,k,3]}\right)\left(-C^{\prime[bbk]}_{\chi^+t}O_L^{ab}\right.\Bigg.\nonumber\\
&&\Bigg.\left.+\left(C^{\prime[aak]}_{\chi^+t}+C_{\chi^+t}^{[aak]}-C_{\chi^+}^{[abk]}\frac{m_t}{m_{\chi^+}^{b}}\right)O_L^{ab}\right)\Bigg)\Bigg]\,,\nonumber\\
T^{\chi^+}_{Z\,R}&&\hspace*{-0.3cm}=-\frac{g^3}{16\pi^2\cos{\theta_{\rm W}}}\sum_{k=1}^{6}\sum_{a=1}^{5}\left(G_{DL}^{*[a,k,2]}-H_{DR}^{*[a,k,2]}\right)\Bigg[\sum_{h=1}^{6}\left(\left(G_{DR}^{[a,h,3]}-H_{DL}^{[a,h,3]}\right)C_{\chi^+top}^{[ahk]}\right.\Bigg.\nonumber\\
&&\Bigg.\left.+\left(G_{DL}^{[a,h,3]}-H_{DR}^{[a,h,3]}\right)C_{\rm top}^{\chi^+[ahk]}\right)\left(T_{3d}\sum_{j=1}^{3}\Gamma_{DL}^{*hj}\Gamma_{DL}^{kj}-Q_d\delta^{hk}\sin^2\theta_{\rm W}\right)\Bigg.\nonumber\\
&&\Bigg.-\sum_{b=1}^{5}\Bigg(\left(G_{DL}^{[b,k,3]}-H_{DR}^{[b,k,3]}\right)O_R^{ab}+\left(G_{DR}^{[b,k,3]}-H_{DL}^{[b,k,3]}\right)
\left(-C^{\prime[bbk]}_{\chi^+t}O_R^{ab}\right.\Bigg.\nonumber\\
&&\Bigg.\left.+\left(C^{\prime[aak]}_{\chi^+t}+C_{\chi^+t}^{[aak]}-C_{\chi^+}^{[abk]}\frac{m_t}{m_{\chi^+}^{b}}\right)O_R^{ab}\right)\Bigg)\Bigg]\,.
\end{eqnarray}

\subsubsection{Neutralino contribution}\label{subsec:Zn}
\begin{eqnarray}
F^{\chi^0}_{Z\,L}&&\hspace*{-0.3cm}=\frac{g^3}{16\pi^2\cos{\theta_{\rm W}}}\sum_{k=1}^{6}\sum_{n=1}^{9}\left(\sqrt{2}G_{UL}^{*[n,k,2]}+H_{UR}^{*[n,k,2]}\right)\Bigg[\Bigg(\left(\sqrt{2}G_{UL}^{[n,k,3]}+H_{UR}^{[n,k,3]}\right)C_{\rm SE}^{\chi^0[nk]}\Bigg.\Bigg.\nonumber\\
&&\Bigg.\Bigg.-\left(\sqrt{2}G_{UR}^{[n,k,3]}-H_{UL}^{[n,k,3]}\right)
\left(C_{\rm SEG}^{\chi^0[nk]}-C_{\rm SEG}^{\chi^0[nk]}(m_t^2=0)\right)\frac{m_{\chi^0}^n}{m_t}\Bigg)\left(T_{3u}-Q_u \sin^2{\theta_{\rm W}}\right)\Bigg.\nonumber\\
&&\Bigg.-\sum_{h=1}^{6}\left(\sqrt{2}G_{UL}^{[n,h,3]}+H_{UR}^{[n,h,3]}\right)C_{\epsilon}^{c\chi^0[nhk]}\left(T_{3u}\sum_{j=1}^{3}\Gamma_{UL}^{*hj}\Gamma_{UL}^{kj}-Q_u\delta^{hk}\sin^2\theta_{\rm W}\right)\Bigg.\nonumber\\
&&\Bigg.+\sum_{m=1}^{9}\Bigg(\left(\sqrt{2}G_{UL}^{[m,k,3]}+H_{UR}^{[m,k,3]}\right)\Big[\left(C_{\epsilon}^{d \chi^0[nmk]}+C_{k^2}^{\chi^0[nmk]}-C_t^{\chi^0[nmk]}\right)O_R^{\prime n m}\Big.\Bigg.\Bigg.\nonumber\\
&&\Bigg.\Bigg.\Big.+C_{\chi^0}^{[nmk]}O_L^{\prime n m}\Big]+\left(\sqrt{2}G_{UR}^{[m,k,3]}-H_{UL}^{[m,k,3]}\right)\left(C^{\prime[nnk]}_{\chi^0 t}O_L^{\prime n m }-C^{\prime[mmk]}_{\chi^0 t}O_R^{\prime n m}\right.\Big.\Bigg.\nonumber\\
&&\Bigg.\Big.\left.-C_{\chi^0}^{[nmk]}O_L^{\prime n m}\frac{m_t}{m_{\chi^0}^{m}}\right)\Bigg)\Bigg]\,,\nonumber\\
F^{\chi^0}_{Z\,R}&&\hspace*{-0.3cm}=-\frac{g^3}{16\pi^2\cos{\theta_{\rm W}}}\sum_{k=1}^{6}\sum_{n=1}^{9}\left(\sqrt{2}G_{UR}^{*[n,k,2]}-H_{UL}^{*[n,k,2]}\right)\Bigg[\Bigg(\left(\sqrt{2}G_{UR}^{[n,k,3]}-H_{UL}^{[n,k,3]}\right)C_{\rm SE}^{\chi^0[nk]}\Big.\Bigg.\nonumber\\
&&\Bigg.\Bigg.-\left(\sqrt{2}G_{UL}^{[n,k,3]}+H_{UR}^{[n,k,3]}\right)
\left(C_{\rm SEG}^{\chi^0[nk]}-C_{\rm SEG}^{\chi^0[nk]}(m_t^2=0)\right)\frac{m_{\chi^0}^n}{m_t}\Bigg)Q_u \sin^2{\theta_{\rm W}}\Bigg.\nonumber\\
&&\Bigg.+\sum_{h=1}^{6}\left(\sqrt{2}G_{UR}^{[n,h,3]}-H_{UL}^{[n,h,3]}\right)C_{\epsilon}^{c\chi^0[nhk]}\left(T_{3u}\sum_{j=1}^{3}\Gamma_{UL}^{*hj}\Gamma_{UL}^{kj}-Q_u\delta^{hk}\sin^2\theta_{\rm W}\right)\Bigg.\nonumber\\
&&\Bigg.-\sum_{m=1}^{9}\Bigg(\left(\sqrt{2}G_{UR}^{[m,k,3]}-H_{UL}^{[m,k,3]}\right)\Big[\left(C_{\epsilon}^{d\chi^0[nmk]}+C_{k^2}^{\chi^0[nmk]}-C_t^{\chi^0[nmk]}\right)O_L^{\prime n m}\Big.\Bigg.\Bigg.\nonumber\\
&&\Bigg.\Bigg.\Big.+C_{\chi^0}^{[nmk]}O_R^{\prime n m}\Big]+\left(\sqrt{2}G_{UL}^{[m,k,3]}+H_{UR}^{[m,k,3]}\right)\left(C^{\prime[nnk]}_{\chi^0 t}O_R^{\prime n m }-C^{\prime[mmk]}_{\chi^0 t}O_L^{\prime n m}\right.\Big.\Bigg.\nonumber\\
&&\Bigg.\Big.\left.-C_{\chi^0}^{[nmk]}O_R^{\prime n m}\frac{m_t}{m_{\chi^0}^{m}}\right)\Bigg)\Bigg]\,,\nonumber\\
T^{\chi^0}_{Z\,L}&&\hspace*{-0.3cm}=\frac{g^3}{16\pi^2\cos{\theta_{\rm W}}}\sum_{k=1}^{6}\sum_{n=1}^{9}\left(\sqrt{2}G_{UR}^{*[n,k,2]}-H_{UL}^{*[n,k,2]}\right)\Bigg[\sum_{h=1}^{6}\left(\left(\sqrt{2}G_{UL}^{[n,h,3]}+H_{UR}^{[n,h,3]}\right)C_{\chi^0\rm top}^{[nhk]}\right.\Bigg.\nonumber\\
&&\Bigg.\left.-\left(\sqrt{2}G_{UR}^{[n,h,3]}-H_{UL}^{[n,h,3]}\right)C_{\rm top}^{\chi^0[nhk]}\right)
\left(T_{3u}\sum_{j=1}^{3}\Gamma_{UL}^{*hj}\Gamma_{UL}^{kj}-Q_u\delta^{hk}\sin^2\theta_{\rm W}\right)\Bigg.\nonumber\\
&&\Bigg.+\sum_{m=1}^{9}\Bigg(\left(\sqrt{2}G_{UR}^{[m,k,3]}-H_{UL}^{[m,k,3]}\right)O_L^{\prime n m}+\left(\sqrt{2}G_{UL}^{[m,k,3]}+H_{UR}^{[m,k,3]}\right)
\left(C^{\prime[mmk]}_{\chi^0 t}O_L^{\prime n m}\right.\Bigg.\Bigg.\nonumber\\
&&\Bigg.\Bigg.\left.-\left(C^{\prime[nnk]}_{\chi^0 t}+C_{\chi^0 t}^{[nnk]}-C_{\chi^0}^{[nmk]}\frac{m_t}{m_{\chi^0}^{m}}\right)O_R^{\prime n m}\right)\Bigg)\Bigg]\,,\nonumber\\
T^{\chi^0}_{Z\,R}&&\hspace*{-0.3cm}=\frac{g^3}{16\pi^2\cos{\theta_{\rm W}}}\sum_{k=1}^{6}\sum_{n=1}^{9}\left(\sqrt{2}G_{UL}^{*[n,k,2]}+H_{UR}^{*[n,k,2]}\right)\Bigg[\sum_{h=1}^{6}\left(\left(\sqrt{2}G_{UR}^{[n,h,3]}-H_{UL}^{[n,h,3]}\right)C_{\chi^0\rm top}^{[nhk]}\right.\Bigg.\nonumber\\
&&\hspace*{-0.5cm}\Bigg.\left.-\left(\sqrt{2}G_{UL}^{[n,h,3]}+H_{UR}^{[n,h,3]}\right)C_{\rm top}^{\chi^0[nhk]}\right)
\left(T_{3u}\sum_{j=1}^{3}\Gamma_{UL}^{*hj}\Gamma_{UL}^{kj}-Q_u\delta^{hk}\sin^2\theta_{\rm W}\right)\Bigg.\nonumber\\
&&\Bigg.+\sum_{m=1}^{9}\Bigg(\left(\sqrt{2}G_{UL}^{[m,k,3]}+H_{UR}^{[m,k,3]}\right)O_R^{\prime n m}+\left(\sqrt{2}G_{UR}^{[m,k,3]}-H_{UL}^{[m,k,3]}\right)
\left(C^{\prime[mmk]}_{\chi^0 t}O_R^{\prime n m}\right.\Bigg.\Bigg.\nonumber\\
&&\Bigg.\Bigg.\left.-\left(C^{\prime[nnk]}_{\chi^0 t}+C_{\chi^0 t}^{[nnk]}-C_{\chi^0}^{[nmk]}\frac{m_t}{m_{\chi^0}^{m}}\right)O_L^{\prime n m}\right)\Bigg)\Bigg]\,,
\end{eqnarray}
\begin{eqnarray}
C_{\epsilon}^{c w [ahk]}&&\hspace*{-0.3cm}=\int_{0}^{1}dx\int_{0}^{1-x}dy\left(\frac{1}{\epsilon}-\gamma-\log{\frac{L^c_{a h k}(w)}{4\pi\Lambda^2}}\right)\,,\nonumber\\
C_{\rm top}^{w[ahk]}&&\hspace*{-0.3cm}=\int_{0}^{1}dx\int_{0}^{1-x}dy\frac{m_t^2x(1-x-y)}{L^c_{a h k}(w)}\,,\nonumber\\
C_{w top}^{[ahk]}&&\hspace*{-0.3cm}=\int_{0}^{1}dx\int_{0}^{1-x}dy\frac{m_t m_{w}^{a}(1-x-y)}{L^c_{a h k}(w)}\,,\nonumber\\
C_{\rm SE}^{w[ak]}&&\hspace*{-0.3cm}=\int_{0}^{1}dx\, x\left(\frac{1}{\epsilon}-\gamma-\log\frac{L^{\rm self}_{a k}(w)}{4\pi\Lambda^2}\right)\,,\nonumber\\
C_{\rm SEG}^{w[ak]}&&\hspace*{-0.3cm}=\int_{0}^{1}dx \left(\frac{1}{\epsilon}-\gamma-\log\frac{L^{\rm self}_{a k}(w)}{4\pi\Lambda^2}\right)\,,\nonumber\\
C_{\epsilon}^{d w [abk]}&&\hspace*{-0.3cm}=\int_{0}^{1}dx\int_{0}^{1-x}dy\left(\frac{1}{\epsilon}-\gamma-1-\log{\frac{L^d_{a b k}(w)}{4\pi\Lambda^2}}\right)\,,\nonumber\\
C_w^{[abk]}&&\hspace*{-0.3cm}=\int_{0}^{1}dx\int_{0}^{1-x}dy\frac{m_w^{a} m_w^{b}}{L^d_{a b k}(w)}\,,\nonumber\\
C_{k^2}^{w[abk]}&&\hspace*{-0.3cm}=\int_{0}^{1}dx\int_{0}^{1-x}dy \frac{k^2 x y}{L^d_{a b k}(w)}\,,\nonumber\\
C_{t}^{w[abk]}&&\hspace*{-0.3cm}=\int_{0}^{1}dx\int_{0}^{1-x}dy \frac{m_t^2 x(1-x-y)}{L^d_{a b k}(w)}\,,\nonumber\\
C_{w t}^{[abk]}&&\hspace*{-0.3cm}=\int_{0}^{1}dx\int_{0}^{1-x}dy\frac{m_w^a m_t(1-x-y)}{L^d_{a b k}(w)}\,,\nonumber\\
C_{w t}^{\prime [abk]}&&\hspace*{-0.3cm}=\int_{0}^{1}dx\int_{0}^{1-x}dy\frac{m_w^a m_t\, x}{L^d_{a b k}(w)}\,,\;\;\;\;w=\tilde{g}, \gamma,Z\,,\label{eq:Cfunct}
\end{eqnarray}
\begin{eqnarray}
L^{\rm self}_{a k}(w)&&\hspace*{-0.3cm}=-m_t^2 x (1-x)+(m_{w}^a)^2(1-x)+m_{\tilde{q}_k}^2 x\,,\nonumber\\
L^{c}_{a h k}(w)&&\hspace*{-0.3cm}=-m_t^2 x (1-x-y)+(m_{w}^a)^2(1-x-y)+m_{\tilde{q}_h}^2 x+m_{\tilde{q}_k}^2y-k^2 x y\,,\nonumber\\
L^d_{a b k}(w)&&\hspace*{-0.3cm}=-m_t^2 x (1-x-y)+(m_w^a)^2 y+(m_w^b)^2 x+m_{\tilde{q}_k}^2(1-x-y)-k^2 x y\,.
\label{eq:Ls}
\end{eqnarray}
Here $\epsilon=2-d/2\,$ ($d$ is the number of dimensions), $C_2[F]=4/3\,(C_2[G]=3)$ is the quadratic Casimir operator of the fundamental (adjoint) representation of $SU(3)_C$ with $\sum_{a\prime}\left(T^{a\prime}T^{a\prime}\right)_{ij}=C_2[F]\delta_{ij}$ ($T^{a\prime}$ are the $SU(2)$ generators in the fundamental representation with normalization ${\rm Tr}\left(T^{a\prime}T^{b\prime}\right)=\delta^{a\prime b\prime}/2$). The indices $a, b$ run over the intermediate chargino mass eigenstates from 1 to 5 while $n, m$ run over the intermediate neutralino mass eigenstates from 1 to 9. The indices $k, h$ are used to represent squarks mass eigenstates, which run from 1 to 6. The arbitrary parameter $\Lambda$ is introduced in Eq.~(\ref{eq:Cfunct}) to make the argument of the logarithm function dimensionless and our results are independent of $\Lambda$. For $w=\tilde{g}$, throughout the Eqs.~(\ref{eq:gg})--(\ref{eq:Ls}), we dropped the indices $a,b$ or $n,m$  carried by the $C$ functions defined above with a generic $w$ dependency, as gluinos are assumed to have identical masses. 

There are two constraints on the form factors that one can check. The first one is the cancellations of infinities when we set $d\to4\, (\epsilon\to 0)$ limit and the second one is the gauge invariance requirement that the coefficient of $\gamma^{\mu}$ term in the $t\to c g,c\gamma$ decays should vanish in the  $k^2\to 0$ limit. Obviously this is not the case for the $t\to cZ$ mode.

Cancellation of divergencies can be realized in two ways in $d$-dimension. Either one can adopt the on-shell renormalization scheme (see for example \cite{Muta:1998vi}), where there is no contribution from self energy diagrams (the diagrams (a) and (b) in Fig.~\ref{FeynFig}), and find counterterms which exactly cancel the $1/\epsilon$ terms, or include all possible diagrams contributing to the process and expect to have overall cancellation of
$1/\epsilon$. We followed the latter. The explicit $\epsilon$ dependencies are included inside the $C$-functions defined in Eq.~(\ref{eq:Cfunct}). As mentioned before, this way of writing the form factors has an advantage over other formulations, such as Passarino-Veltman, since it makes the cancellation of $1/\epsilon$ terms more apparent. In the expressions given above, there are, for example, some combinations of $C$-functions appearing repeatedly in which cancellation of $1/\epsilon$ terms requires no computation at all. Note that by using the 
Gordon identity one can decompose  Eq.~(\ref{eq:vertex}) into an alternative form which contains the Dirac structures $\gamma^{\mu}$ and $\sigma^{\mu\nu}k_{\nu}$, where $\sigma^{\mu\nu}=(i/2)[\gamma^\mu,\gamma^\nu]$. So, there will be a $\gamma^{\mu}$ component coming from $2p^{\mu}$ part and while we are checking the cancellation of divergencies, the combination $F_{V\,L(R)}^w+T_{V\,R(L)}^w$ should be considered. However, none of the $T_{V\,R(L)}^w$ form factors has $1/\epsilon$ dependency so that this point is going to be important only in the discussion of gauge invariance. Cancellation of $1/\epsilon$ happens in a nontrivial way but is relatively straightforward to check for the $t\to c g$ and $t\to c \gamma$ decay modes with respect to the case for the $t\to c Z$ decay, which requires use of some further properties of $\Gamma_{QL,QR}$,
\begin{eqnarray}
\sum_{i=1}^3\left(\Gamma_{QL}^{*hi}\Gamma_{QL}^{ki}+\Gamma_{QR}^{*hi}\Gamma_{QR}^{ki}\right)&&\hspace*{-0.3cm}=\delta^{hk}\,,\nonumber\\
\sum_{k=1}^6\Gamma_{QL,QR}^{*ki}\Gamma_{QL,QR}^{kj}&&\hspace*{-0.3cm}=\delta^{ij}\,,
\end{eqnarray} 
with unitarity properties of the $(U^*,V)$ and $N^{\prime}$ matrices defined in Appendix~\ref{sec:AppendixMasses} in the chargino and neutralino sectors, respectively.

Gauge invariance requires vanishing the $\gamma^{\mu}$ terms for massless gauge bosons. Thus, we should expect to have $F_{V\,L(R)}^w+T_{V\,R(L)}^w=0$ for only $V=g, \gamma$ in each $w=\tilde{g},\chi^+, \chi^0$ cases. Using the identities given in Ref.~\cite{Couture:1997ep} for $w=\tilde{g}$ together with the similar ones for $w=\chi^+$ and $\chi^0$ the vanishing of $F_{V\,L(R)}^w+T_{V\,R(L)}^w$ for each $w$ is straightforward.

The branching ratio of the decay $t\to c\,V$ can be given,  neglecting the mass of the charm quark
\begin{eqnarray}
{\rm BR}(t\to c\,Z)&&\hspace*{-0.3cm}=\frac{m_t\left(1-\eta\right)^2}{32\pi\Gamma_t(t\to bW)}\sum_{w=\tilde{g},\chi^+,\chi^0}\Bigg[\left(\eta-2+\frac{1}{\eta}\right)\Big(|T_{ZL}^w|^2+|T_{ZR}^w|^2\Big)\Bigg.\nonumber\\
&&\Bigg.+2\left(-1+\frac{1}{\eta}\right){\rm Re}\Big[T_{ZL}^w F_{ZR}^{*w}+T_{ZR}^{*w} F_{ZL}^{w}\Big]+\left(2+\frac{1}{\eta}\right)\Big(|F_{ZL}^w|^2+|F_{ZR}^w|^2\Big) \Bigg],\nonumber\\
{\rm BR}(t\to c\,V)&&\hspace*{-0.3cm}=\frac{m_t}{16\pi\Gamma_t(t\to bW)}\sum_{w=\tilde{g},\chi^+,\chi^0}\Bigg(|T_{VL}^w|^2+|T_{VR}^w|^2\Bigg)\,,\;\;\;\;V=g,\gamma,
\end{eqnarray}
where $\eta=m_Z^2/m_t^2$ and $\Gamma_t(t\to bW)$ is taken as the total decay width of the top quark.

\section{Numerical Results}\label{sec:Numreslt}
In this section we present some numerical results for the $\rm BR$ of the decays $t\to c\,V\,(V=g,\gamma,Z)$ including the contributions from gluino-up-squark, chargino-down squark and neutralino-up-squark loops within the context of both flavor diagonal and non-diagonal scenarios. We prefer to discuss each contributions separately, as this is missing from the literature.
Within the framework we have chosen, there is a large set of parameters that need to be fixed. Throughout our numerical computations, the SM parameters are taken as $m_b=4.5$ GeV, $m_t=173.5$ GeV, $m_W=80.425$ GeV, $m_Z=91.187$ GeV, $\alpha_s(m_Z)=0.1172$ and $\sin^2\theta_{\rm W}=0.2312$. The elements of CKM matrix are taken as $K_{\rm CKM}^{ui}=\left(0.973, 0.220,3.67\times10^{-3}\right),\,K_{\rm CKM}^{ci}=
\left(-0.224, 0.986, 41.3\times10^{-3}\right),\,K_{\rm CKM}^{ti}=\left(0.01, -0.05, 0.997\right)$. The free SUSY parameters in the LR symmetric framework\footnote{The LRSUSY is assumed not embedded into some larger supersymmetric grand unified theories like $SO(10)$, $E_6$. Otherwise, one can relate some of these parameters by using the relations among them at the unification scale with the use of renormalization group equations. Such framework, however, would lead to more complicated particle spectra.} are $M_R,~g_R,~v_{\Delta_R}$ and $v_{\delta_R}$ in addition to the usual ones, $M_L,~M_V,~M_{\rm susy},~A,~\mu$, and $\tan\beta$ (we assume  bilinear and trilinear scalar couplings in the soft symmetry breaking Lagrangian to be flavor diagonal, $\mu_{ij}\equiv \mu, A_{ij}=A$). The mass of the gluino needs to be fixed as well. In addition to these, there are new flavor changing parameters, $(\delta_{U(D)}^{AB})_{ij},\,\,i\!\ne\! j=2,3,~A,B=L,R$ appearing in both the up-type and down-type squark mass squared matrices defined in Eq.~(\ref{MUFC}). These are relevant only in the flavor non-diagonal case and we vary them in $(0,1)$ interval.

The $5\times5$ chargino and $9\times9$ neutralino mass matrices are diagonalized numerically and the following relations are assumed $M_L=M_R, M_V=M_L/2, g_R=g_L$ and we set $M_L=150$ GeV, $v_{\Delta_R}=v_{\delta_R}=1$ TeV, $\mu=200$ GeV, and $\tan\beta=10$ throughout our analysis.\footnote{We also considered large $\tan\beta$ values and discuss it at the end of Section \ref{sec:FViol}.} There are experimental lower bounds on the masses of charginos, neutralinos as well as gluinos and squarks \cite{Eidelman:2004wy}. For the values of the parameters given above, the lightest chargino is around $130$ GeV and the lightest neutralino $90$ GeV, which are consistent with the experimental lower bounds \cite{Eidelman:2004wy}. To determine the squark masses, we need to further fix $M_{\rm susy}$ and the coupling $A$. We set $A=M_{\rm susy}$ and vary them within $(100,1000)$ GeV in the flavor diagonal case, and set $M_{\rm susy}=300,400,1000$ GeV in the flavor non-diagonal case. For example, the lightest up-type (down-type) squark mass is around $250\,(290)$ GeV for $M_{\rm susy}=300$ GeV in the flavor diagonal context. When the flavor changing parameters are turned on, the lightest up-type (down-type) squark is around $100 (200)$ GeV for intermediate values of $\delta_U^{RR,LL}=\delta_U^{RL,LR}(\delta_D^{RR,LL}=\delta_D^{RL,LR})$. When the flavor violation comes only from the RR or LL sector, we are getting slightly larger squark mass values. We now discuss in the following subsections the flavor diagonal and non-diagonal cases separately. 

\subsection{Flavor Diagonal Case}\label{sec:FD}
We  discuss first the flavor diagonal case and  analyse the flavor changing effects based on the flavor diagonal results. We assume that the SM contribution to the $t\to c V~(V=g,\gamma,Z)$ decays is negligible with respect to the contributions from the LRSUSY as the total branching ratio is concerned. In some cases, chargino or neutralino loop contributions might be comparable with the SM values. However, we mainly concentrate on the relative contributions from each loop involving supersymmetric particles. In all our considerations, the gluino loop dominates both chargino and neutralino loops.
\begin{figure}[htb]
\vspace{-2.6in}  
    \centerline{ \epsfxsize 5.5in {\epsfbox{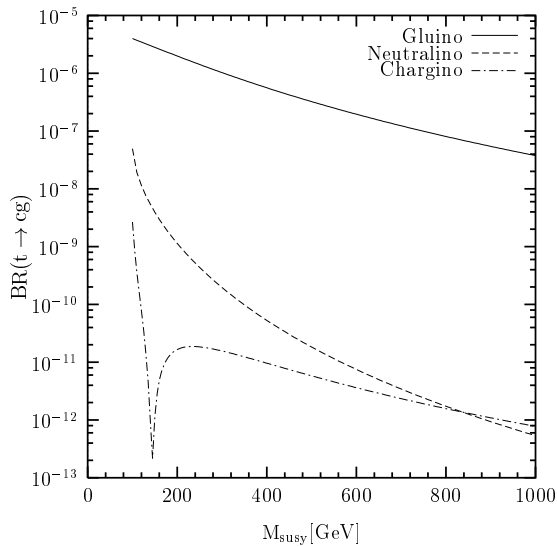}} }
\vskip -2.25in
    \caption{The gluino, chargino and neutralino contributions to the  ${\rm BR}(t\to c g)$ as a function of $M_{\rm susy}$ in the flavor diagonal scenario for $m_{\tilde{g}}=300$ GeV, $\tan\beta=10$ together with the parameter values defined at the beginning of section~\ref{sec:Numreslt}. The solid, dashed, dotted-dashed curves as coded represent gluino, neutralino and chargino contributions, respectively.} \label{fig:tcgFD}
\end{figure}

We first investigate the $t\to c g$ branching ratio as a function of $M_{\rm susy}$ within the range $(100,1000)$ GeV for $\tan\beta=10$ and $m_{\tilde{g}}=300$ GeV. Fig.~{\ref{fig:tcgFD}} shows the dependency including the gluino, chargino and neutralino contributions separately. Neutralino dominates the chargino contribution for the most part of the parameter space (except for large $M_{\rm susy}$ values, where they are roughly equal). There is a two orders of magnitude difference between them for $M_{\rm susy}\sim 200$ GeV which gets smaller as $M_{\rm susy}$ gets larger. The gluino contribution is around three orders of magnitude larger than the neutralino contribution for $M_{\rm susy}\sim 200$ and the difference gets even bigger, reaching $4\times10^{-6}$ when $M_{\rm susy}$ is around 100 GeV. Overall both the neutralino and chargino contributions are practically negligible. In general, the neutralino contribution is bigger than the chargino contribution in the flavor diagonal scenario for $t\to c\,V$ (with the notable exception of the $t\to cZ$ decay mode and the $t\to c\gamma$ decay for certain $M_{\rm susy}$ values) and this is consistent with the findings of Lopez {\it et al.} \cite{Lopez:1997xv}. Note that this picture is reversed for the rare decays involving down-type quarks like bottom-strange quark transitions ($b$ decays). There the chargino contribution is always larger than the corresponding neutralino one. In $b$ decays, the chargino couples  with up-type squarks while here it couples with down-type squarks, whereas the neutralino does the opposite. So, the features seen here are consistent with the characteristics of the down-type rare decays \cite{Frank:2002nj}.
\begin{figure}[htb]
\vspace{-2.7in}  
    \centerline{ \epsfxsize 5.5in {\epsfbox{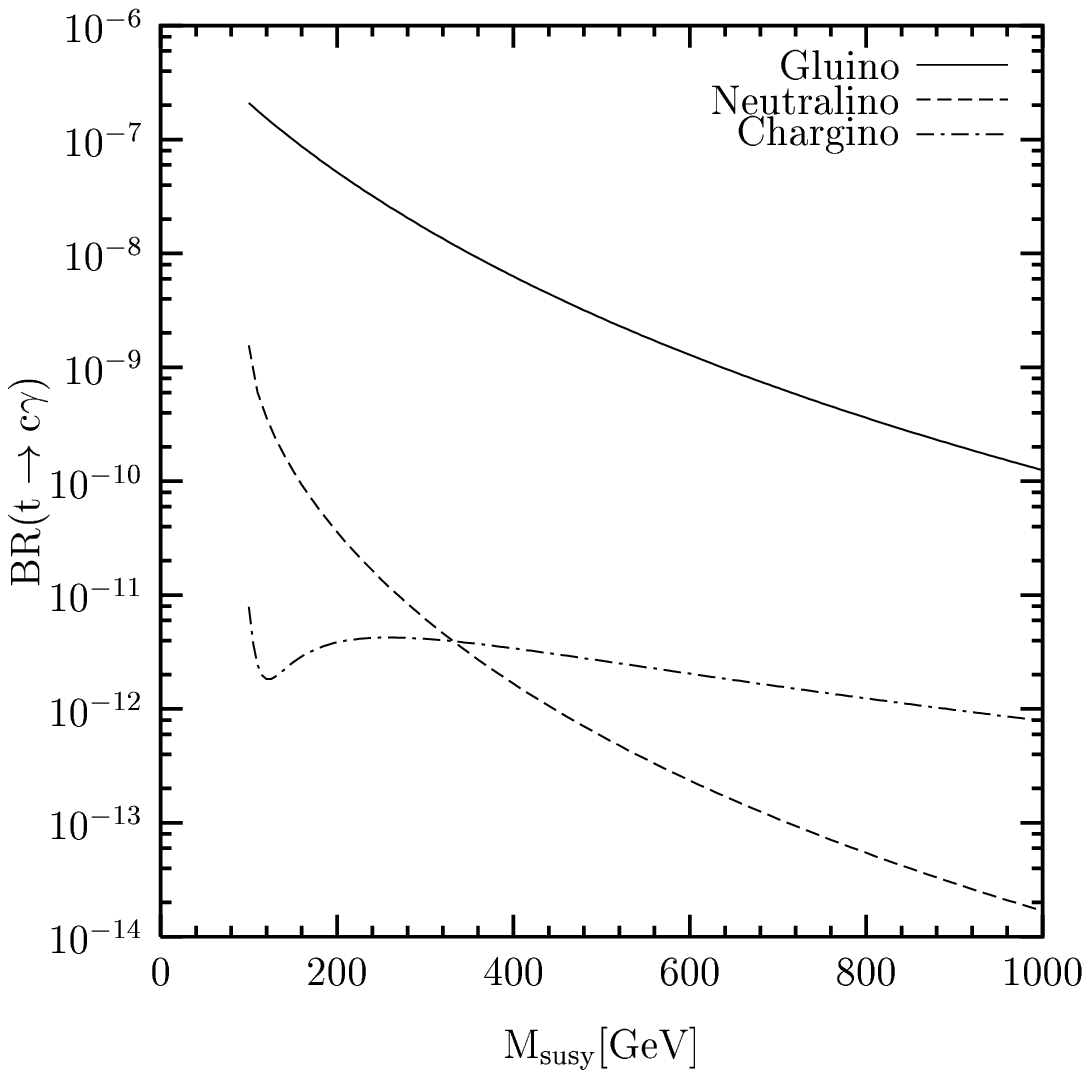}} }
\vskip -2.25in
    \caption{The same as Fig.~\ref{fig:tcgFD} but for the $t\to c\gamma$ decay mode.} \label{fig:tcgammaFD}
\end{figure}

In Fig.~{\ref{fig:tcgammaFD}}, we present the $M_{\rm susy}$ dependency of the branching ratio of the decay $t\to c\gamma$ for the same parameter values chosen above. The gluino contribution is one to two orders of magnitude smaller than the one in the $t\to c g$ decay and it reaches $2\times10^{-7}$ at the maximum level. The neutralino contribution is two to four orders of magnitude smaller than gluino. Even though, after $M_{\rm susy}\sim 300$ GeV, the chargino contribution is larger than the neutralino, both are still negligible in the entire parameter space considered.

\begin{figure}[b]
\vspace{-2.5in}  
    \centerline{ \epsfxsize 5.5in {\epsfbox{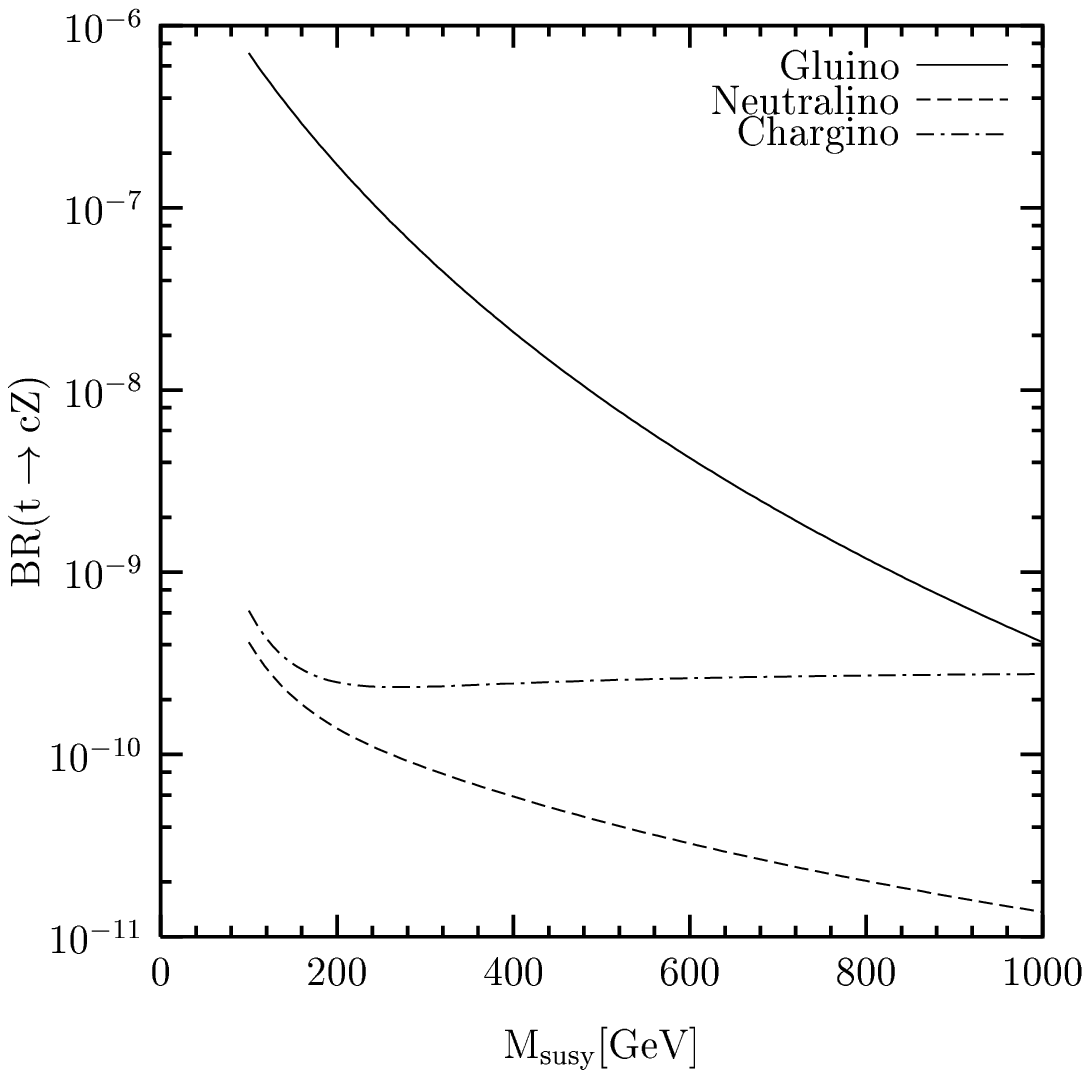}} }
\vskip -2.25in
    \caption{The same as Fig.~\ref{fig:tcgFD} but for the $t\to c\gamma$ decay mode.}\label{fig:tczFDWW2}
\end{figure}

The final figure of this subsection, Fig.~\ref{fig:tczFDWW2}, shows the same dependency, but for $t\to c Z$ decay. The general pattern is the same except that chargino dominates neutralino everywhere in the $(100,1000)$ GeV range for ${\rm M_{susy}}$ (one order of magnitude larger than the neutralino) and furthermore the chargino loop contribution is almost constant as ${\rm M_{susy}}$ changes. The gluino contribution takes maximum values of $7\times10^{-7}$ but decreases sharply as ${\rm M_{susy}}$ gets bigger and even becomes comparable to the chargino contribution.

When we consider the total branching ratios of the three decay modes, in the flavor diagonal case, the largest one is $t\to c g$ as expected and the smallest is the $t\to c \gamma$ decay mode. For instance, we have ${\rm BR}(t\to cg)\sim10^{-6},\,{\rm BR}(t\to cZ)\sim 5\times10^{-8},\,{\rm BR}(t\to c\gamma)\sim10^{-8}$  for $m_{\tilde{g}}=300$ GeV and the intermediate ${\rm M_{susy}}=300$ GeV value . It is also possible to get one to two orders of magnitude larger values for some smaller gluino mass. In all three cases, both the chargino and neutralino contributions  are quite suppressed and negligible for practical purposes.

Before discussing the flavor non-diagonal effects, we would like to comment on the current experimental limits on these decay modes. The best bound for $t\to c\gamma$ decay mode is from CDF \cite{Abe:1997fz}, which looked for flavor changing top quark interactions  in $p\bar{p}$ at $1.8$ TeV center of mass energy. The bound is ${\rm BR}(t\to c\gamma)\le 3.2\times10^{-2}\,(95\%{\rm CL})$. The bound for the $t\to cZ$ decay channel is weaker, ${\rm BR}(t\to cZ)\le 0.137\,(95\%{\rm CL})$ for $m_t=174$ GeV, from the OPAL experiment \cite{Abbiendi:2001wk} which searches for single top quark production in the $e^+e^-\to \bar{t}c$ reaction at around $200$ GeV center of mass energy.\footnote{There is a slightly weaker bound from the ALEPH experiment \cite{ALEPH:2000}.} For the top quark decaying into the charm quark and gluon, there is no current experimental bound available since background problems make detection of such channel difficult \cite{Giammanco:2005ex}.  

These bounds will hopefully be improved by LHC in near future \cite{ATLAS:99}. Both LHC and the future LC have an advantage over some other colliders for detecting rare top decays because of having better statistics capabilities with lower backgrounds. For example, the foreseen sensitivities to these channels can be as small as \cite{Giammanco:2005ex,ATLAS:99}
\begin{eqnarray}
{\rm BR}(t\to c g)&&\hspace*{-0.3cm}\le 7.4\times10^{-3}\,,\nonumber\\
{\rm BR}(t\to c \gamma)&&\hspace*{-0.3cm}\le 1.0\times10^{-4}\,,\nonumber\\
{\rm BR}(t\to c Z)&&\hspace*{-0.3cm}\le 1.1\times10^{-4}\,,
\end{eqnarray}
so that consideration of the flavor non-diagonal effects are essential, since the enhancement from such effects makes it possible to reach the experimentally accessible range for at least some of the modes considered here. Theoretical estimates for top flavor changing neutral couplings have been performed by using parton-level simulations \cite{Han:1995pk}. Further references and a description of the processes involved can be found in \cite{Aguilar-Saavedra:2004wm}.  
\subsection{Flavor Non-Universal Case}\label{sec:FViol}
he remaining part of our study is devoted to discussing the effects of flavor changing mixings in both the up- and down-type squark mass matrices on the rare $t\to c\,V~(V=g,\gamma, Z)$ decays. As motivated in section \ref{sec:FCLRSUSY}, we only concentrate on mixings between the second and third generations and neglect any kind of mixing involving the first generation. Furthermore, unlike some previous studies where the so-called ``mass insertion" method has been used, we follow the general ``mass eigenstate" formalism.  In addition to the parameters in the flavor diagonal case, we have eight more, essentially unknown, parameters in each sector as given in Eq.~(\ref{MUFC}) (that is, there are totally sixteen parameters in both the up and down sectors). 

In order to reduce the set, we should conservatively make some further assumptions and consider some limiting cases. First of all, we set the flavor and chirality diagonal elements of the squark mass matrix to a common SUSY scale, ${\rm M_{susy}}$, which enables us to define dimensionless flavor changing parameters, $(\delta_Q^{AB})_{ij},\;i\!\!\ne\!\!j=2,3,\;A,B=L,R$. In addition to that, we are going to restrict our consideration to a LRSUSY model with Hermitian or symmetric trilinear couplings so that the previously defined mixing parameters satisfy $(\delta_Q^{AB})_{ij}=(\delta_Q^{AB})_{ji},\;\,i\!\!\ne\!\!j=2,3,\;A,B=L,R$. Furthermore, from the LR symmetry we expect $(\delta_Q^{LR})_{ij}=(\delta_Q^{*RL})_{ij}$ to hold, though we will not always assume it. Note that this is in general not necessarily true (MSSM is an example in this respect). Therefore, we are left with four additional parameters, $(\delta_U^{LL})_{23},(\delta_U^{RR})_{23},(\delta_U^{LR})_{23},(\delta_U^{RL})_{23}$, in the up squark sector, and similarly for the down squark sector. 

We investigate five limiting cases: the case where the dominant mixing effects come from only the LL, or RR, or LL+LR, or RR+RL, or LR+RL 
blocks of the squark matrices. However, only the RR, RR+RL and LR+RL cases are presented here\footnote{We concentrate on mainly the RR sector since the model has an additional $SU(2)_R$ group, which makes the right-handed sector more interesting.} since the ones with LL are very similar but not identical. This is basically because  the flavor conserving LR mixing in the second generation is not equal to the one in the third generation. We don't consider limiting cases between $\delta$'s and flavor conserving but chirality changing entries of the matrix in Eq.~(\ref{MUFC}). One reason is that we kept the trilinear soft terms set to $A\!=\!{\rm M_{susy}}$ which makes such limiting cases very similar to the flavor diagonal case which we have discussed in the previous section. So, we proceed to investigate the five limiting cases in each decay. 
\subsubsection{$t\to cg$}\label{subsec:tcgFC}
\begin{figure}[htb]
\vspace{-2.1in}  
    \centerline{ \epsfxsize 4in {\epsfbox{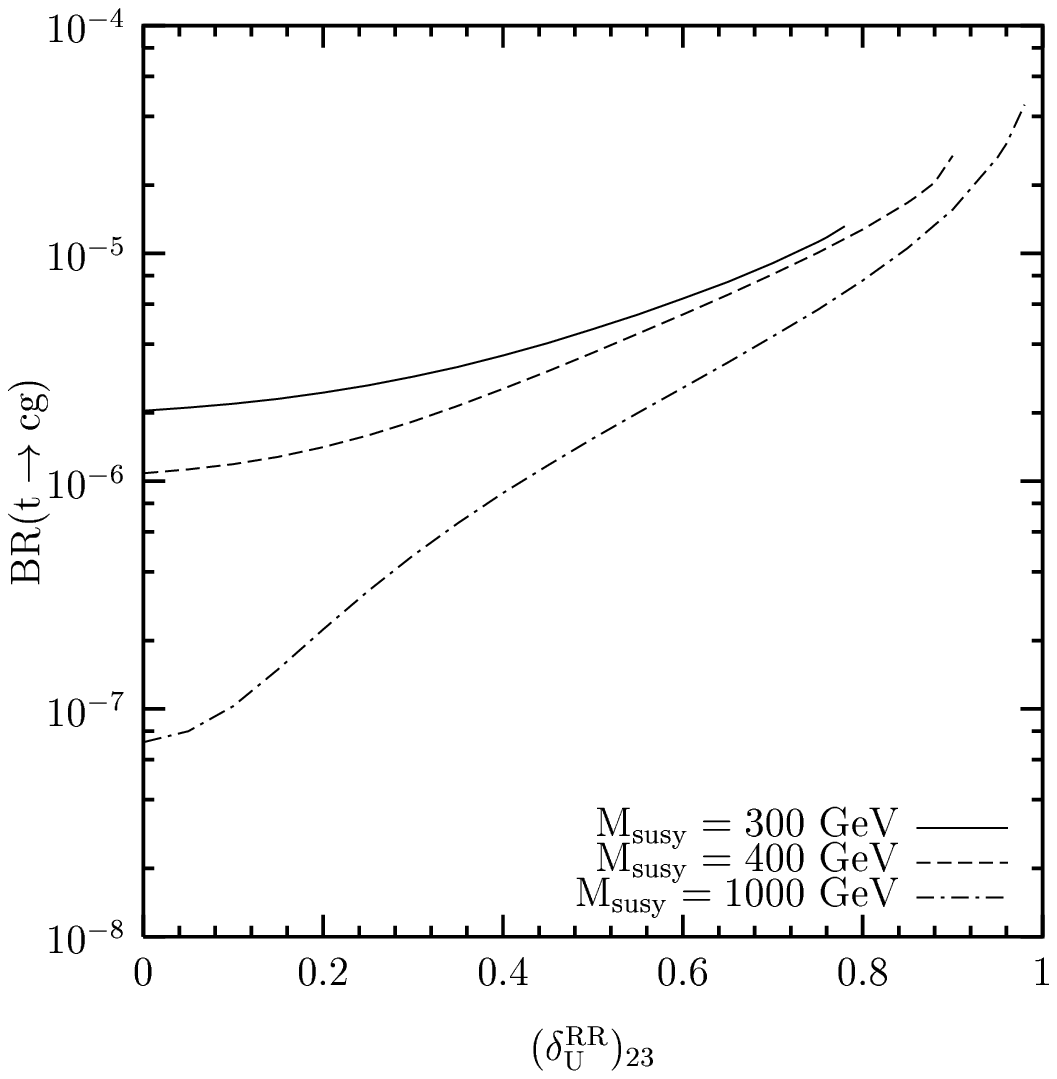}} \hspace{-4.95cm} \epsfxsize 4in {\epsfbox{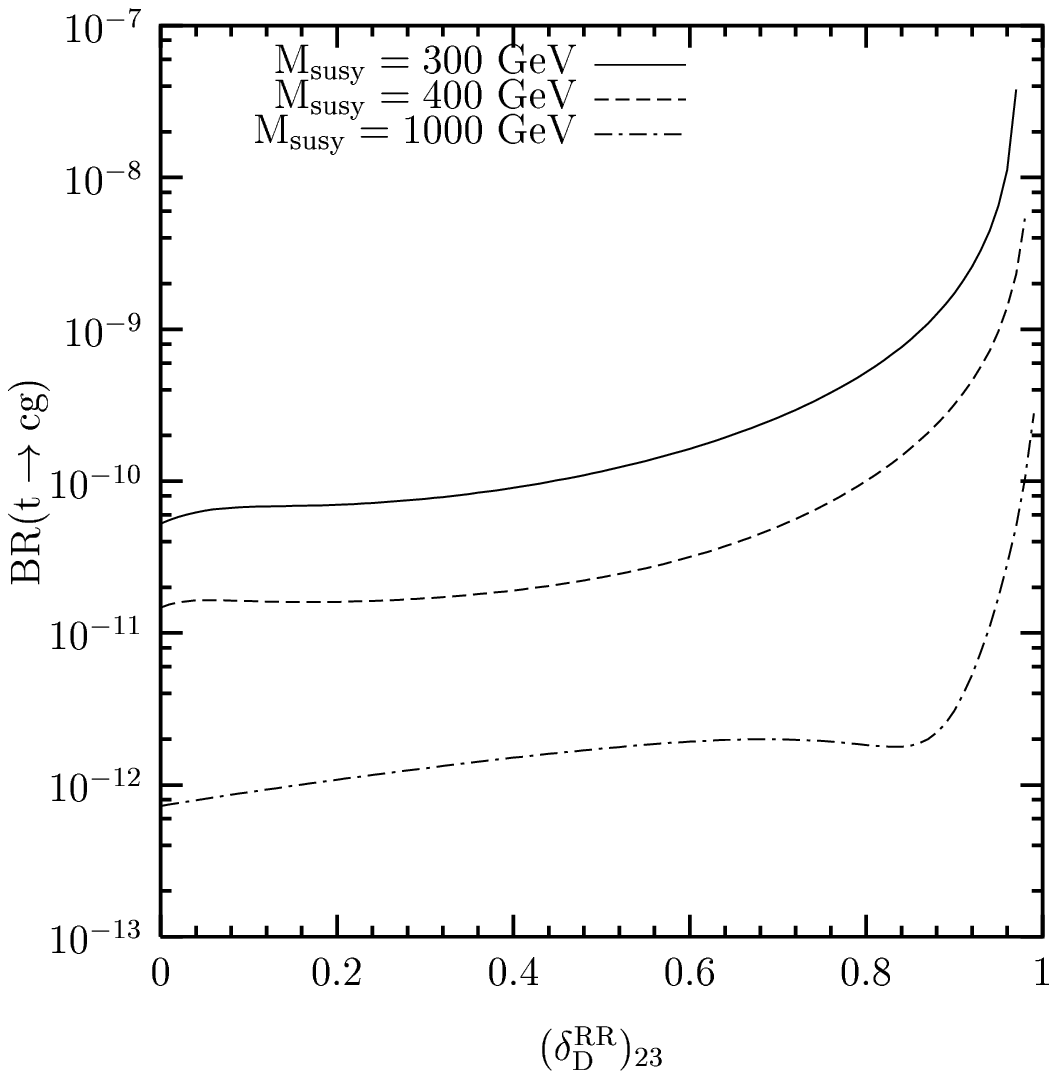}} \hspace{-4.95cm} \epsfxsize 4in {\epsfbox{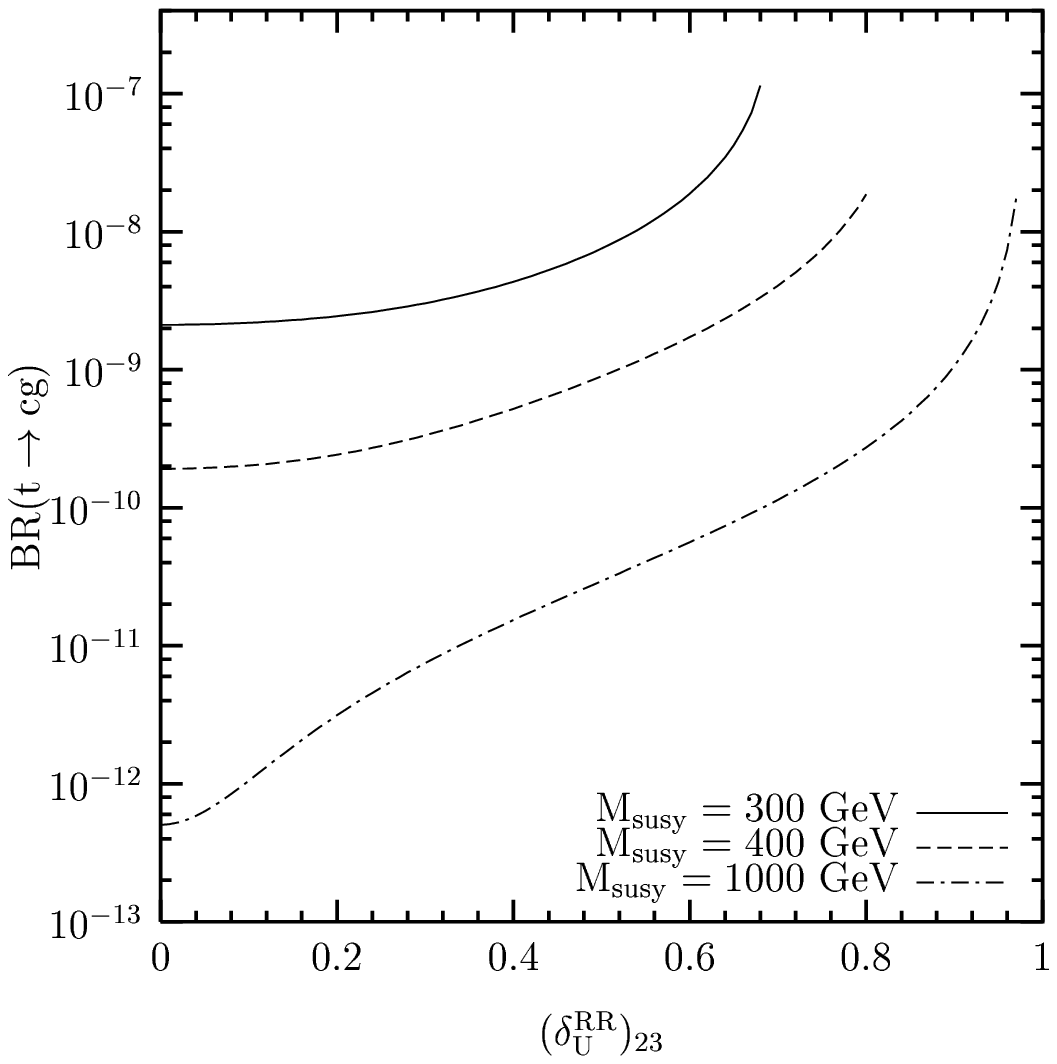}}}
\vskip -1.7in
    \caption{The gluino, chargino and neutralino contributions to the ${\rm BR}(t\to c g)$ as function of $(\delta_{U(D)}^{RR})_{23}$ for the case where the flavor changing effects come from the mixing in the RR sector only. The graph on the left(right) denotes the gluino(neutralino) contribution for the up RR mixing. The one in the middle is for the chargino as a function of $(\delta_D^{RR})_{23}$. Three representative ${\rm M_{susy}}$  values (300,400,1000) GeV are chosen with $m_{\tilde{g}}=300$ GeV, $\tan\beta=10$, $A={\rm M_{susy}}$ and $\mu=200$ GeV.}\label{fig:tcgRR}
\end{figure}

In Fig.~\ref{fig:tcgRR}, we plot in a set of three graphs the gluino, chargino and neutralino contributions to the ${\rm BR}(t\to cg)$ as a function of $(\delta_U^{RR})_{23}$, for ${\rm M_{susy}}=300,400,1000$ GeV, assuming all other mixings to be zero; $m_{\tilde{g}}=300$ GeV and $\tan\beta=10$ are assumed. As seen from the graph on the left, the gluino contribution to the branching ratio depends strongly on $(\delta_U^{RR})_{23}$ and gets enhanced as $(\delta_U^{RR})_{23}$ takes larger values. When we compare  with the flavor diagonal case (Fig.~\ref{fig:tcgFD}), for ${\rm M_{susy}}=300$ GeV, there is at least a factor of five enhancement for intermediate values of $(\delta_U^{RR})_{23}$ and of one order of magnitude for larger values. As ${\rm M_{susy}}$ gets larger, the enhancement is two to three orders of magnitude for intermediate-large values of $(\delta_U^{RR})_{23}$. Two comments are in order. The first one is that the comparison with the flavor diagonal case is not completely right. Here we assume all diagonal elements in the main diagonal identical (to ${\rm M_{susy}}$), which leads to larger values for the branching ratio. That is, the enhancement would be slightly different if we relaxed this condition. The second one is that the full (0,1) interval chosen for the flavor changing parameters is not always available since  squark masses become unphysical when $\delta$'s exceed certain critical values, which also depend on ${\rm M_{susy}}$.

In the same figure, for the same parameter values, we depict the chargino and neutralino contributions to the ${\rm BR}(t\to cg)$ in the second and third graphs, respectively. The chargino contribution is not very sensitive to $\delta$ for $(\delta_D^{RR})_{23}\le 0.8$ and the enhancement with respect to the flavor diagonal case is one order of magnitude in that range but becomes three orders of magnitude after. When we compare it with the gluino contribution, it remains suppressed even under extreme conditions (i.e., ($\delta_U^{RR})_{23}=0,(\delta_D^{RR})_{23}\sim1$). So, the flavor mixing effects from down squark sector do not contribute to the total $\rm BR$ of the $t\to c g$ decay nearly as much as the ones from the up squark sector. For the neutralino case, even though there is three to four orders of magnitude enhancement for intermediate or large $(\delta_D^{RR})_{23}$ values, it is at least one order of magnitude smaller than the gluino contribution. Neutralino contribution still dominates the chargino one unless  $(\delta_U^{RR})_{23}=0,(\delta_D^{RR})_{23}\sim1$ occurs.
\begin{figure}[htb]
\vspace{-2.1in}  
    \centerline{ \epsfxsize 4in {\epsfbox{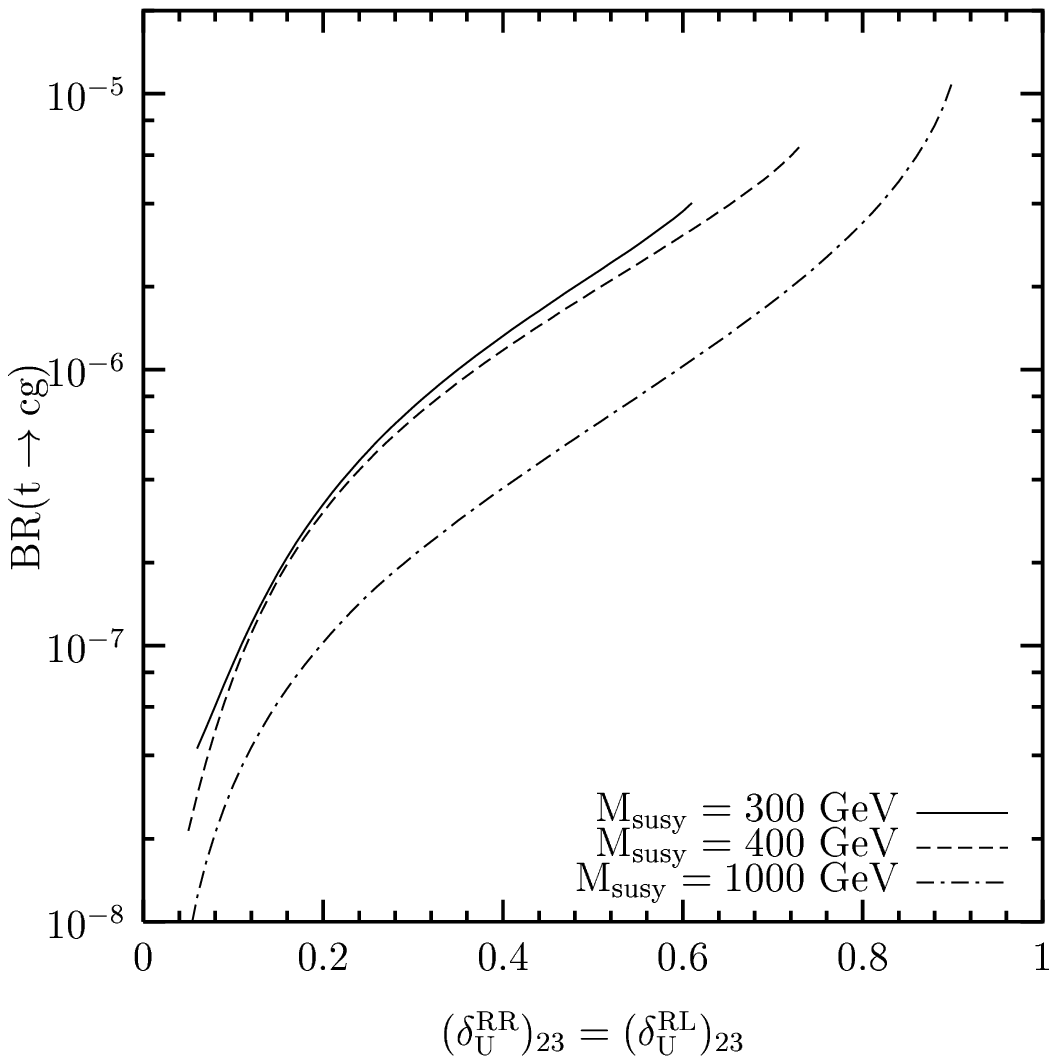}} \hspace{-4.95cm} \epsfxsize 4in {\epsfbox{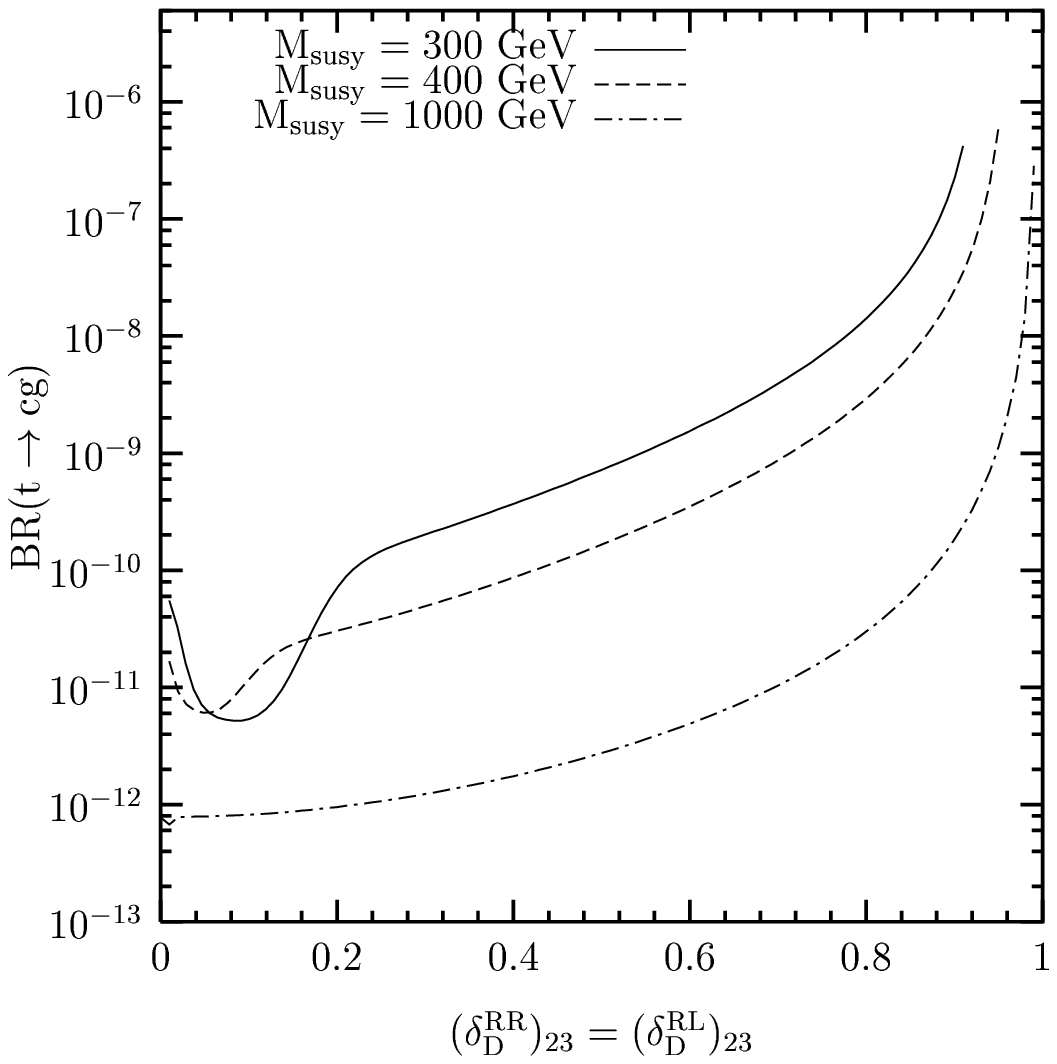}} \hspace{-4.95cm} \epsfxsize 4in {\epsfbox{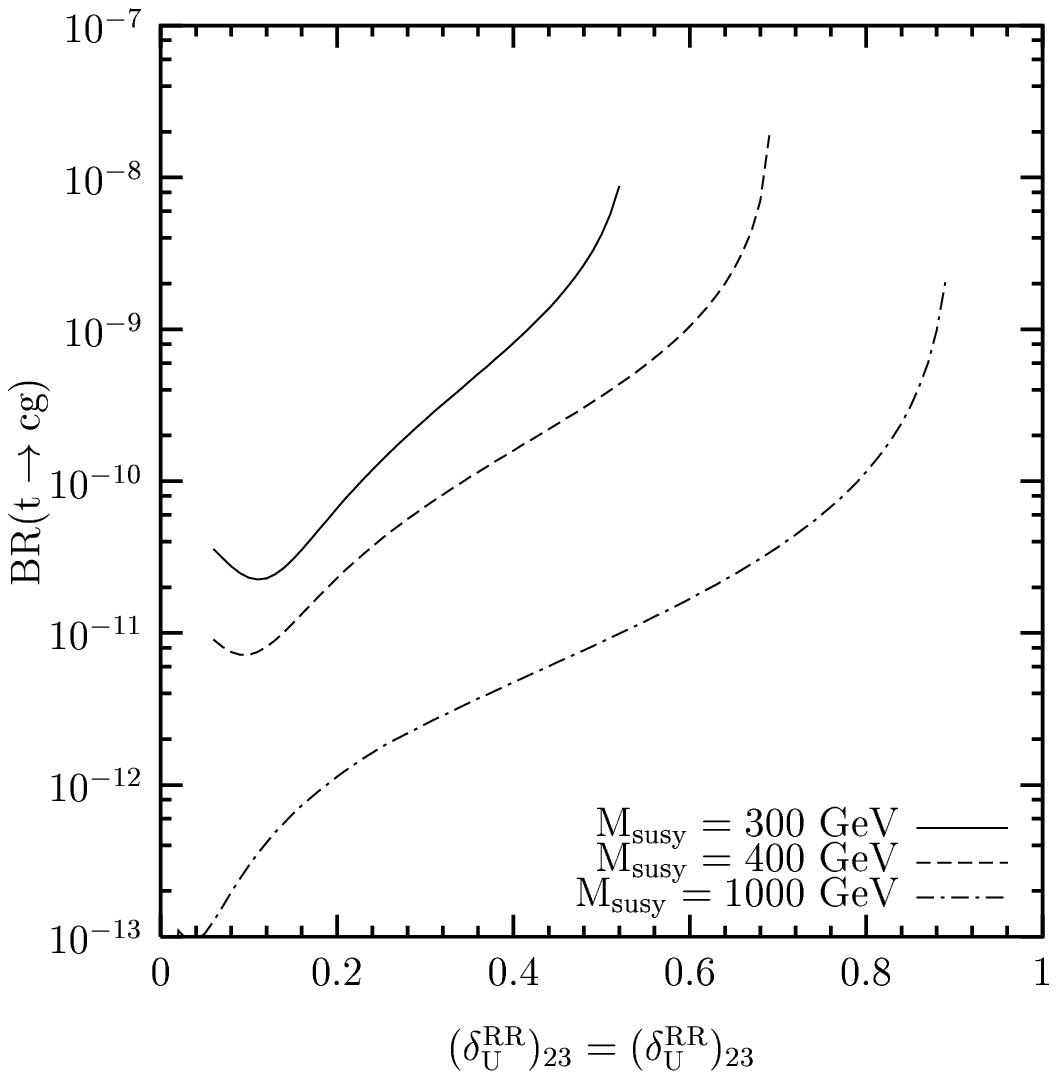}}}
\vskip -1.7in
    \caption{The same as Fig.\ref{fig:tcgRR} but for the case where both $(\delta_{U(D)}^{RR})_{23}$ and $(\delta_{U(D)}^{RL})_{23}$ contribute, with the assumption $(\delta_{U(D)}^{RR})_{23}=(\delta_{U(D)}^{RL})_{23}$.}\label{fig:tcgRRRL}
\end{figure}

In Fig.~\ref{fig:tcgRRRL}, we show the same as Fig.~\ref{fig:tcgRR} but this time additionally the mixing in the LR sector is turned on. For simplicity we present the case $(\delta_{U(D)}^{RR})_{23}=(\delta_{U(D)}^{RL})_{23}$. With respect to the case where only the RR mixing is turned on, except the chargino contribution, the gluino and neutralino contributions are suppressed about one order of magnitude. However, in the down sector, the chargino contribution depends strongly on $(\delta_D^{RR})_{23}$ and reaches $10^{-6}$ at maximum level. In the rest of the interval, it is still very suppressed in the total ${\rm BR}(t\to cg)$.
\begin{figure}[htb]
\vspace{-2.1in}  
    \centerline{ \epsfxsize 4in {\epsfbox{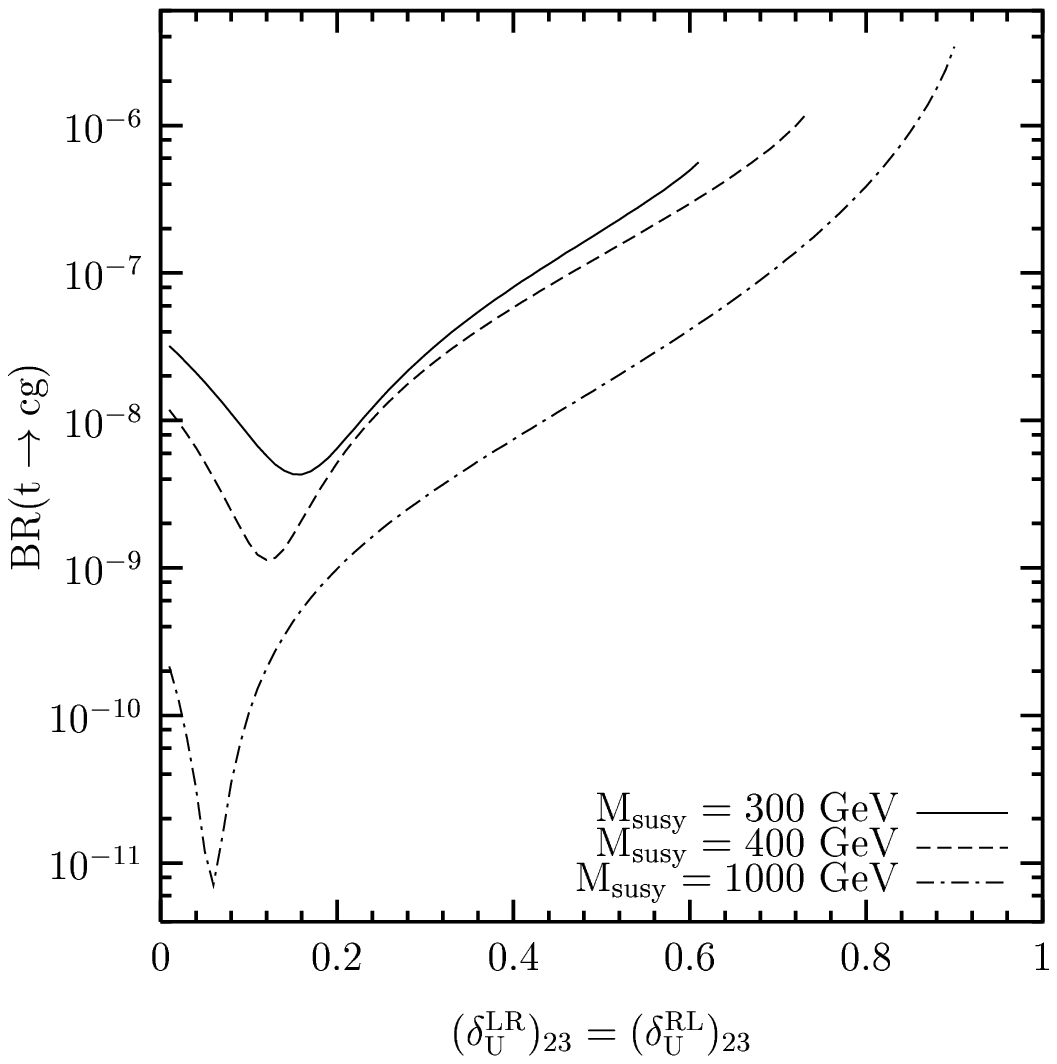}} \hspace{-4.95cm} \epsfxsize 4in {\epsfbox{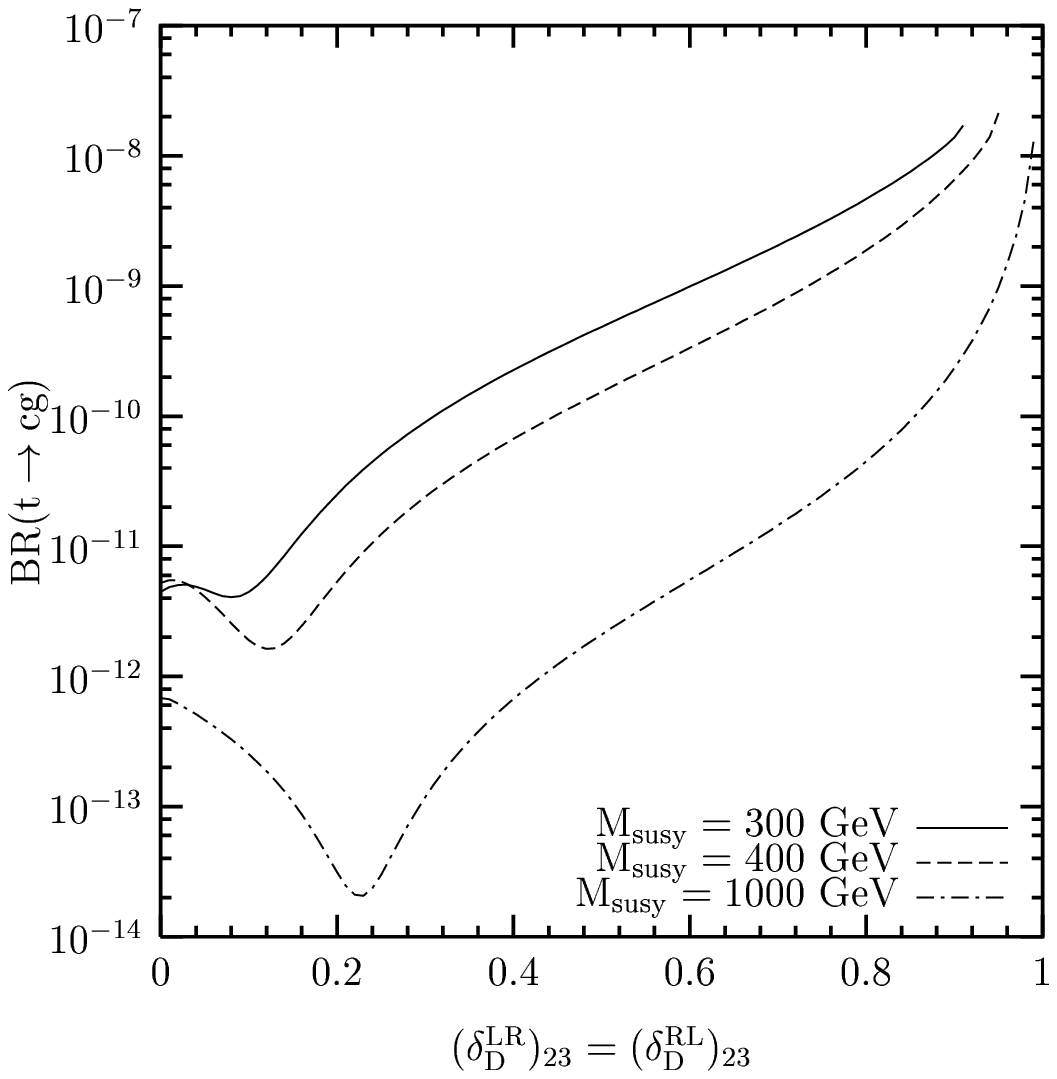}} \hspace{-4.95cm} \epsfxsize 4in {\epsfbox{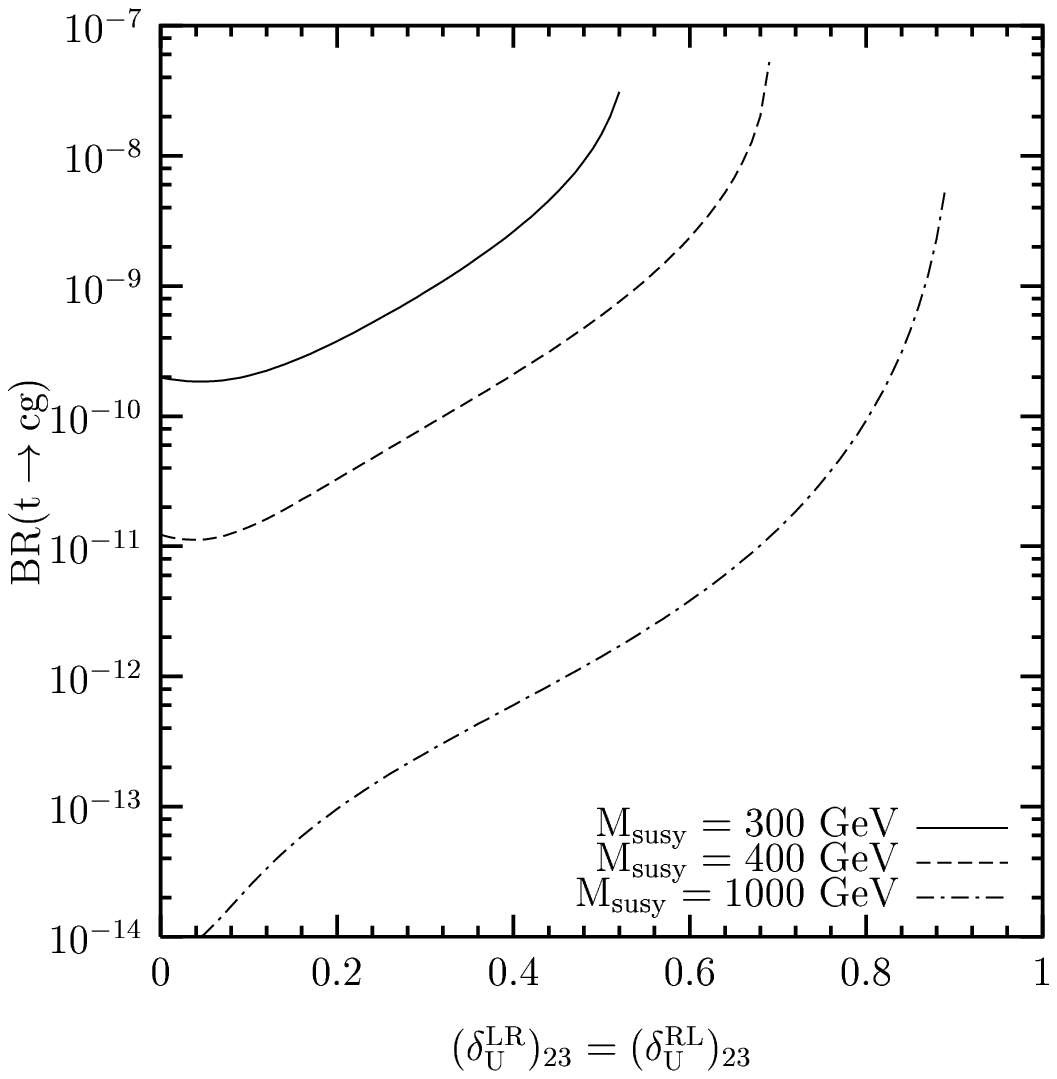}}}
\vskip -1.7in
    \caption{The same as Fig.\ref{fig:tcgRR} but for the case where both $(\delta_{U(D)}^{LR})_{23}$ and $(\delta_{U(D)}^{RL})_{23}$ contribute, with the assumption $(\delta_{U(D)}^{LR})_{23}=(\delta_{U(D)}^{RL})_{23}$.}\label{fig:tcgLRRL}
\end{figure}

The last case for the ${\rm BR}(t\to cg)$, as a function of $(\delta_{U(D)}^{LR})_{23}=(\delta_{U(D)}^{RL})_{23}$, is depicted in Fig.~\ref{fig:tcgLRRL}. We have smaller contributions with respect to both the RR and RR+RL cases except for the neutralino contribution, which reaches values obtained in the RR+RL case for upper values of $(\delta_U^{LR})_{23}$ at different ${\rm M_{susy}}$ values. So, overall, the gluino gives the largest contribution among three in all limiting cases considered above, and the total BR can get up to a few times $10^{-5}$ for $m_{\tilde{g}}=300$ GeV.
\subsubsection{$t\to c\gamma$}\label{subsec:tcgammaFC}
\begin{figure}[htb]
\vspace{-2.1in}  
    \centerline{ \epsfxsize 4in {\epsfbox{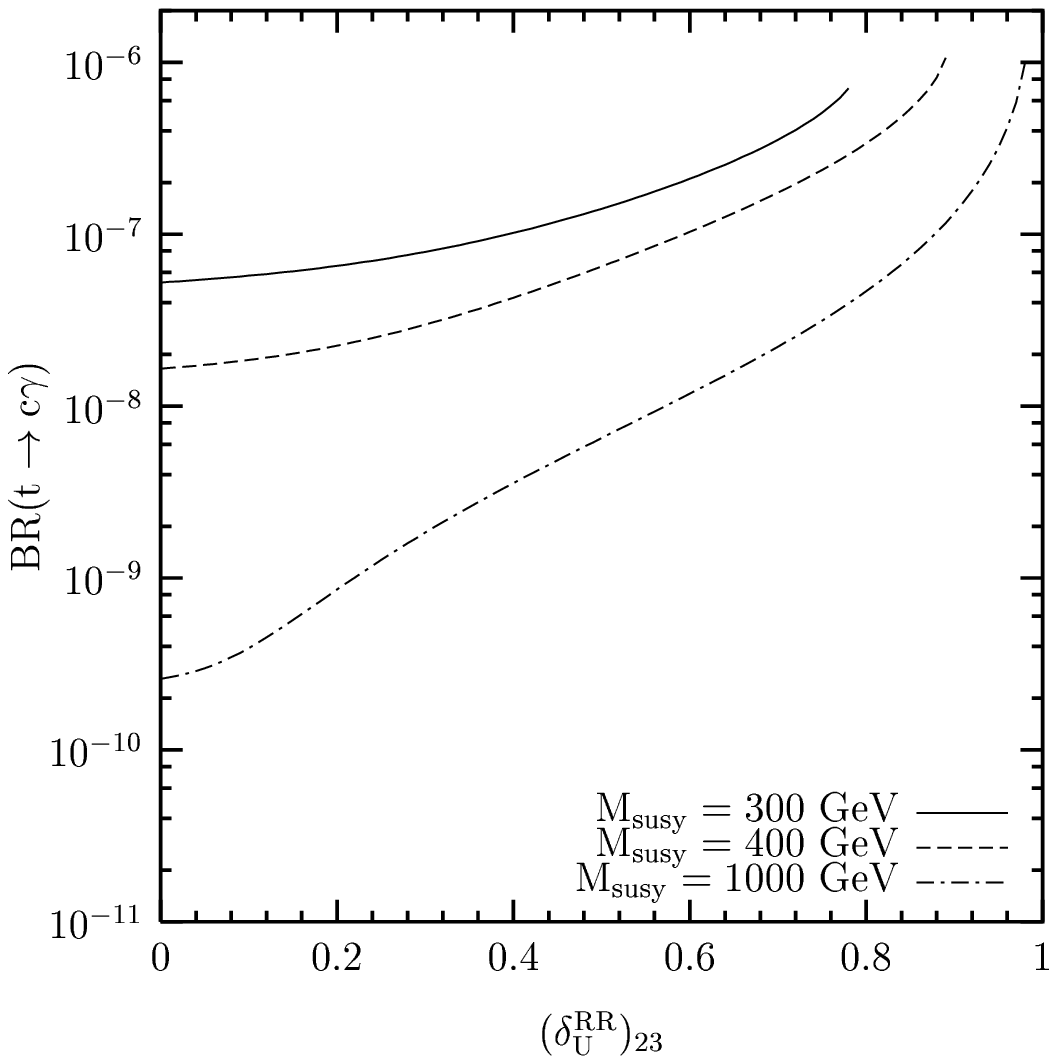}} \hspace{-4.95cm} \epsfxsize 4in {\epsfbox{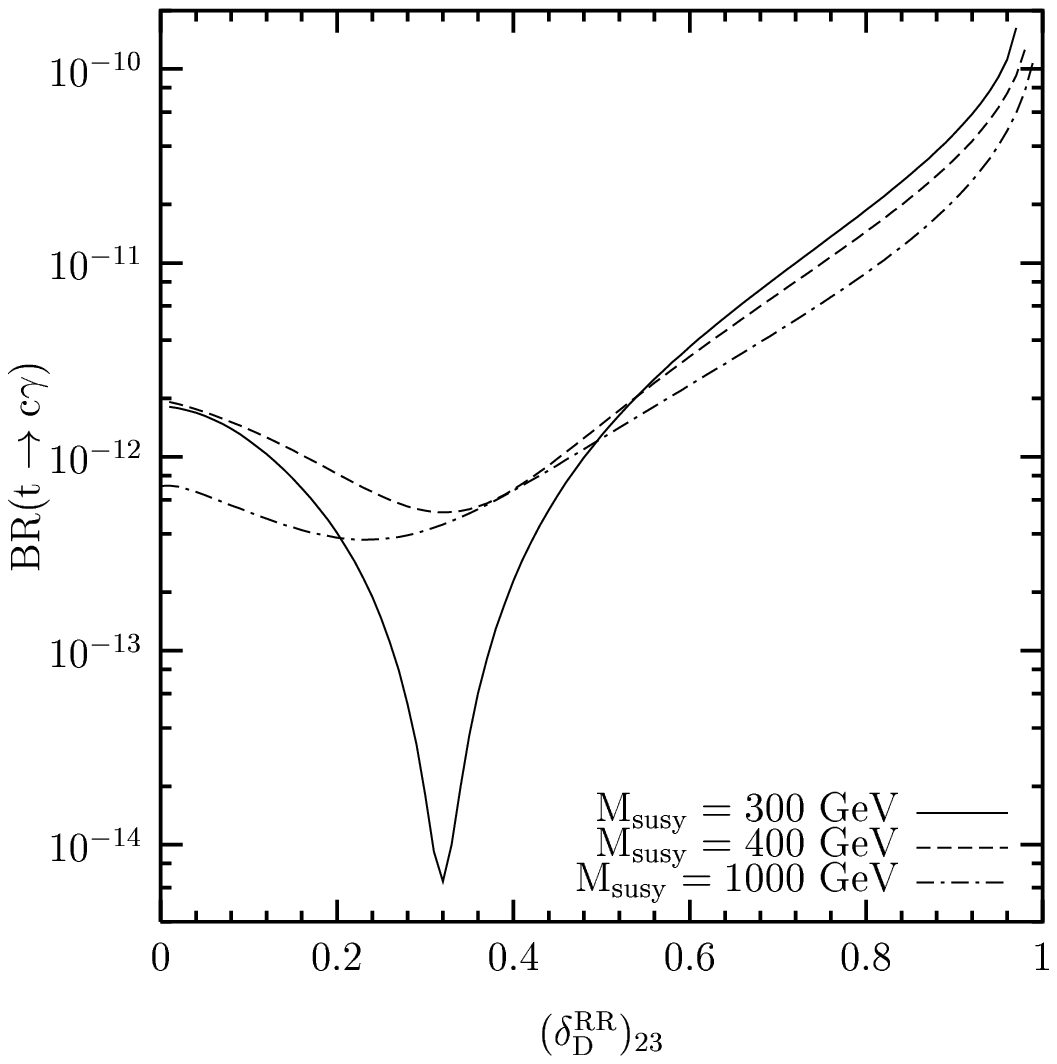}} \hspace{-4.95cm} \epsfxsize 4in {\epsfbox{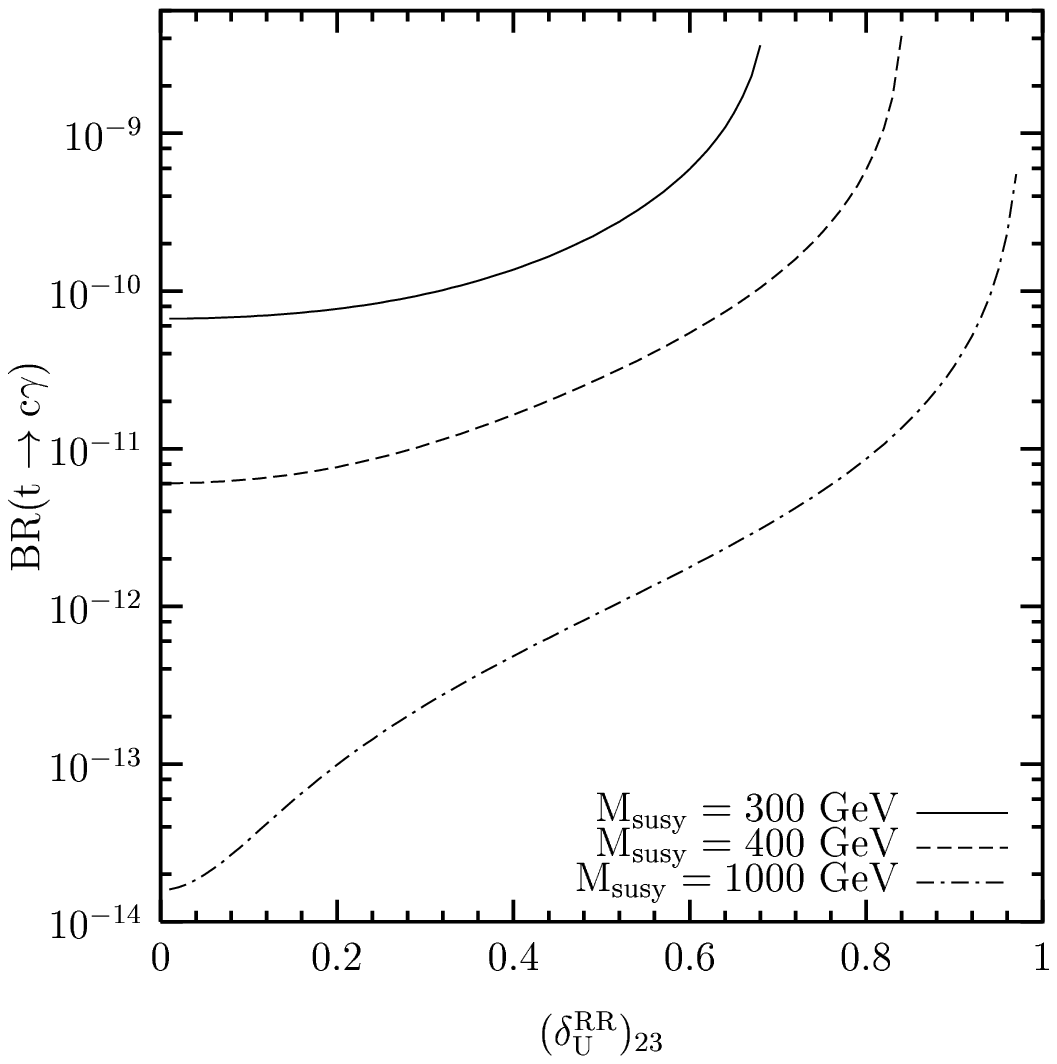}}}
\vskip -1.7in
    \caption{The gluino, chargino and neutralino contributions to the ${\rm BR}(t\to c \gamma)$ as function of $(\delta_{U(D)}^{RR})_{23}$ for the case where the flavor changing effects come from the mixing in the RR sector only. The graph on the left(right) denotes the gluino(neutralino) contribution for the up RR mixing. The one in the middle is for the chargino as a function of $(\delta_D^{RR})_{23}$. Three representative ${\rm M_{susy}}$  values (300,400,1000) GeV are chosen, with $m_{\tilde{g}}=300$ GeV, $\tan\beta=10$, $A={\rm M_{susy}}$ and $\mu=200$ GeV.}\label{fig:tcgammaRR}
\end{figure}

We analyse the $t\to c \gamma$ decay in the same order as we did the $t\to cg$ decay and we present only the RR and RR+RL cases. In Fig.~\ref{fig:tcgammaRR}, we show the dependence of the gluino, chargino and neutralino contributions in the ${\rm BR}(t\to c\gamma)$ to the parameter $(\delta_{U(D)}^{RR})_{23}$ at different ${\rm M_{susy}}$ values, $300,400,1000$ GeV for $\tan\beta=10$. The gluino contribution is now one order of magnitude bigger than in the flavor diagonal case for ${\rm M_{susy}}=300$ GeV and $(\delta_U^{RR})_{23}\sim 0.5$ and it can go up to three orders larger  for ${\rm M_{susy}}=1000$ GeV with maximum possible mixing in the up RR sector, or for smaller ${\rm M_{susy}}$ with slightly smaller mixings, when it reaches $10^{-6}$. In the chargino case, in the down sector, the dependence of the BR to ${\rm M_{susy}}$ is very weak and enhancement with respect to the flavor-diagonal case varies from two to three times to about two orders of magnitude. Here the curve for ${\rm M_{susy}}=300$ GeV has a minimum which is peculiar only to this ${\rm M_{susy}}$ value and occurs due to some precise cancellations in the loop functions. For example, we don't have such behavior in the down LL mixing case which we haven't discussed here explicitly. It is, however, possible to get  bigger enhancements for smaller ${\rm M_{susy}}$ values. For the neutralino contribution, bigger enhancements ranging between two to four orders of magnitude occur, depending on the value of ${\rm M_{susy}}$ and the amount of mixing allowed in the up RR sector. Unlike the flavor diagonal case, the neutralino contribution always dominates the chargino one through the entire interval scanned here; while it is still suppressed with respect to the gluino contribution.
\begin{figure}[htb]
\vspace{-2.1in}  
    \centerline{ \epsfxsize 4in {\epsfbox{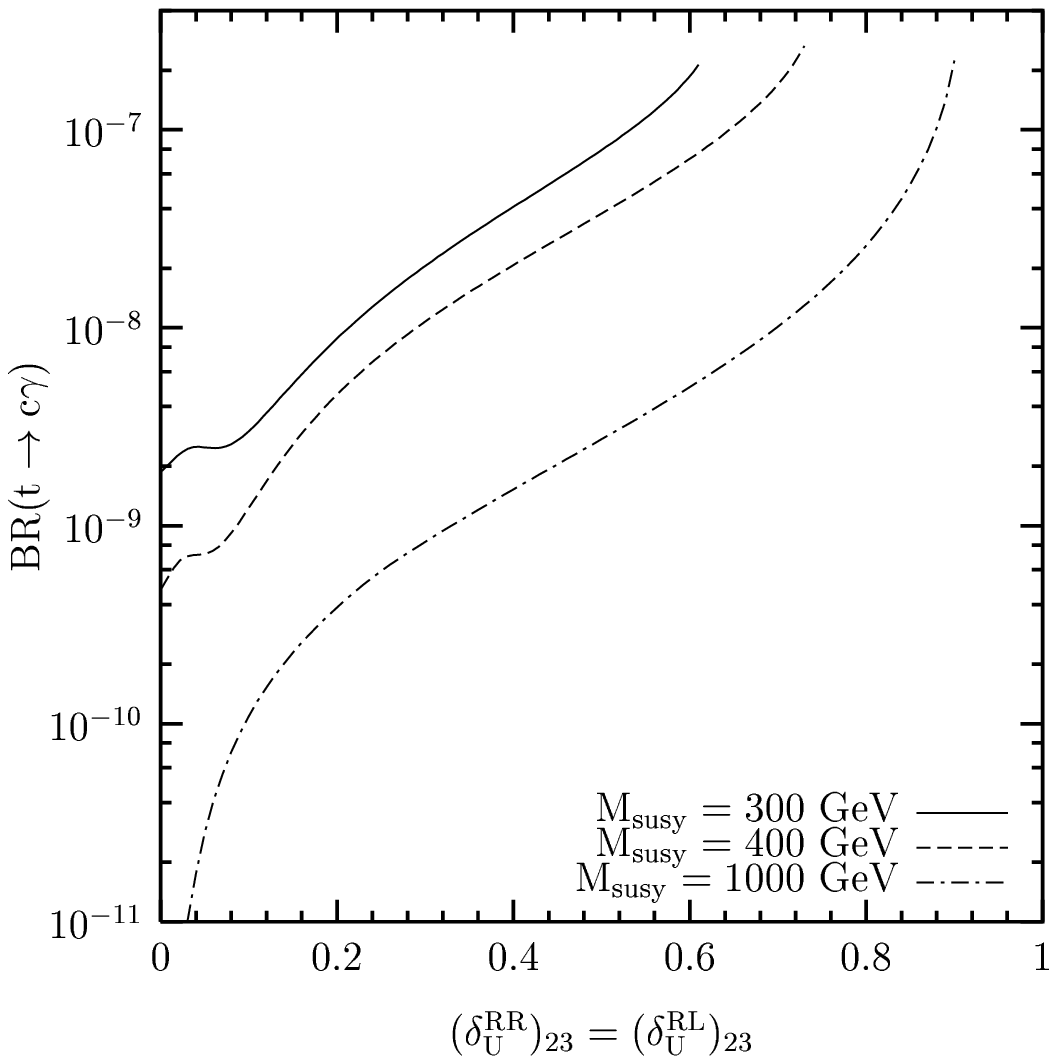}} \hspace{-4.95cm} \epsfxsize 4in {\epsfbox{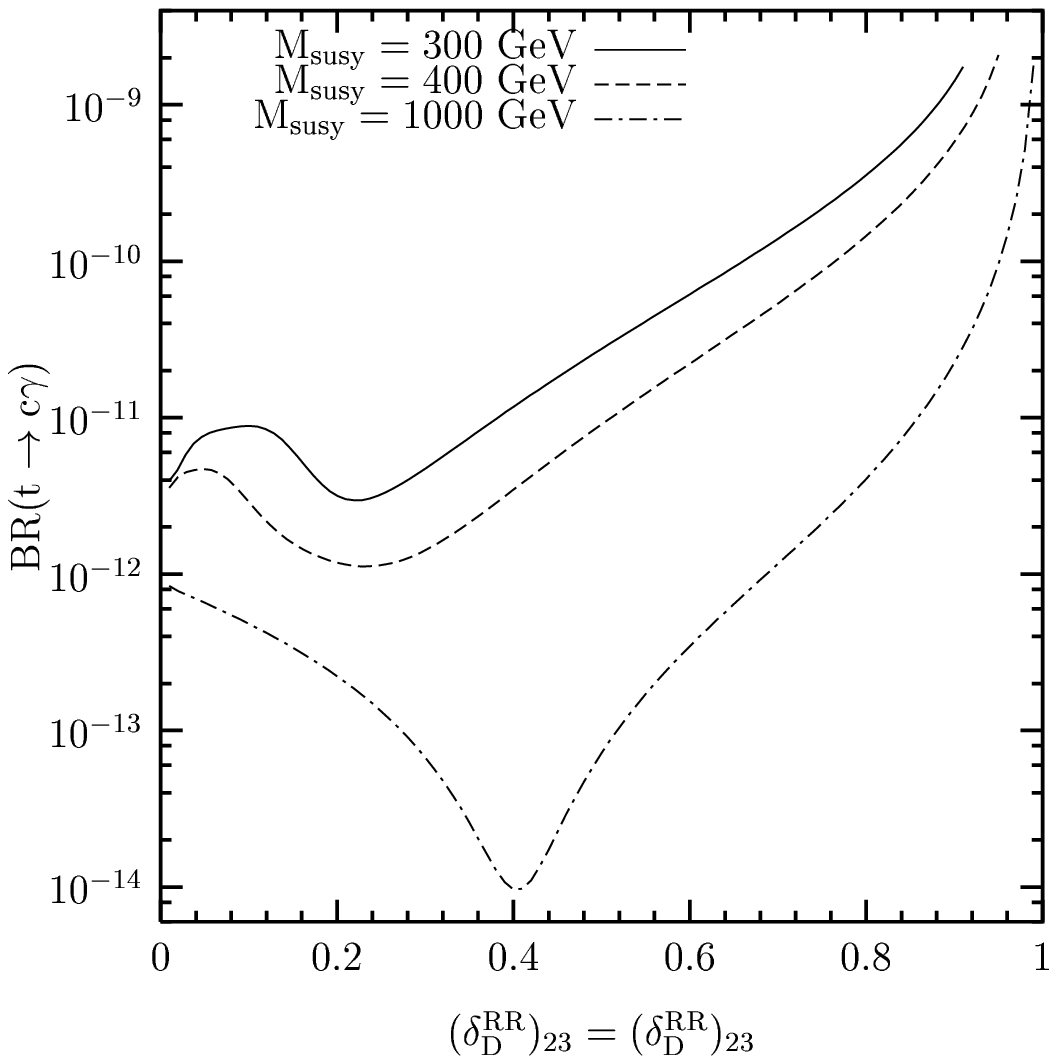}} \hspace{-4.95cm} \epsfxsize 4in {\epsfbox{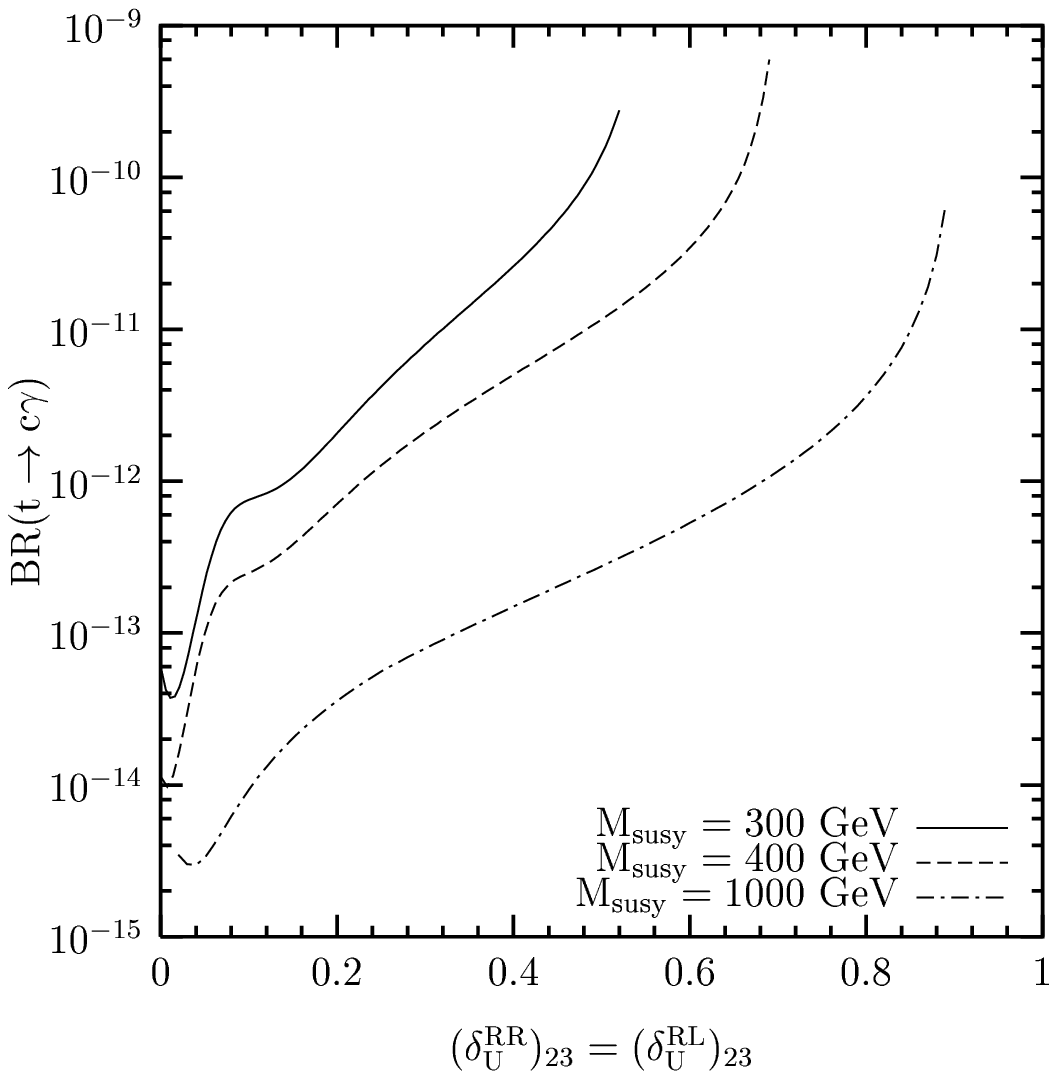}}}
\vskip -1.7in
    \caption{The same as Fig.\ref{fig:tcgammaRR} but for the case where both $(\delta_{U(D)}^{RR})_{23}$ and $(\delta_{U(D)}^{RL})_{23}$ contribute, with the assumption $(\delta_{U(D)}^{RR})_{23}=(\delta_{U(D)}^{RL})_{23}$.}\label{fig:tcgammaRRRL}
\end{figure}

The RR+RL case ($(\delta_{U(D)}^{RR})_{23}=(\delta_{U(D)}^{RL})_{23}$) for the ${\rm BR}(t\to c\gamma)$ is similar and shown in Fig.~\ref{fig:tcgammaRRRL}. The gluino contribution remains always one order smaller than in the RR mixing case. This is true for neutralino as well even though there is a sharp dependency on the mixing parameters in the up sector. In this case, like in the flavor diagonal case, the chargino curves can still cross the neutralino curves and become larger with larger down type RR+RL mixing. If one considers the mixing in the up and the one in the down sector to be completely independent, one could end up with different conclusions.
\subsubsection{$t\to c Z$}\label{subsec:tczFC}
\begin{figure}[htb]
\vspace{-2.1in}  
    \centerline{ \epsfxsize 4in {\epsfbox{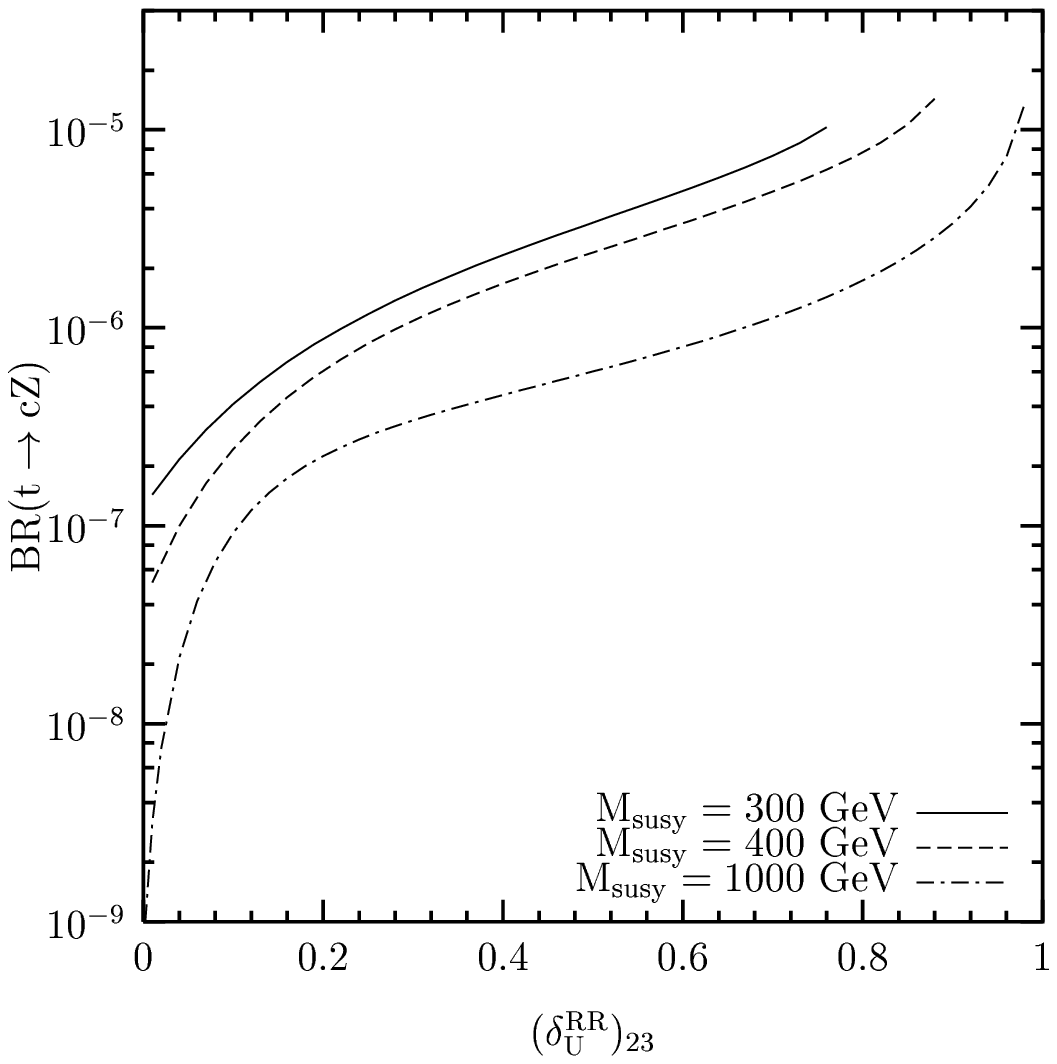}} \hspace{-4.95cm} \epsfxsize 4in {\epsfbox{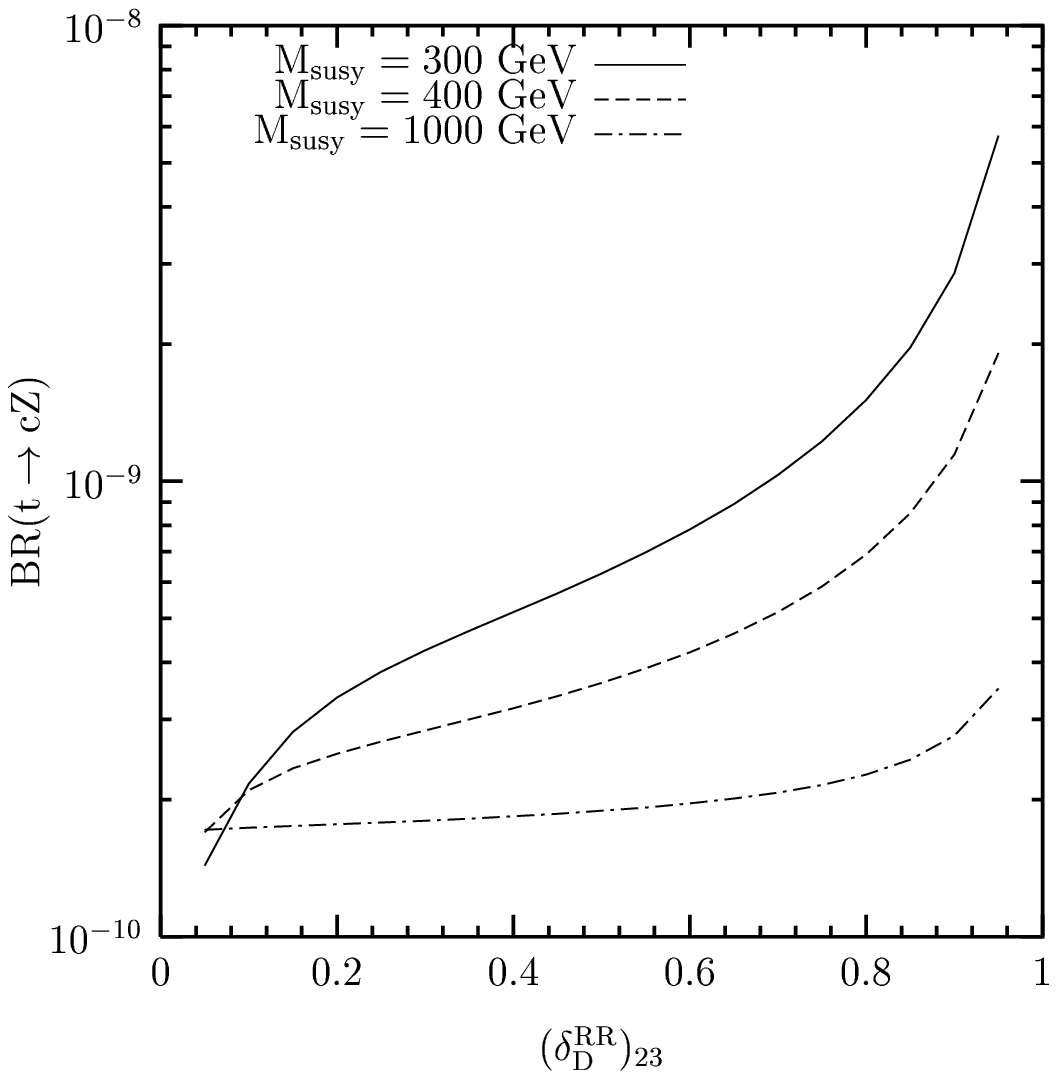}} \hspace{-4.95cm} \epsfxsize 4in {\epsfbox{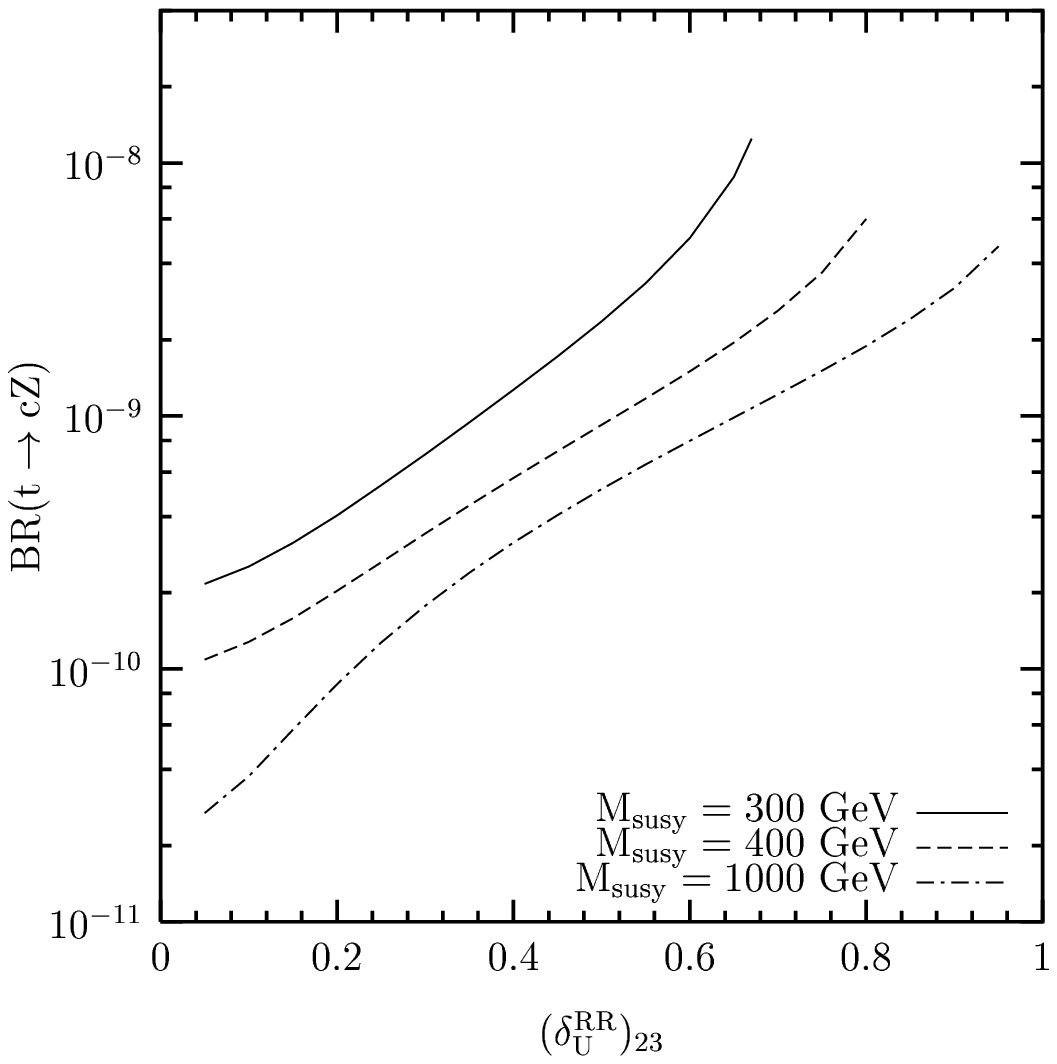}}}
\vskip -1.7in
    \caption{The gluino, chargino and neutralino contributions to the ${\rm BR}(t\to c Z)$ as function of $(\delta_{U(D)}^{RR})_{23}$ for the case where the flavor changing effects come from the mixing in the RR sector only. The graph on the left(right) denotes the gluino(neutralino) contribution for the up RR mixing. The one in the middle is for the chargino as a function of $(\delta_D^{RR})_{23}$. Three representative ${\rm M_{susy}}$  values (300,400,1000) GeV are chosen with $m_{\tilde{g}}=300$ GeV, $\tan\beta=10$, $A={\rm M_{susy}}$ and $\mu=200$ GeV.}\label{fig:tczRR}
\end{figure}
\begin{figure}[b]
\vspace{-2.1in}  
    \centerline{ \epsfxsize 4in {\epsfbox{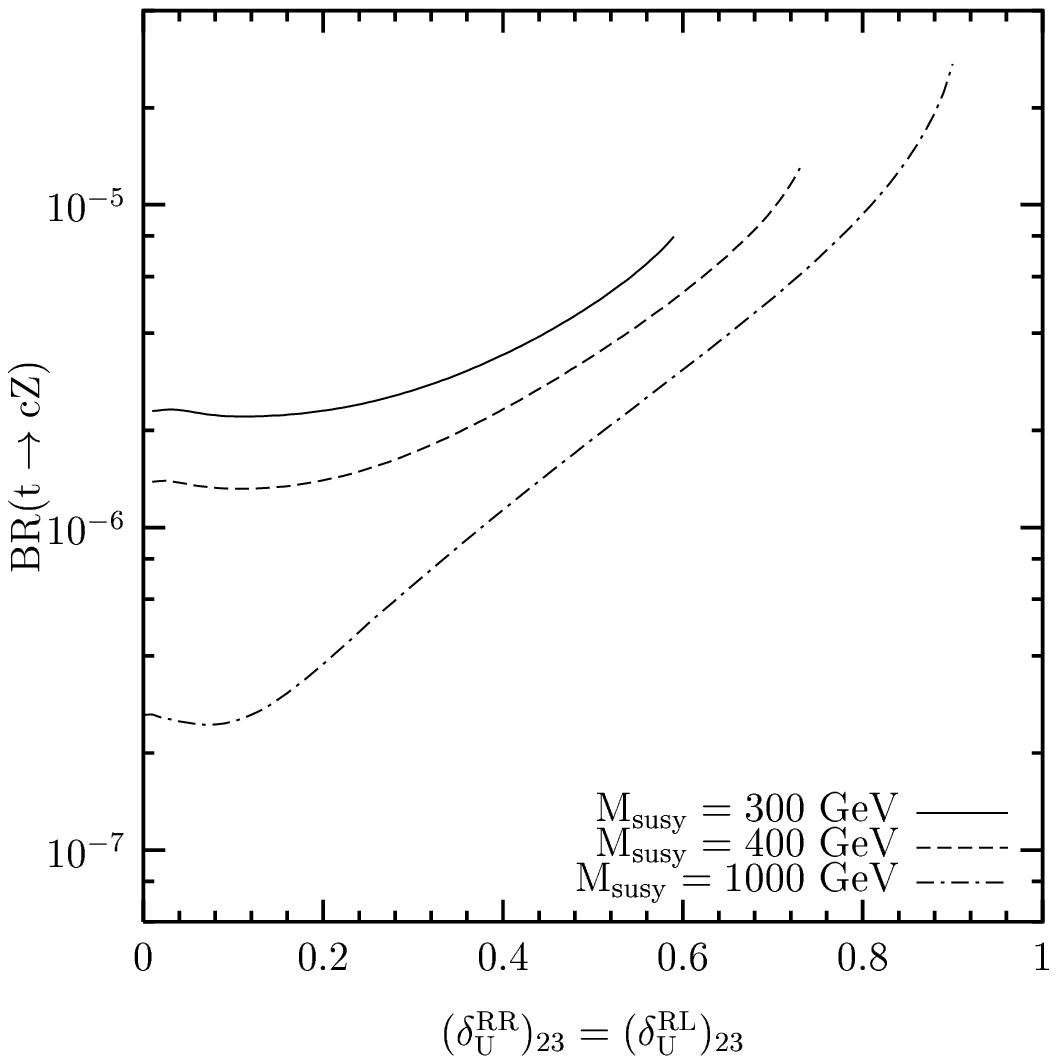}} \hspace{-4.95cm} \epsfxsize 4in {\epsfbox{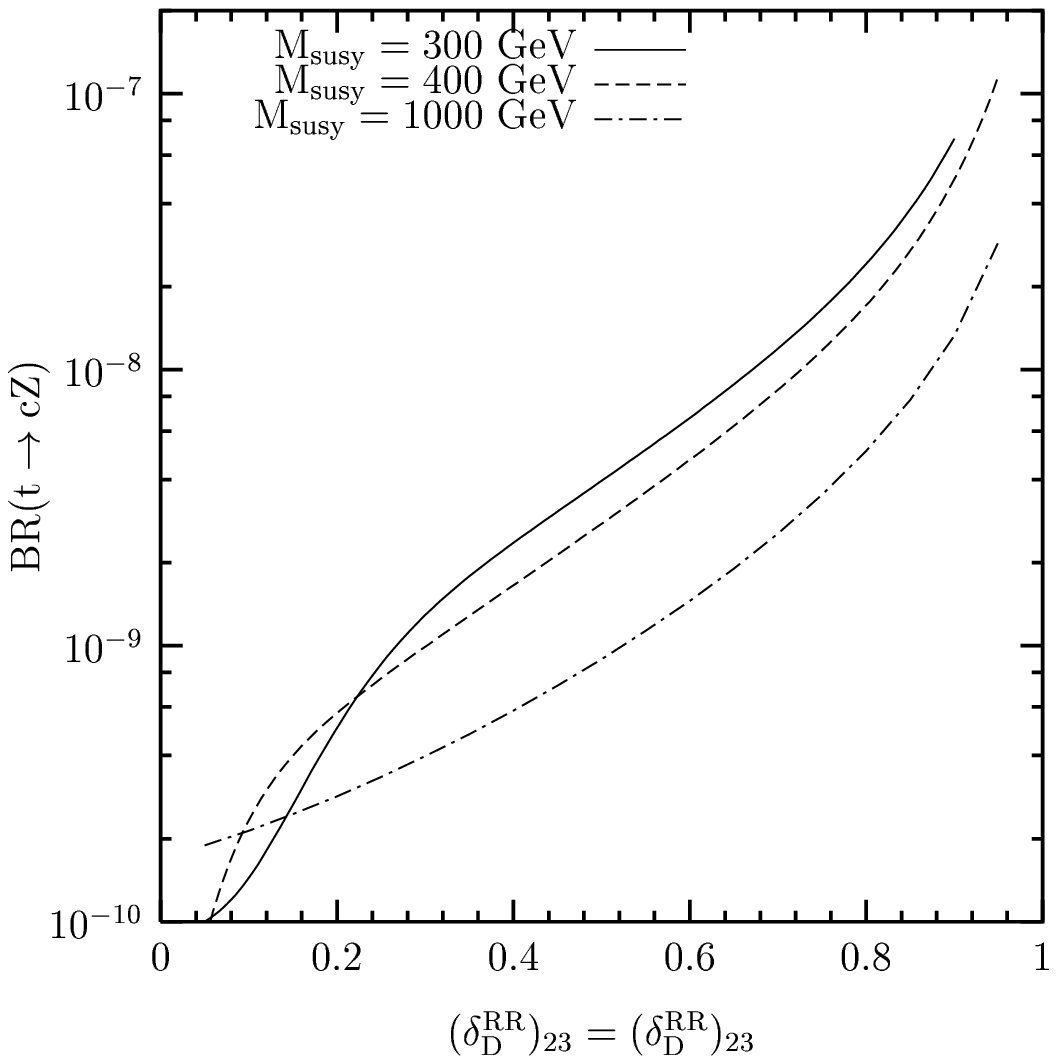}} \hspace{-4.95cm} \epsfxsize 4in {\epsfbox{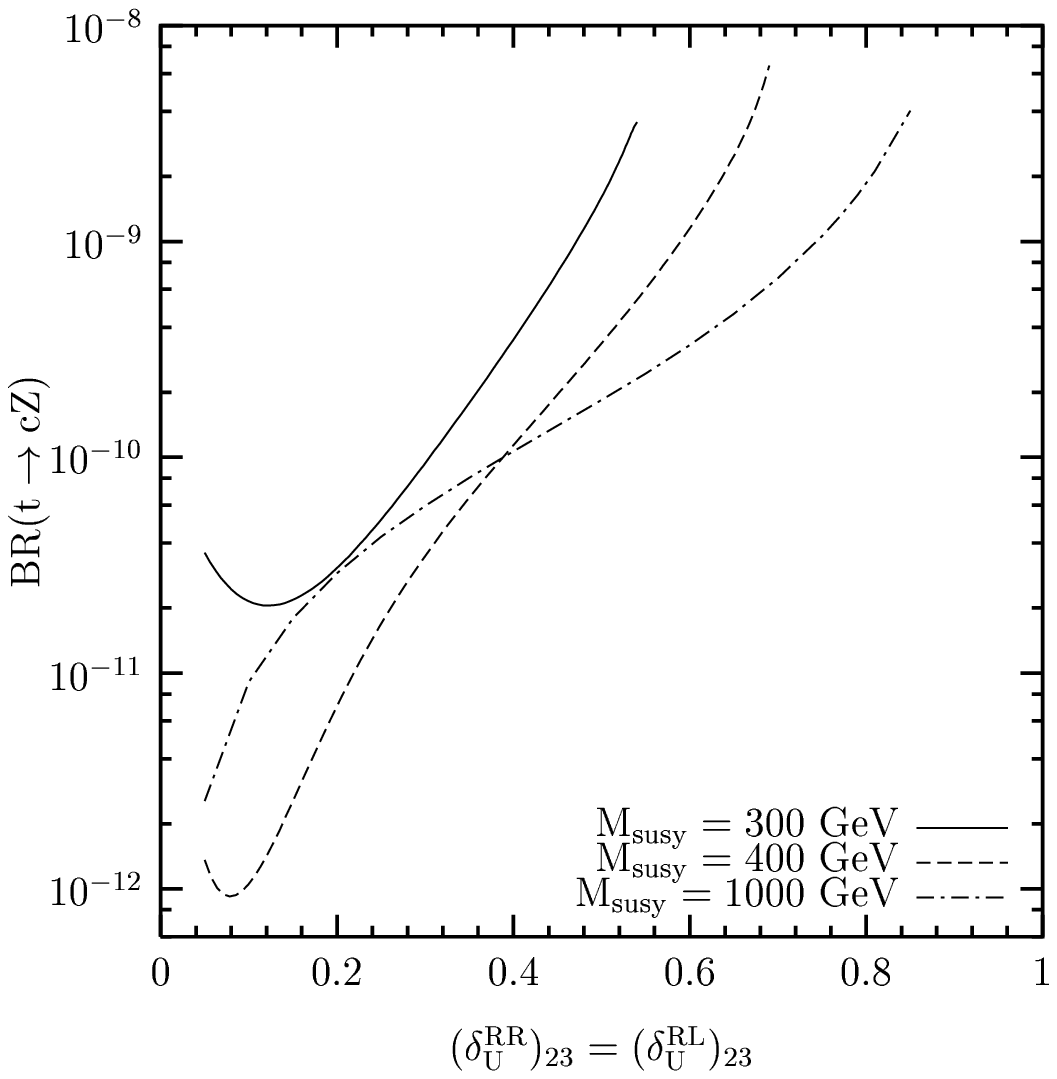}}}
\vskip -1.7in
    \caption{The same as Fig.\ref{fig:tczRR} but for the case where both $(\delta_{U(D)}^{RR})_{23}$ and $(\delta_{U(D)}^{RL})_{23}$ contribute, with the assumption $(\delta_{U(D)}^{RR})_{23}=(\delta_{U(D)}^{RL})_{23}$.}\label{fig:tczRRRL}
\end{figure}

Finally, we present the final decay channel, $t\to cZ$. Fig.~\ref{fig:tczRR} shows the branching ratio of $t\to cZ$ as a function of $(\delta_{U(D)}^{RR})_{23}$ when the rest of flavor changing parameters are set to be zero. The gluino contribution becomes enhanced one to more than four orders of magnitude  compared to the case where all flavor changing parameters are turned off.  A $10^{-5}$ branching ratio seems to be reachable. We can the discuss neutralino contributions and directly compare with the gluino contributions since, unlike the chargino contribution, they vary with respect to the same mixing parameter. The neutralino contribution is more than two orders of magnitude enhanced over the flavor diagonal case but still suppressed with respect to the gluino contribution. The effect in the chargino sector is tiny.

The RR+RL case, presented in Fig.~\ref{fig:tczRRRL}, is very similar to the above case for both the gluino and neutralino contributions but the mixing in the RR+RL down sector contributes to the chargino loop more effectively and increases its contribution to one order of magnitude with respect to the RR mixing case considered above. The chargino contribution, like in the flavor diagonal case, is still dominant over the neutralino one.

Consideration of flavor changing effects in either sectors is quite important for rare top decays $t\to c\,V~(V=g,\gamma,Z)$ not only for the expected enhancements, but also for revealing the details of the SUSY flavor changing mechanism.

Before closing this section, we would like to comment on the large 
$\tan\beta$ effects in both flavor diagonal and non-diagonal 
scenarios. As can be seen from the Feynman rules for gluino, the 
gluino loop for each decay is not sensitive to $\tan\beta$ value. 
Basically, the branching ratios are slightly bigger for $\tan\beta$ 
values smaller than $10$ and almost constant after $10$. This indeed 
makes the total branching ratios of the decays insensitive to 
$\tan\beta$. Since we discuss contributions from each loop 
separately, it is worth mentioning the $\tan\beta$ dependencies of 
chargino and neutralino loops. We comment on specifically the $t\to 
c\gamma$ decay, but these comments apply to other decay modes as 
well. In the flavor diagonal scenario, unlike neutralino case, 
$\tan\beta$ dependency of the chargino contribution to $t\to c\gamma$ 
is quite strong. If one scans the $\rm BR(t\to c\gamma)$ in 
$\tan\beta$ in the $0-50$ interval, the chargino contribution can 
reach $2\times 10^{-10}(3\times 10^{-11})$ for $\rm 
M_{susy}=300,\;400\;(1000)$ GeV, which is about 40 (30) times bigger 
than $\tan\beta=10$ case and three orders of magnitude bigger with 
respect to very small $\tan\beta$ values. For the neutralino case, 
the branching ratio of $t\to c\gamma$ is almost constant as 
$\tan\beta$ changes for $\rm M_{susy}=300,400$ GeV. As $\rm M_{susy}$ 
gets bigger, there is a slight increase in the branching ratio as 
$\tan\beta$ increases. So, for $\rm M_{susy}=1000$ GeV, we have one 
order of magnitude enhancement at maximum value of the branching 
ratio in the interval $\tan\beta\in(0,50)$ and five times bigger 
branching ratios when we compare the case for $\tan\beta=50$ with the 
one for $\tan\beta=10$. Even though the chargino contribution becomes 
larger for large $\tan\beta$, it is still quite suppressed with 
respect to the gluino contribution. For example, for $\tan\beta=50$, 
the chargino contribution is around $75;40;4$ times smaller for $\rm 
M_{susy}=300;400;1000$ GeV. Where the chargino contribution was one 
to two orders of magnitude smaller than the neutralino contribution 
for small $\tan\beta$ values ($\tan\beta\le 5$), it becomes more than 
one (two) order(s) bigger then the neutralino contribution for $\rm 
M_{susy}=300,\;400\;(1000)$ GeV  as $\tan\beta$ takes its maximum 
value considered here.

For the flavor non-diagonal case, we consider again as an example the 
$t\to c\gamma$ decay with mixing only in the RR sector. The gluino 
contribution is very similar to the flavor diagonal case. The 
chargino contribution is again most sensitive to $\tan\beta$ among 
the three contributions. The sensitivity in this case is slightly 
less, because we have assumed for simplicity the diagonal entries of 
the squark mass matrices identical to $\rm M_{susy}$. For the flavor 
mixing parameter $(\delta_D^{RR})_{23}=0.5$, the chargino 
contribution to the branching ratio becomes 35; 25; 10 times bigger 
for $\rm M_{susy}=300; 400;1000$ GeV when we compare the cases 
$\tan\beta=10$ and $\tan\beta=50$. For the neutralino case, the 
discussion is very similar to the flavor diagonal case with a reduced 
dependency on $\tan\beta$.  In summary, when we add up all 
contributions the branching ratio is not sensitive to the $\tan\beta$ 
parameter because the gluino loop which dominates all contributions 
and it is not sensitive to $\tan\beta$. When we look at the 
individual contributions to the branching ratio, the chargino 
contribution has the strongest dependency on $\tan\beta$ both in 
flavor diagonal and non-diagonal scenarios. Thus, our overall 
analysis is valid as one varies $\tan\beta$ in the phenomenologically 
relevant interval.


\section{Conclusion}\label{sec:conclusion}
We have presented a complete analysis of the two-body flavor violating decays of the top quark in a fully left-right supersymmetric model. The model has a left-right scalar quark sector, as well as a right-handed gaugino in both neutral and charged sectors. 

We have first evaluated the branching ratios in the flavor diagonal case, which corresponds to the constrained LRSUSY, where explicit flavor violating terms do not exist in the squark mass matrix, and the only source of flavor violation is the CKM matrix. The enhancement due to the right handed sector proves to be minimal, and top quark FCNC decay is somewhat below the detectable level at future collider experiments. We present the results for the gluino, chargino and neutralino contributions to the decays $t \rightarrow c g, c \gamma$ and $c Z$ explicitly and separately, unlike previous analyses. We confirm that branching ratios can reach $10^{-5}$ for the gluon, and $10^{-6}$ for the photon and $Z$, in agreement with previous results.

We then analysed the branching ratios for $t \rightarrow c V$ in the unconstrained LRSUSY, where flavor changing elements in the squark mass matrix are allowed to be arbitrarily large, though for the mixing between the second and third generation only. This mixing is rather unconstrained in the top-charm squark sector, and only very weakly constrained in the bottom-strange squark sector from $b$ decays. We analyse again separately contributions from gluino, chargino and neutralino to $t \rightarrow c g, c \gamma$ and $c Z$ and look at results for the branching ratios when flavor violation is driven by the right-right, right-left and left-right, or both, squark mixings. 
As expected, enhancements of the branching ratio could occur here, and we obtain, under the best circumstances, ${\rm BR}(t \rightarrow c g)$ close to $10^{-4}$, while ${\rm BR}(t \rightarrow c \gamma)$ and ${\rm BR}(t \rightarrow c Z)$ can reach $10^{-6}$ and $10^{-5}$ respectively. These enhancements are comparable to other studies in the unconstrained MSSM, so while $t$ decays are important as a probe of SUSY FCNC couplings, they do not appear to differentiate between the MSSM and LRSUSY, unless there are clear indications that it is the right-handed scalar quark sector which is responsible for flavor violation. In this case LRSUSY provides large enhancements, at least competitive with ones driven by the left-handed scalar quark mixings in MSSM. 

Comparing $t$ to $b$ decays, it appears that there the chargino contribution is more evident and sometimes comparable to the gluino; while this is never the case for $t\to cV$ and thus here the effect of the right-handed gaugino is relatively obscured.

\begin{acknowledgments}
This work was funded by NSERC of Canada under the grant number 0105354.
\end{acknowledgments}

\appendix
\section{Chargino and Neutralino Mixings}\label{sec:AppendixMasses}
In this appendix we present the chargino and neutralino mass matrices and the way of their diagonalizations.
\subsection{Chargino Mixing}\label{subsec:CMixing}

The terms relevant to the masses of charginos in the Lagrangian are
\begin{equation}
{\cal L}_C=-\frac{1}{2}(\psi^{+T}, \psi^{-T}) \left ( \begin{array}{cc}
                                                        0 & X^T \\
                                                        X & 0
                                                      \end{array}
                                              \right ) \left (
\begin{array}{c}
                                                               \psi^+ \\
                                                               \psi^-
                                                               \end{array}
                                                        \right ) + {\rm H.c.} \ ,
\end{equation}
where $\psi^{+T}=(-i \lambda^+_L, -i \lambda^+_R, \tilde{\phi}_{1u}^+,
\tilde{\phi}_{1d}^+, \tilde{\Delta}_R^+)$
and $\psi^{-T}=(-i \lambda^-_L, -i \lambda^-_R, \tilde{\phi}_{2u}^-,
\tilde{\phi}_{2d}^-, \tilde{\delta}_R^-)$, and
\begin{equation}
X=\left( \begin{array}{ccccc}
                            M_L & 0 & g_L
\kappa_u & 0 & 0
\\
           0 & M_R & g_R\kappa_u & 0
&\sqrt{2}g_Rv_{\delta_R}
\\
0 & 0 & 0 & -\mu & 0\\
    g_L\kappa_d & g_R \kappa_d &-\mu & 0 & 0
\\
       0 & \sqrt{2} g_R v_{\Delta_R} & 0 & 0 & -\mu
               \end{array}
         \right )
\end{equation}
where we have taken, for simplification, $\mu_{ij}=\mu$. Note that although there are six charginos, $\tilde{\Delta}^-_L$ and $\tilde{\delta}^+_L$ decouple from the spectrum because we have chosen the VEV's of $\Delta_L$ and $\delta_L$ to be zero.

 The chargino mass
eigenstates $\chi_i$ are obtained by
\begin{eqnarray}
\chi_i^+=V_{ij}\psi_j^+, \ \chi_i^-=U_{ij}\psi_j^-, \ i,j=1, \ldots 5,
\end{eqnarray}
with $V$ and $U$ unitary matrices satisfying
\begin{equation}
U^* X V^{-1} = X_D,
\end{equation}
The diagonalizing matrices $U^*$ and $V$ are obtained by
computing the eigenvectors corresponding
to the eigenvalues of $X X^{\dagger}$ and $X^{\dagger} X$, respectively.
to the eigenvalues of $X X^{\dagger}$ and $X^{\dagger} X$, respectively.

\subsection{Neutralino Mixing}\label{subsec:NMixing}

The terms relevant to the masses of neutralinos in the Lagrangian are
\begin{equation}
{\cal L}_N=-\frac{1}{2} {\psi^0}^T Z\, \psi^0  + {\rm H.c.} \ ,
\end{equation}
where $\psi^{0\,T}=(-i \lambda_L^0, -i \lambda_R^0, -i \lambda_V,
\tilde{\phi}_{1u}^0, \tilde{\phi}^0_{2d},\tilde{\Delta}_R^0,
\tilde{\delta}_R^0,
\tilde{\phi}_{1d}^0, \tilde{\phi}^0_{2u} ) $,
and
\begin{equation}
\displaystyle
Z=\left( \begin{array}{ccccccccc}
               M_L & 0 & 0 & \frac{g_L \kappa_u}{\sqrt{2}} &
-\frac{g_L \kappa_d}{\sqrt{2}} &0 & 0 & 0 & 0 \\
          0 & M_R & 0 & \frac{g_R \kappa_u}{\sqrt{2}} & -\frac{g_R
\kappa_d}{\sqrt{2}} & -\sqrt{2}g_Rv_{\Delta_R} & -\sqrt{2}g_R v_{\delta_R} &
0 & 0
\\
               0 & 0 & M_V & 0 & 0 & 2\sqrt{2}g_V v_{\Delta_R} & 2\sqrt{2} g_V
v_{\delta_R} & 0 &0
\\
               \frac{g_L \kappa_u}{\sqrt{2}} & \frac{g_R \kappa_u}{\sqrt{2}}
&
0 & 0 & -\mu & 0 & 0 & 0 & 0  \\
               -\frac{g_L \kappa_d}{\sqrt{2}} & -\frac{g_R \kappa_d}{\sqrt{2}} & 0 &-\mu
& 0 & 0 & 0 & 0 & 0 \\
              0&-\sqrt{2}g_Rv_{\Delta_R} &2\sqrt{2}g_Vv_{\Delta_R}& 0& 0& 0 &-\mu & 0 & 0
\\
  0 & -\sqrt{2}g_R v_{\delta_R} & 2\sqrt{2}g_V v_{\delta_R} & 0 & 0 & -\mu & 0 &0 &0
   \\
               0 & 0 & 0 & 0 & 0 & 0 & 0 & 0 &
-\mu \\
  0 & 0 & 0 & 0 & 0 & 0 & 0 & -\mu & 0
               \end{array}
         \right ).
\end{equation}
As in the case of charginos, $\tilde{\Delta}^0_L$ and $\tilde{\delta}^0_L$ decouple from the spectrum because we have chosen the VEV's of $\Delta_L$ and $\delta_L$ to be zero.
The mass eigenstates are defined by
\begin{equation}
\chi^0_i=N_{ij} \psi^0_j \ (i,j=1,2, \ldots 9),
\end{equation}
where $N$ is a unitary matrix chosen such that
\begin{equation}
N Z N^T = Z_D,
\label{equationN}
\end{equation}
and $Z_D$ is a diagonal matrix with non-negative entries.\footnote{The positivity of the entries of $Z_D$ can be achieved by multiplying some of the rows of $N$ with $i$. This freedom comes from the fact that in order to determine $N$ the square of Eq.~(\ref{equationN}) needs to be considered. We have a similar situation for the chargino sector. For example, see \cite{Haber:1984rc} for details in the MSSM.} 

One can further switch to a basis involving the photino, left and right zino states
$$\psi^{0\prime\,T}=(-i \lambda_{\tilde \gamma}, -i \lambda_{\tilde Z_L}, -i \lambda_{\tilde Z_R},
\tilde{\phi}_{1u}^0, \tilde{\phi}^0_{2d},\tilde{\Delta}_R^0,
\tilde{\delta}_R^0,
\tilde{\phi}_{1d}^0, \tilde{\phi}^0_{2u} ), $$
where the photino and left and right zino states are defined as
\begin{eqnarray}
\left(\begin{array}{c}
\lambda_{\tilde \gamma}\\
\lambda_{\tilde Z_L}\\
\lambda_{\tilde Z_R}\end{array}\right)=\left( \begin{array}{ccc}
\sin\theta_{\rm W} & \sin\theta_{\rm W}                             & \sqrt{\cos 2\theta_{\rm W}}                     \\
\cos\theta_{\rm W} & -\sin\theta_{\rm W}\tan\theta_{\rm W}          & -\sqrt{\cos 2\theta_{\rm W}}\tan\theta_{\rm W} \\
  0                & \sqrt{\cos 2\theta_{\rm W}}\sec\theta_{\rm W}  & -\tan\theta_{\rm W}     
\end{array}\right)\left(\begin{array}{c}
\lambda_L^0\\
\lambda_R^0\\
\lambda_V\end{array}\right).
\end{eqnarray}
The mass matrix in the above basis becomes
\begin{equation}
\displaystyle
Z=\left( \begin{array}{ccccccccc}
               m_{\tilde\gamma} & m_{\tilde\gamma \tilde Z_L} & m_{\tilde\gamma \tilde Z_R} & \sqrt{2}e\kappa_u& 
-\sqrt{2}e\kappa_d & \sqrt{2}e v_{\Delta_R} & \sqrt{2}e v_{\delta_R} & 0 & 0 \\
  m_{\tilde\gamma \tilde Z_L} & m_{\tilde Z_L} & m_{\tilde Z_L \tilde Z_R} & A_u & - A_d &
 C_{\Delta_R}& C_{\delta_R} & 0 & 0  \\
 m_{\tilde\gamma \tilde Z_R} & m_{\tilde Z_L \tilde Z_R} & m_{\tilde Z_R} & E_u & -E_d & 
 B_{\Delta_R} & B_{\delta_R} & 0 &0 \\
               \sqrt{2}e\kappa_u & A_u & E_u & 0 & -\mu & 0 & 0 & 0 & 0  \\
               -\sqrt{2}e\kappa_d & -A_d & -E_d & -\mu & 0 & 0 & 0 & 0 & 0 \\
          \sqrt{2}ev_{\Delta_R}  & C_{\Delta_R} & B_{\Delta_R} & 0 & 0& 0 & -\mu & 0 & 0 \\     
            \sqrt{2}ev_{\delta_R}  & C_{\delta_R} & B_{\delta_R} & 0 & 0& -\mu & 0 & 0 & 0\\
         0 & 0 & 0 & 0 & 0 & 0 & 0 & 0 &-\mu \\
         0 & 0 & 0 & 0 & 0 & 0 & 0 & -\mu & 0 \\      
               \end{array}
         \right ).
\end{equation}
with the functions
\begin{eqnarray}
m_{\tilde\gamma}&&\hspace*{-0.3cm}=\left(M_L+M_R\right)\sin^2\theta_{\rm W}+M_V \cos2\theta_W,\nonumber\\
m_{\tilde\gamma Z_L}&&\hspace*{-0.3cm}=M_L\cos\theta_{\rm W}\sin\theta_{\rm W}-\left(M_R\sin^2\theta_{\rm W}+M_V\cos2\theta_{\rm W}\right)\tan\theta_{\rm W},\nonumber\\
m_{\tilde\gamma Z_R}&&\hspace*{-0.3cm}=\left(M_R-M_V\right)\sqrt{\cos2\theta_{\rm W}}\tan\theta_{\rm W},\nonumber\\
m_{\tilde Z_L}&&\hspace*{-0.3cm}=M_L\cos^2\theta_{\rm W}+M_R\sin^2\theta_{\rm W}\tan^2\theta_{\rm W}+M_V\cos2\theta_{\rm W}\tan^2\theta_{\rm W},\nonumber\\
m_{\tilde Z_L \tilde Z_R}&&\hspace*{-0.3cm}=-\left(M_R-M_V \right)\sqrt{\cos2\theta_{\rm W}}\tan^2\theta_{\rm W},\nonumber\\
m_{\tilde Z_R}&&\hspace*{-0.3cm}=M_R\left(1-\tan^2\theta_{\rm W}\right)+M_V\tan^2\theta_{\rm W},\nonumber\\
A_d&&\hspace*{-0.3cm}=\sqrt{2}e\kappa_d \cot2\theta_{\rm W}\,,\;\;\;\;\;\;\;\;A_u=\sqrt{2}e\kappa_u \cot2\theta_{\rm W},\nonumber\\
B_{\Delta_R}&&\hspace*{-0.3cm}=\frac{-\sqrt{2}g_Lv_{\Delta_R}}{\cos\theta_{\rm W}\sqrt{\cos2\theta_{\rm W}}}\,,\;\;\;\;B_{\delta_R}=\frac{-\sqrt{2}g_Lv_{\delta_R}}{\cos\theta_{\rm W}\sqrt{\cos2\theta_{\rm W}}},\nonumber\\
C_{\Delta_R}&&\hspace*{-0.3cm}=-\sqrt{2}e v_{\Delta_R} \tan\theta_{\rm W}\,,\;\;\;\;\;\;\;\;C_{\delta_R}=-\sqrt{2}e v_{\delta_R} \tan\theta_{\rm W},\nonumber\\
E_d&&\hspace*{-0.3cm}=\frac{g_L\sqrt{\cos2\theta_{\rm W}}}{\sqrt{2}\cos\theta_{\rm W}}\kappa_d\,,\;\;\;\;\;\;\;\;E_u=\frac{g_L\sqrt{\cos2\theta_{\rm W}}}{\sqrt{2}\cos\theta_{\rm W}}\kappa_u,\nonumber\\
\end{eqnarray}
where we have used $g_L=g_R$ and $g_V=e/\sqrt{\cos2\theta_{\rm W}}$. Furthermore, the unitary matrix $N$ can be expressed in the rotated basis as
\begin{eqnarray}
\left(\begin{array}{c}
N^\prime_{j1}\\
N^\prime_{j2}\\
N^\prime_{j3}\end{array}\right)&&\hspace*{-0.3cm}=\left( \begin{array}{ccc}
\sin\theta_{\rm W} & \sin\theta_{\rm W}                             & \sqrt{\cos 2\theta_{\rm W}}                     \\
\cos\theta_{\rm W} & -\sin\theta_{\rm W}\tan\theta_{\rm W}          & -\sqrt{\cos 2\theta_{\rm W}}\tan\theta_{\rm W} \\
  0                & \sqrt{\cos 2\theta_{\rm W}}\sec\theta_{\rm W}  & -\tan\theta_{\rm W}     
\end{array}\right)\left(\begin{array}{c}
N_{j1}\\
N_{j2}\\
N_{j3}\end{array}\right),\nonumber\\
N^\prime_{jk}&&\hspace*{-0.3cm}=N_{jk},\;\;\;\;\;\;j=1,...,9\,,\;\;\;k=4,...,9\,.
\end{eqnarray}

\section{Feynman Rules}\label{sec:AppendixFynmVert}
In this appendix, we give the Feynman rules involving SUSY particles. 
The ones for the SM particles can be found in many textbooks.
\begin{itemize}
    \item Gauge boson-squark-squark interaction
\vskip -1.0cm
\begin{figure}[h]
\begin{center}
\begin{picture}(200,90)(0,0)
\hspace*{-3cm}
\SetWidth{0.8}  
\Vertex(40,0){2}
\DashArrowLine(40,0)(0,40){3}
\DashArrowLine(0,-40)(40,0){3}
\Photon(40,0)(80,0){3}{5}
\Text(85,25)[l]{$g^{a\prime}_{\mu}$}
\Text(85,0)[l]{$\gamma_\mu$}
\Text(85,-25)[l]{$Z_\mu$}
\Text(6,-49)[c]{\bf $\tilde{q}_{h}$} 
\Text(6,47)[c]{\bf $\tilde{q}_{k}$}
\Text(12,-18)[c]{$q$}
\Text(12,17)[c]{$q^\prime$}
\Text(150,25)[l]{	
$:-ig_s T^{a\prime}\left(q+q^\prime\right)_\mu \delta_{hk}$}
\Text(150,0)[l]{
$:-ieQ_q\left(q+q^\prime\right)_\mu\delta_{hk}$}
\Text(150,-25)[l]{
$\displaystyle :-\frac{ig \left(q+q^\prime\right)_\mu}{\cos\theta_{\rm W}}\left[T_{3Q}\sum_{i=1}^{3}\Gamma_{QL}^{*hi}\Gamma_{QL}^{ki}-Q_d\sin^2\theta_{\rm W}\delta_{hk}\right]$}
\end{picture}
\end{center}
\vskip 1cm
\end{figure}
    \item Gluon-gluino-gluino interaction
\vskip -1cm
\begin{figure}[h]
\begin{center}
\begin{picture}(200,90)(0,0)
\hspace*{-3cm}
\SetWidth{0.8}  
\Vertex(40,0){2}
\ArrowLine(40,0)(0,40)
\ArrowLine(0,-40)(40,0)
\Photon(40,0)(80,0){3}{5}
\Text(85,0)[l]{$g^{a\prime}_{\mu}$}
\Text(6,-46)[c]{\bf $\tilde{g}^{c^\prime}$} 
\Text(6,50)[c]{\bf $\tilde{g}^{b^\prime}$}
\Text(150,0)[l]{
$:-ig_sf_{a^\prime b^\prime c^\prime}\gamma_\mu$}
\end{picture}
\end{center}
\end{figure}
\vskip 1cm
    \item Gauge boson-chargino-chargino interaction
\vskip -1cm
\begin{figure}[h]
\begin{center}
\begin{picture}(200,90)(0,0)
\hspace*{-3cm}
\SetWidth{0.8}  
\Vertex(40,0){2}
\ArrowLine(40,0)(0,40)
\ArrowLine(0,-40)(40,0)
\Photon(40,0)(80,0){3}{5}
\Text(85,10)[l]{$\gamma_\mu$}
\Text(85,-10)[l]{$Z_\mu$}
\Text(6,-46)[c]{\bf $\chi_b^\pm$} 
\Text(6,50)[c]{\bf $\chi_a^\mp$}
\Text(150,10)[l]{
$:-ie\gamma_\mu$}
\Text(150,-10)[l]{
$:-\frac{ig}{\cos\theta_{\rm W}}\gamma_\mu\left[O^{ab}_LP_L+O^{ab}_RP_R\right]$}
\Text(100,-60)[l]{$O^{ab}_L=-V_{a1}V^*_{b1}-\frac{1}{2}V_{a3}V^*_{b3}-\frac{1}{2}V_{a4}V^*_{b4}+\delta_{ab}\sin^2\theta_{\rm W}$}
\Text(100,-80)[l]{$O^{ab}_R=-U^*_{a1}U_{b1}-\frac{1}{2}U^*_{a3}U_{b3}-\frac{1}{2}U^*_{a4}U_{b4}+\delta_{ab}\sin^2\theta_{\rm W}$}
\end{picture}
\end{center}
\end{figure}
\vskip 2.5cm
    \item Z-neutralino-neutralino interaction
\vskip -2.5cm
\begin{figure}[h]
\begin{center}
\begin{picture}(200,90)(0,0)
\hspace*{-3cm}
\SetWidth{0.8}  
\Vertex(40,0){2}
\ArrowLine(40,0)(0,40)
\ArrowLine(0,-40)(40,0)
\Photon(40,0)(80,0){3}{5}
\Text(85,0)[l]{$Z_\mu$}
\Text(6,-46)[c]{\bf $\chi_m^0$} 
\Text(6,50)[c]{\bf $\chi_n^0$}
\Text(150,0)[l]{
$:\frac{ig}{\cos\theta_{\rm W}}\gamma_\mu\left[O^{\prime nm}_LP_L+O^{\prime nm}_RP_R\right]$}
\Text(100,-60)[l]{$O^{\prime nm}_L=\frac{1}{2}\left(N^{\prime}_{n4}N^{\prime*}_{m4}-N^{\prime}_{n5}N^{\prime*}_{m5}+N^{\prime}_{n6}N^{\prime*}_{m6}-N^{\prime}_{n7}N^{\prime*}_{m7}\right)$}
\Text(100,-80)[l]{$O^{\prime nm}_R=-O^{\prime* nm}_L$}
\end{picture}
\end{center}
\vspace*{-5cm}
\end{figure}
\vspace*{6.9cm}
    \item $\tilde g,\chi^0,\chi^+$-quark-squark interaction
\vskip -1.0cm
\begin{figure}[h]
\begin{center}
\begin{picture}(200,90)(0,0)
\hspace*{-3cm}
\SetWidth{0.8}  
\Vertex(40,0){2}
\ArrowLine(40,0)(80,0)
\DashArrowLine(40,0)(0,40){3}
\DashArrowLine(0,-40)(40,0){3}
\Text(85,24)[l]{$g^{a\prime}_{\mu}$}
\Text(85,0)[l]{$\chi^0_n$}
\Text(85,-24)[l]{$\left(\chi^+_a\right)$}
\Text(1,23)[c]{\bf $(\tilde d_k)$}
\Text(6,-46)[c]{\bf $u_i$} 
\Text(6,50)[c]{\bf $\tilde u_k$}
\Text(115,24)[l]{	
$:-i\sqrt{2}g_sT^{a^\prime}\left(\Gamma_{UL}^{*ki}P_L-\Gamma_{UR}^{ki}P_R\right)$}
\Text(115,0)[l]{
$:-ig\left[\left(\sqrt{2}G_{UL}^{[n,k,i]}+H_{UR}^{[n,k,i]}\right)P_L-\left(\sqrt{2}G_{UR}^{[n,k,i]}-H_{UL}^{[n,k,i]}\right)P_R\right]$}
\Text(115,-24)[l]{
$:-ig\left[\left(G_{DL}^{[a,k,i]}-H_{DR}^{[a,k,i]}\right)P_L+\left(G_{DR}^{[a,k,i]}-H_{DL}^{[a,k,i]}\right)P_R\right]$}
\end{picture}
\end{center}
\end{figure}
\vspace*{1cm}
\hspace{-1.0cm} where
\begin{eqnarray}
\hspace*{-1cm}G_{UL}^{[n,k,i]}&&\hspace*{-0.3cm}=\left[Q_u \sin\theta_{\rm W}N^{\prime*}_{n1}+\frac{1}{\cos\theta_{\rm W}}\left(T_{3u}-Q_u\sin^2\theta_{\rm W}\right)N^{\prime*}_{n2}-\frac{Q_u+Q_d}{2}\frac{\sin\theta_{\rm W}\tan\theta_{\rm W}}{\sqrt{\cos2\theta_{\rm W}}}N^{\prime*}_{n3}\right]\Gamma_{UL}^{ki},\nonumber\\
\hspace*{-1cm}G_{UR}^{[n,k,i]}&&\hspace*{-0.3cm}=\left[Q_u \sin\theta_{\rm W}N^{\prime*}_{n1}-\frac{Q_u\sin^2\theta_{\rm W}}{\cos\theta_{\rm W}}N^{\prime*}_{n2}+\frac{\sqrt{\cos2\theta_{\rm W}}}{\cos\theta_{\rm W}}\left(T_{3u}-\frac{Q_u+Q_d}{2}\frac{\sin^2\theta_{\rm W}}{\sqrt{\cos2\theta_{\rm W}}}\right)N^{\prime*}_{n3}\right]\Gamma_{UR}^{ki},\nonumber\\
\hspace*{-1cm}H_{UL}^{[n,k,i]}&&\hspace*{-0.3cm}=\frac{1}{\sqrt{2}m_{\rm W}}\left[\frac{m_{u_i}}{\sin\beta}N^{\prime}_{n5}+\frac{m_{d_i}}{\cos\beta}N^{\prime}_{n7}\right]\Gamma_{UL}^{ki},\nonumber\\
\hspace*{-1cm}H_{UR}^{[n,k,i]}&&\hspace*{-0.3cm}=\frac{1}{\sqrt{2}m_{\rm W}}\left[\frac{m_{u_i}}{\sin\beta}N^{\prime*}_{n5}+\frac{m_{d_i}}{\cos\beta}N^{\prime*}_{n7}\right]\Gamma_{UR}^{ki},\nonumber\\
\hspace*{-1cm}G_{DL}^{[a,k,i]}&&\hspace*{-0.3cm}=\sum_{j=1}^{3}K_{\rm CKM}^{*ij}\Gamma_{DL}^{kj}U^*_{a1},\nonumber\\
\hspace*{-1cm}G_{DR}^{[a,k,i]}&&\hspace*{-0.3cm}=\sum_{j=1}^{3}K_{\rm CKM}^{*ij}\Gamma_{DR}^{kj}V_{a2},\nonumber\\
\hspace*{-1cm}H_{DL}^{[a,k,i]}&&\hspace*{-0.3cm}=\frac{1}{\sqrt{2}m_{\rm W}}\sum_{j=1}^{3}\left[\frac{m_{u_i}}{\sin\beta}V_{a4}+\frac{m_{d_j}}{\cos\beta}V_{a3}\right]K_{\rm CKM}^{*ij}\Gamma_{DL}^{kj},\nonumber\\
\hspace*{-1cm}H_{DR}^{[a,k,i]}&&\hspace*{-0.3cm}=\frac{1}{\sqrt{2}m_{\rm W}}\sum_{j=1}^{3}\left[\frac{m_{u_i}}{\sin\beta}U^*_{a4}+\frac{m_{d_j}}{\cos\beta}U^*_{a3}\right]K_{\rm CKM}^{*ij}\Gamma_{DR}^{kj}.\nonumber
\end{eqnarray}
\end{itemize}


\end{document}